%% file: SOS.tex
\pdfoutput=1

\documentclass[11pt]{article}

\usepackage{acl}
\usepackage{times}
\usepackage{latexsym}
\usepackage{graphicx}
\usepackage{url}
\usepackage{amsmath} 
\usepackage{algorithm}
\usepackage{algpseudocode}
\usepackage{colortbl}
\usepackage{pgf}
\usepackage{color}
\usepackage{xcolor}
\usepackage{tabularray}
\usepackage[T1]{fontenc}

\usepackage[utf8]{inputenc}

\usepackage{microtype}
\usepackage{marvosym}
\usepackage{multirow}
\usepackage{graphicx}
\usepackage{subcaption}
\usepackage{booktabs} 

\definecolor{myblue}{RGB}{137, 196, 225} 
\definecolor{myred}{RGB}{149, 117, 222} 
\definecolor{mywhite}{RGB}{248, 246, 244}

\newcommand{\ApplyGradient}[2]{%
    \pgfmathsetmacro{\diff}{#2-#1}
    \pgfmathsetmacro{\PercentColor}{abs(\diff)/25*100}
    \pgfmathsetmacro{\PercentColor}{min(\PercentColor,100)} 
    \ifdim\diff pt > 0pt
        \edef\x{\noexpand\cellcolor{myred!\PercentColor!white}}\x #2%
    \else
        \edef\x{\noexpand\cellcolor{myblue!\PercentColor!white}}\x #2%
    \fi
}

\newcommand{\ApplyGradientd}[2]{%
    \pgfmathsetmacro{\diff}{#2-#1}
    \pgfmathsetmacro{\PercentColor}{abs(\diff)/60*100}
    \pgfmathsetmacro{\PercentColor}{min(\PercentColor,100)} 
    \ifdim\diff pt > 0pt
        \edef\x{\noexpand\cellcolor{myred!\PercentColor!white}}\x #2%
    \else
        \edef\x{\noexpand\cellcolor{myblue!\PercentColor!white}}\x #2%
    \fi
}

\title{Spiral of Silence: How is Large Language Model Killing Information Retrieval?---A Case Study on Open Domain Question Answering}

\author{
$\text{Xiaoyang Chen}^{1,2} \text{, Ben He}^{1,2}\textsuperscript{\Letter} \text{, Hongyu Lin}^{2}\textsuperscript{\Letter} \text{, Xianpei Han}^{2}\text{,}$ \\ $\text{\textbf{Tianshu Wang}}^{2,3}
\text{\textbf{, Boxi Cao}}^{2,1} \text{\textbf{, Le Sun}}^{2}\textsuperscript{\Letter} \text{\textbf{, Yingfei Sun}}^{1}\textsuperscript{\Letter}$\\
  $^{1}\text{University of Chinese Academy of Sciences}$ \\
  $^{2}\text{Chinese Information Processing Laboratory, Institute of Software, Chinese Academy of Sciences}$ \\
  $^{3}\text{Hangzhou Institute for Advanced Study, University of Chinese Academy of Sciences}$\\
  \texttt{chenxiaoyang19@mails.ucas.ac.cn}, 
  \texttt{\{benhe, yfsun\}@ucas.ac.cn} \\
  \texttt{\{hongyu, xianpei, tianshu2020, boxi2020, sunle\}@iscas.ac.cn} \\
  }
\begin{document}
\maketitle
\begin{abstract}
The practice of Retrieval-Augmented Generation (RAG), which integrates Large Language Models (LLMs) with retrieval systems, has become increasingly prevalent. However, the repercussions of LLM-derived content infiltrating the web and influencing the retrieval-generation feedback loop are largely uncharted territories. In this study, we construct and iteratively run a simulation pipeline to deeply investigate the short-term and long-term effects of LLM text on RAG systems. Taking the trending Open Domain Question Answering (ODQA) task as a point of entry, our findings reveal a potential digital ``Spiral of Silence'' effect, with LLM-generated text consistently outperforming human-authored content in search rankings, thereby diminishing the presence and impact of human contributions online. This trend risks creating an imbalanced information ecosystem, where the unchecked proliferation of erroneous LLM-generated content may result in the marginalization of accurate information. We urge the academic community to take heed of this potential issue, ensuring a diverse and authentic digital information landscape.\footnote{We release the resources at \url{https://github.com/VerdureChen/SOS-Retrieval-Loop}}
\end{abstract}

\input{parts/introduction}
\input{parts/related_works}
\input{parts/pipeline}

\input{parts/experiments}
\input{parts/results_effects}
\input{parts/analysis}
\section{Conclusion}
In this study, we initiate our research from empirical observations, aiming to investigate the implications of progressively integrating LLM-generated text into RAG systems. 
To this end, We employ the ODQA task as a case study to examine both the immediate and extended impacts of LLM text on these systems. 
Our simulation has revealed the emergence of a ``Spiral of Silence'' effect, suggesting that without appropriate intervention, human-generated content may progressively diminish its influence within RAG systems. 
Further investigation into this phenomenon reveals that unchecked accumulation of erroneous LLM-generated information could lead to the overlooking of correct information by IR systems, resulting in harm. We urge the academic community to be vigilant and take measures to prevent the potential misuse of LLM-generated data.
\input{parts/limit}
\section*{Acknowledgements}
This work is supported by the National Natural Science Foundation of China (62122077/62272439/62106251), Beijing Municipal Science and Technology Project (No. Z231100010323002), and the Fundamental Research Funds for the Central Universities.


\bibliography{custom}
\bibliographystyle{acl_natbib}
\clearpage
\appendix
\input{parts/appendix}

\end{document}

%% file: parts/introduction.tex
\section{Introduction}
The integration of Large Language Models (LLMs)~\citep{openai_chatgpt,DBLP:journals/corr/abs-2303-08774,DBLP:journals/corr/abs-2302-13971,gemini} is reshaping the online information landscape, making text generation easier, increasing content production, enhancing personalized knowledge assistance, and enabling advanced fake news creation.
\citet{schick2020deep} suggest that by 2026, synthetic content could dominate up to 90\% of the web. CounterCloud\footnote{\url{https://countercloud.io/?page_id=307}} shows that a single developer can create an AI fake news factory cheaply and convincingly. 
AI-driven content generation is rapidly becoming commonplace, impacting how content is produced and shared~\citep{DBLP:journals/corr/abs-2301-04246,DBLP:conf/emnlp/PanPCNKW23,DBLP:conf/acl/DaiLZ0HL23}. These developments pose novel challenges and opportunities for information retrieval (IR) and generation, especially for Retrieval-Augmented Generation (RAG) systems, which combine both capabilities~\citep{REALM,RAG,Retro,Atlas}. As text produced by large language models continues to flood the internet and is indexed by search systems, the enduring effects of such text on the retrieval-generation process grow more ambiguous, and the future landscape of the information environment is yet to be determined.

\begin{figure}[t]
    \centering
    \setlength{\belowcaptionskip}{-0.4cm}
    \includegraphics[width=1\linewidth]{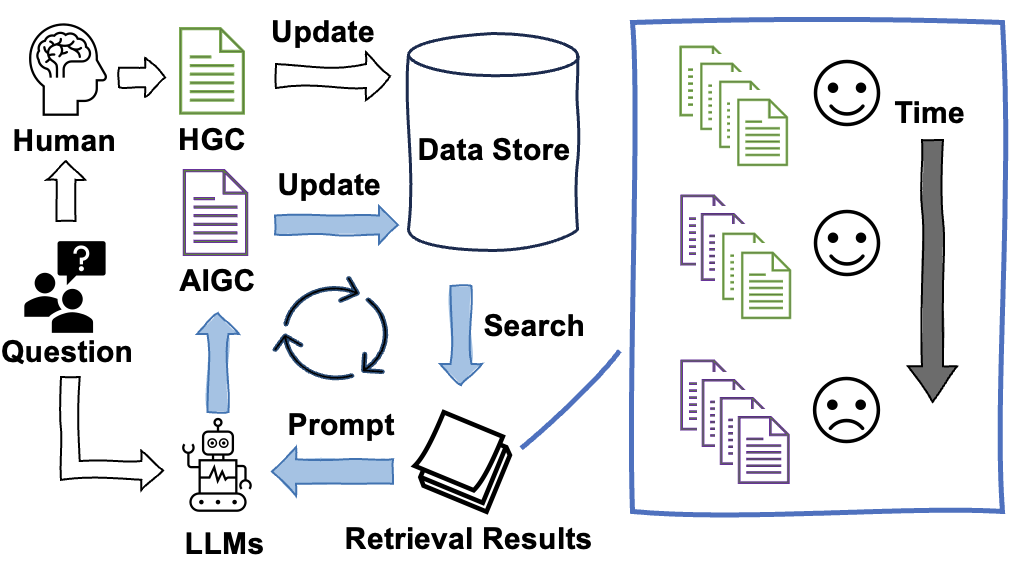}
    \caption{The evolution of RAG systems after introducing LLM-generated texts, where the ``Spiral of Silence'' effect gradually emerges.}
    \label{fig:spiral}
\end{figure}
In our research, we focus on the \textbf{effects of LLM-generated text on RAG systems}. 
As shown in Figure~\ref{fig:spiral}, we construct a \textbf{pipeline that simulates the continuous influx of LLM-generated text into web datasets} and assess its impact on the performance of RAG through iterative runs. 
To evaluate the RAG performance in the simulation process, we adopt the Open Domain Question Answering (ODQA) task as our evaluative benchmark due to its recent surge in research popularity as an effective test of both retrieval accuracy and generation quality~\citep{DBLP:conf/emnlp/PanPCNKW23,DBLP:journals/corr/abs-2309-01431}.
We employ widely used retrieval and re-ranking methods to supply the context necessary for LLMs to generate answer documents. Upon evaluating these documents, we integrate them into the text corpus for subsequent retrieval-generation cycles. This process is repeated multiple times to monitor and assess the emerging patterns. Experimental results show that LLM-generated text has an immediate effect on RAG systems, generally improving retrieval outcomes while producing varied effects on QA performance. However, over the long term, a marked decrease in retrieval effectiveness emerges, while the QA performance remains unaffected. 

Further examination reveals a bias in search systems towards LLM-generated texts, which consistently rank higher than human-written content. As LLM-generated texts increasingly dominate the search results, the visibility and influence of human-authored web content diminish, fostering a digital ``\textbf{Spiral of Silence}'' effect. This effect aptly explains what we observe in our simulations and reveals the potential negative impact of LLM-generated texts on the information ecosystem: while LLM-generated texts sometimes provide a more effective IR experience in the short term, in the long term they may lead to the invisibility of human-authored content, the homogenization of search results, and the inaccessibility of certain accurate information, thereby adversely affecting public knowledge acquisition and decision-making.

The contributions of this paper are threefold: 1) We propose an iterative pipeline to investigate the short-term and long-term impacts of LLM-generated text on RAG systems. 2) We study the potential emergence of a ``Spiral of Silence'' phenomenon within RAG systems. 3) We analyze the implications of this phenomenon, offering a new perspective on the dynamic interplay between LLM-generated content and RAG systems. 


%% file: parts/related_works.tex
\section{Related Works}
\textbf{Retrieval Augmented Generation.} 
RAG systems have been extensively analyzed, demonstrating retrieval's role in enhancing language model efficacy~\citep{REALM,RAG,Retro,Atlas,DBLP:journals/corr/abs-2302-00083}.
These systems also curtail LLMs' hallucinations during text generation~\citep{DBLP:journals/csur/JiLFYSXIBMF23,DBLP:journals/corr/abs-2311-05232} and reduce knowledge obsolescence~\citep{DBLP:journals/corr/abs-2301-00303}. 
Applied in ODQA~\citep{DBLP:conf/eacl/IzacardG21,DBLP:conf/acl/TrivediBKS23,DBLP:conf/emnlp/LiuLSJLLW23} and other tasks~\citep{DBLP:conf/naacl/CaiWBTLLS19,DBLP:conf/iclr/Zhou0XJN23}, current research explores LLMs' output accuracy against specific contexts~\citep{DBLP:journals/corr/abs-2307-16877}, robustness to extraneous information~\citep{DBLP:journals/corr/abs-2309-01431}, and the effects of output integration strategies~\citep{DBLP:journals/corr/abs-2308-12574}. 
Our study aims to provide a novel perspective to observe and predict the potential trajectory and impact of its future development.

\textbf{Effects of AIGC.}
Advances in Artificial Intelligence Generated Content (AIGC) have significantly impacted society and technology.
LLMs facilitate creating content to combat misinformation~\citep{DBLP:conf/mm/XuFK23,DBLP:journals/corr/abs-2311-05656} but can also produce damaging content~\citep{DBLP:journals/corr/abs-2310-06987}. The potential biases and discrimination in AIGC have garnered widespread attention~\citep{DBLP:conf/icml/LiangWMS21,DBLP:journals/corr/abs-2301-12867}. ~\citet{DBLP:journals/corr/abs-2305-17493} and ~\citet{DBLP:journals/corr/abs-2307-01850} show that LLMs trained on self-generated data degrade without fresh real-world input. ~\citet{DBLP:conf/emnlp/PanPCNKW23} investigated the impact of erroneous information generated by LLMs on ODQA systems. ~\citet{DBLP:journals/corr/abs-2310-20501} indicated that AI-modified texts might rank higher in search results, potentially affecting the fairness of those outcomes. Our research aims to further explore the short-term and long-term effects on RAG systems when AIGC text is continuously integrated into the search system's datasets.

\textbf{Spiral of Silence.}
The ``Spiral of Silence'' theory~\citep{noelle1974spiral}, is a seminal theory within the field of communications that 
describes how people may suppress their views to avoid isolation, thus often reinforcing dominant public opinions~\citep{scheufle2000twenty,DBLP:conf/wsdm/LiuLZXT19,DBLP:journals/tkde/LinLTX22}. 
We shift focus to a novel ``passive human silence'' influenced by LLMs, where rapid AI content production and biased search algorithms potentially marginalize human contributions in public discourse. 
This theory stands apart from concepts such as ``echo chambers", ``filter bubbles", and ``degenerate feedback loops'' prevalent in recommendation systems ~\citep{DBLP:journals/corr/abs-2112-05084,DBLP:conf/wsdm/ChitraM20,DBLP:conf/aies/JiangCLGK19}. While these terms describe the narrowing of informational scope as users engage with algorithmic systems, the ``Spiral of Silence'' theory proposes a scenario where human users are compelled into silence in public discourse due to the influence of LLMs and IR systems. This phenomenon goes beyond mere selective exposure or algorithmic recommendations.
Our study explores how LLM-generated text might induce a ``Spiral of Silence'' in RAG systems over time. For further discussion regarding the rationale for applying this theory to our study, please see Appendix~\ref{sec:discussion}.

%% file: parts/pipeline.tex
\section{Pipeline Construction}\label{sec:pipeline}
In this section, a simulation framework is designed to explore the potential impacts that texts generated by LLMs may have on RAG systems. This framework models a simplified process that tracks how RAG systems gradually adjust their responses as they accumulate LLM-generated text over time. 

\subsection{Preliminaries}\label{prelim}
An RAG system \( f \) can be formalized as \( f: (Q \times D \times K) \rightarrow S \), where \( Q \) is the set of queries, \( D \) represents a large collection of documents, \( K \) is the knowledge within the LLM, and \( S \) is the set of text outputs generated by the system.
For a particular query \( q \in Q \), the goal of the RAG system is to find a mapping \( f(q, D, K) = s \) that produces a response text \( s \in S \) satisfying the query \( q \). This process involves two stages:

\textbf{Retrieval Stage,} executed by the retrieval function \( R \), is formally defined as \( R: (Q \times D) \rightarrow D' \), where \( D' \subseteq D \) represents the subset of documents judged by \( R \) to be most relevant to the query \( q \).

\textbf{Generation Stage,} executed by the generation function \( G: (P \times Q \times D' \times K) \rightarrow S \). Its task is to utilize the prompt \(p \in P\), the query \( q\in Q \), the related document subset \( D' \), and the knowledge of LLMs \( K \) to construct the answer \( s \).

Within the entire RAG system \( f \), the functions \( R \) and \( G \) act in series to form a process expressed as \( f(q, D, K) = G(p, q, R(q, D), K) \). 
In this manner, the RAG system integrates the precision of IR with the richness of LLMs to provide information-rich content when answering questions.

\subsection{Simulation Process}~\label{sec:simulation}
Our simulation process starts with a pure human-authored text dataset and gradually introduces the LLM-generated text, observing how this change over time affects the RAG system. Adhering to the specifications outlined in Section~\ref{prelim}, the RAG architecture is instantiated and expanded with additional details.
In the \textbf{retrieval stage}, we apply sparse and dense retrieval strategies to obtain a candidate document set that is relevant to the query. Additionally, we also have the option to perform a re-ranking of the candidate documents to further optimize the process.
In the \textbf{generation stage}, we use the LLMs which are widely used to generate responses. To accurately simulate the evolution process of the RAG system, we specifically use an \textbf{iteratively updated indexing structure} that supports incorporating the newly generated LLM text into the index in each iteration, keeping the dataset updated for subsequent retrieval and evaluation.

Specifically, the iterative simulation process unfolds as follows:
    1) \textbf{Baseline Establishment}: Utilizing an initial dataset comprised of human-authored text unaffected by LLM (\( D_0 \)), ascertain the performance of a benchmark RAG pipeline.
    2) \textbf{Zero-shot Text Introduction}: The baseline dataset \( D_0 \) is enriched with text set \( T_{\text{LLM}}^{(\text{zero-shot})} \) generated by LLMs in zero-shot manner, yielding \( D_1 = D_0 \cup T_{\text{LLM}}^{(\text{zero-shot})} \). This simulates the evolution of users' application of LLMs from initial zero-shot deployments to sophisticated RAG configurations. 
    3) \textbf{Retrieval and Re-ranking}: For each query \( q \), a subset of documents \( D'_i \) is retrieved from the dataset \( D_i \) through a retrieval and optional re-ranking step \( R(q, D_i) \rightarrow D'_i \). The retrieval function \( R \) remains constant throughout the experimental process to control variables. 
    4) \textbf{Generation Phase}: Answers \( S \) are generated using the LLMs (\( G(p, D', q, K) \rightarrow s \)) with a uniform prompt \(p\) in the experiment.
    5) \textbf{Post-processing Phase}: Post-process \( S \) to obtain \( S' \), removing text fragments that may expose the identity of the LLMs. 
    6) \textbf{Index Update}: Integrate \( S' \) into \( D_i \) to update the dataset to \( D_{i+1} \).
    7) \textbf{Iterative Operation}: Repeat steps 3 to 6 for each new dataset \( D_{i+1} \),  until the required number of iterations \( t \) is reached.

The pseudo-code for this process is presented in Appendix~\ref{code:pseudo}.
Through the simulation process, we observe how LLM-generated text influences the RAG systems and how this impact evolves with data accumulation. While the main simulation assumes that the LLMs remain static due to their relatively infrequent update cycles, we also conduct experiments on the effects of LLM evolution over time in Appendix~\ref{sec:evolution}.
For prompt and post-processing details, see Appendix~\ref{sec:prompt} and \ref{sec:postprocess}.

%% file: parts/experiments.tex
\section{Experiment}\label{sec:exp}

\textbf{Datasets and Metrics.} We conduct experiments on commonly used ODQA datasets, including \textbf{NQ}~\citep{nq}, \textbf{WebQ}~\citep{webq}, \textbf{TriviaQA}~\citep{tqa}, and \textbf{PopQA}~\citep{popqa}. We preprocess the datasets following~\citet{DBLP:conf/iclr/0002IWXJ000023} and~\citet{llm-embedder}.
Given the constraints on experimental resources, we randomly select 200 samples from each test set.
When evaluating the retrieval phase, we utilize \textbf{Acc@5} and \textbf{Acc@20} following \citet{DBLP:conf/emnlp/KarpukhinOMLWEC20}. These metrics assess the proportion of questions where the correct answers appear in the top 5 or top 20 retrieval results, respectively.
For the answer quality of the LLM output for each iteration,  we follow~\citet{DBLP:journals/corr/abs-2309-01431} by applying the \textbf{Exact Match (EM)} metric, which checks if the correct answer is fully contained within the generated text. 
Furthermore, in Section~\ref{sec:results}, we adopt a holistic perspective to examine the RAG pipeline, with a focus on the interaction between the retrieval and generation phases and how the ranking of human-generated texts changes over time. 

\textbf{Retrieval and Re-ranking Methods.} In our experiments, we employ a variety of retrieval methods, including the sparse model BM25, the contrastive learning-based dense retriever Contriever~\citep{contriever}, the advanced BGE-Base~\citep{BGE} retriever, and the LLM-Embedder~\citep{llm-embedder} designed for LLMs. 
For the results retrieved using BM25 and BGE-Base, we separately apply the T5-based~\citep{T5} re-ranking model MonoT5-3B~\citep{Monot5}, the UPR-3B~\citep{UPR} which uses the unsupervised capabilities of T0-3B~\citep{T0}, and the BGE-Reranker~\citep{BGE}, which is based on the XLM-RoBERTa-Large~\citep{xlm-roberta}.

\textbf{Generative Models.} Considering the complexity and variability of real-world environments, the text that is continuously integrated into the system may be generated by a variety of LLMs. Our iterative experiments incorporate text produced by a suite of prevalent LLMs. These include GPT-3.5-Turbo~\citep{openai_chatgpt}, LLaMA2-13B-Chat~\citep{DBLP:journals/corr/abs-2302-13971}, Qwen-14B-Chat~\citep{qwen}, Baichuan2-13B-Chat~\citep{baichuan}, and ChatGLM3-6B~\citep{chatglm}. This enables the RAG systems to blend varied linguistic styles and knowledge, leading to results that more closely replicate real-world scenarios.
For more implementation details, please refer to Appendix~\ref{sec:implement}.

%% file: parts/results_effects.tex
\input{parts/tables/short_term_ret}

\input{parts/tables/short_term_figure_sim}
\section{Results}\label{sec:results}
In this section, we examine both initial and extended iterations within the simulation framework. We define the short-term effect as the immediate effects observed in the first iteration, while the long-term effect is analyzed from the second to the tenth iteration. We investigate the occurrence of the ``Spiral of Silence'' effect and how RAG systems respond.
Under the task settings of ODQA, we analyze the potential influence of the ``Spiral of Silence'' on RAG systems' response patterns.

\subsection{Short-Term Effects on RAG Performance}
When comparing RAG system results using different retrieval methods on the original dataset versus the augmented one in the first iteration, we observe that:
1) \textbf{Immediate Impact of LLM-Generated Text on the RAG System:} The introduction of a minimal amount of LLM-generated text produces immediate effects on both retrieval and QA performance of the RAG system, as shown in Table~\ref{tab:combined_ret} and Figure~\ref{fig:Short-Term}. Specifically, both retrieval and generation performance exhibit noticeable fluctuations. These changes highlight the sensitivity of the system to even small modifications made by LLM-generated text.
2) \textbf{LLM-Generated Text Generally Improves Retrieval Accuracy:} Table~\ref{tab:combined_ret} reveals that adding LLM-generated responses to a dataset typically enhances the accuracy of retrieval systems, as measured by Acc@5 and Acc@20 metrics. For example, using the BM25 on TriviaQA resulted in accuracy improvements of 31.2\% and 19.1\% respectively. However, a slight decline in Acc@5 is also observed in certain cases. This suggests a primarily positive, yet complex, impact of LLM-generated text on retrieval accuracy.
3) \textbf{The Impact on QA Performance is Mixed:} 
Due to space constraints, we only present the results of four retrieval methods.
As shown in Figure~\ref{fig:Short-Term}, while the RAG system's QA performance typically surpasses the zero-shot LLM outputs, the addition of LLM text can either enhance or impair QA performance depending on the dataset and retrieval strategy. It appears to enhance performance for TriviaQA, but for NQ and PopQA, the effect is detrimental with non-BM25 retrieval methods, suggesting that without significant retrieval enhancement, LLM text inclusion might be counterproductive.
\input{parts/tables/long_term_figure}

\subsection{Long-term Effects on RAG Performance} \label{sec:long-term}
In this section, we investigate whether the short-term effects are predictive of the long-term behavior of the system. We present the results on NQ and PopQA in Figure~\ref{fig:Long-Term}. For results on other datasets, please refer to Figure~\ref{fig:Long-Term-app} in Appendix~\ref{sec:app_fig_wt}, where we observe consistent patterns across these datasets. We find that:
1) \textbf{Decreased Retrieval Effectiveness Over Time:} Figures~\ref{fig:nq_loop_retrieval} and \ref{fig:pop_loop_retrieval} show a general decline in Acc@5 across successive iterations for most methods, with an average drop of 21.4\% for NQ and 19.4\% for PopQA from the first iteration to the last, except for a temporary improvement in BM25 during the second iteration on PopQA. This trend signals that the retrieval quality boost provided by LLM-generated text may be transient, with a propensity for degradation over time.
2) \textbf{Stability in QA Performance Despite Retrieval Decline:} Contrary to expectations, the QA performance does not mirror the retrieval accuracy's decrease. As shown in Figure~\ref{fig:nq_loop_qa} and \ref{fig:pop_loop_qa}, the EM exhibit slight variations but generally maintain their level throughout the iterations.
While a diminished retrieval accuracy intuitively seems to undermine the system's capacity to output correct answers, this does not unequivocally translate into a decline in QA efficacy.
In subsequent sections, we will delve deeper into the reasons behind these observations and examine the complex dynamic relationship that may exist between retrieval and QA performance.

\subsection{Spiral of Silence}\label{sec:sos}
In the context of LLM-augmented RAG systems, we have observed a rapid shift in response to the integration of LLM-generated text, a decline in retrieval performance over time, and stability in QA performance despite retrieval decline. To explain these phenomena, we draw on the theory of the ``Spiral of Silence'' as posited by \citet{noelle1974spiral}, extending its principles to the behavior of RAG systems enhanced by LLMs. 
To explore the presence of a ``Spiral of Silence'' phenomenon, we propose \textbf{three Hypotheses} for investigation.
\textbf{(H1): Dominance of LLM-Generated Texts:} Retrieval models are more likely to prioritize LLM-generated text in search results, which could result in LLM-generated text taking a dominant position in the retrieval hierarchy. 
\textbf{(H2): Marginalization of Human-Generated Content:} If human-authored text consistently loses ranking prominence through successive iterations, it may be excluded from the top results until it becomes invisible, thus creating silence.
\textbf{(H3): Homogenization of Opinions:} The preferential ranking of LLM-generated text could culminate in a uniformity of displayed perspectives by the RAG system, potentially sidelining the accuracy or variety of the information.
\input{parts/tables/dominance}

To verify \textbf{(H1)}, we analyze Iteration 1 where LLM-generated texts are first introduced to the retrieval system. We calculate the proportion of these texts appearing in the top 5 search results:
\begin{equation}
   \text{P} = \frac{\sum_{q \in Q} c_{q}^{LLM}}{\sum_{q \in Q} (c_{q}^{LLM} + c_{q}^{Human})} \times 100\% 
\end{equation}
\noindent where $c_{q}^{LLM}$ is the count of LLM-generated texts and $c_{q}^{Human}$ is the count of human-generated texts in the top 5 search results for query $q$. 
Table~\ref{tab:llm_dominance} reveals that, even with a modest inclusion of LLM-generated texts, most retrieval models often rank them at the top. This behavior supports the findings of ~\citet{DBLP:journals/corr/abs-2310-20501}, where LLM-rewritten texts are preferred by retrieval models over the originals. Our study extends this by directly generating query-specific texts with LLMs. The preference might stem from inherent biases within the system or the actual relevance of the LLM-produced content. \textbf{This suggests retrieval systems tend to favor LLM-generated texts, making them more prominent in search results, which can rapidly influence an RAG system's behavior. }

\input{parts/tables/percentage_figure}
\input{parts/tables/bleu}
To validate \textbf{(H2)}, we incorporate a \textbf{temporal dimension}, observing the percentage change of texts generated by various LLMs and humans within the top 50 search results across different datasets over time. 
As shown in Figure~\ref{fig:Percentage}, after ten iterations, the percentage of human-generated texts significantly decreased, falling below 10\% for all datasets. \textbf{This pattern suggests a sustained diminishing impact of human-contributed texts and hints at the possibility of their eventual exclusion from search results if the trend continues.}

\input{parts/tables/context}
To explore \textbf{(H3)}, we examine the risk of potential viewpoint homogenization in the RAG system from both the \textbf{diversity} and \textbf{accuracy} dimensions during the simulation.
\textbf{Diversity} is quantified using Self-BLEU ~\citep{DBLP:conf/sigir/ZhuLZGZWY18}, which works by comparing a generated text with other generated texts to measure similarity. High similarity results in a high Self-BLEU score, indicating lower diversity and suggesting a convergence of viewpoints.
As shown in Figure~\ref{fig:BLEU}, upon introducing zero-shot LLM-generated texts (Iteration 1), the Self-BLEU scores across different datasets experience varying degrees of change.
However, over more iterations, the Self-BLEU scores for the top 5 results consistently rise and plateau across all datasets, \textbf{indicating a significant reduction in textual diversity with each iterative cycle}.
Subsequently, we assessed whether the \textbf{accuracy} of the top documents returned by the IR system tends toward uniformity over time. Figure~\ref{fig:nq_context} charts the number of documents with the correct answer in the top 5 results ("Context Right Num") against the number of queries LLM answers correctly or incorrectly, across the Iteration1, 2, 5, 10. For simplicity, we showcase only the NQ dataset's averaged outcomes, but we find the same trends across other datasets.
It indicates that in the initial iterations, fewer correct answer documents (e.g., ``Context Right Num'' of 0, 1, or 2) typically correlate with a greater number of LLM-answered queries being incorrect (EM=0). Despite this, a significant fraction of the top documents still include the correct answer. 
When the LLM correctly answers a query (EM=1), the correct answer documents within the top results can range anywhere from 1 to 5. As the iterations continue, the frequency of having 1 to 4 correct answer documents in the top 5 results for each query diminishes, and by the Iteration10, contexts for LLM-correct queries almost always contain the correct answer, whereas contexts for LLM-incorrect answers almost always do not. \textbf{This pattern demonstrates a trend towards polarization and uniformity in the accuracy of provided contexts as the RAG system iterates.}

At this point, we have confirmed through our experiments the presence of the ``Spiral of Silence'' phenomenon, as outlined in three tested hypotheses. Moreover, the dashed lines in Figure~\ref{fig:nq_context} represent the total number of correct and incorrect LLM answered queries, along with the Acc@5 retrieval metric over various iterations. 
The LLM's rate of correct answers remains constant through the iterations, aligning to Section~\ref{sec:long-term}. However, as iterations advance, correct answers diminish within top documents for LLM-incorrect queries, reducing their contribution to the Acc@5 and thus decreasing retrieval performance. In contrast, for LLM-correct queries, more retrieved documents containing the correct answer do not affect Acc@5 or EM. \textbf{Thus, the pattern discussed in Section~\ref{sec:long-term} can be explained by the ``Spiral of Silence'' theory, which accounts for the observed dip in IR results and the sustained QA performance.}
\subsection{Effects of ``Spiral of Silence'' on ODQA}
\input{parts/tables/change_index}
\input{parts/tables/rank}
We will delve into a more nuanced discussion of the impact of the ``Spiral of Silence'' within the context of ODQA. It is important to note that the influence of the phenomenon is not confined to this scenario; it may also be pertinent across all settings that involve knowledge retrieval, generation, and the influx of text from LLMs.
Specifically, our analysis is structured around two dimensions: the query level and the document level.

At the \textbf{query level}, 
Figure~\ref{fig:change_index} signifies the average count of queries shifting between consecutive iterations from incorrect to correct and vice versa, respectively. 
Notably, during the 1->2 iteration, there is an initial surge in both metrics, which subsequently experience a sharp decline as the iterations continue. This suggests that the LLM-generated text's initial introduction catalyzes a more dynamic state, likely due to the correction of existing errors or the introduction of new inaccuracies. Over time, however, the ``Spiral of Silence'' effect seems to guide the system towards a state of equilibrium where the transition rate stabilizes to less than 1\% per 200 queries. 
\textbf{This means most queries maintain their status as either correct or incorrect, indicating that individual query QA results become fixed.}

At the \textbf{document level}, we compute the average rank shifts of the first documents containing the correct answer within retrieval results, under different LLM answer states. In Figure~\ref{fig:rank}, we observe that:
1) \textbf{Different Trends of Correct Answer Rankings Based on Source:} In instances where EM=0, correct documents from all sources (``First Right From ALL Sources'') and from humans (``First Right From Human") both tend to be ranked lower over time. When EM=1, the rankings for correct documents from all sources improve slightly, while rankings for correct answers from humans continue to decline. This suggests that the LLM's correct texts gain prominence in retrieval rankings over time, overshadowing correct texts from human-generated texts.
2) \textbf{Gradual Dysfunction of the IR System in Incorrect LLM Responses:} When the LLM provides incorrect answers (EM=0), there's a risk that documents that once rose to the top with accurate information might increasingly be obscured by the growing mass of LLM-generated content. This can lead to a scenario where the IR system, originally intended to help users find precise information, becomes less reliable. If it prioritizes and disseminates the LLM's inaccuracies, a feedback loop could ensue, solidifying these errors. This concerning trend highlights the critical need for ongoing adjustments and improvements to the IR systems to uphold their purpose. 

%% file: parts/tables/short_term_ret.tex
\begin{table*}[htbp]
\centering
\setlength{\tabcolsep}{3pt}
\resizebox{0.97\textwidth}{!}{
\begin{tabular}{l|ll|ll|ll|ll|ll|ll|ll|ll}
\toprule
\multirow{2}{*}{Model} & \multicolumn{4}{c|}{NQ} & \multicolumn{4}{c|}{WebQ} & \multicolumn{4}{c|}{TriviaQA} & \multicolumn{4}{c}{PopQA} \\
\cline{2-17}
& \multicolumn{2}{c|}{Acc@5} & \multicolumn{2}{c|}{Acc@20} & \multicolumn{2}{c|}{Acc@5} & \multicolumn{2}{c|}{Acc@20} & \multicolumn{2}{c|}{Acc@5} & \multicolumn{2}{c|}{Acc@20} & \multicolumn{2}{c|}{Acc@5} & \multicolumn{2}{c}{Acc@20} \\
& Ori. & +LLM$_{Z}$ & Ori. & +LLM$_{Z}$ & Ori. & +LLM$_{Z}$ & Ori. & +LLM$_{Z}$ & Ori. & +LLM$_{Z}$ & Ori. & +LLM$_{Z}$ & Ori. & +LLM$_{Z}$ & Ori. & +LLM$_{Z}$ \\
\toprule
BM25                  & 49.0   & \ApplyGradient{49.0}{57.5} & 67.0 & \ApplyGradient{67.0}{73.5} & 41.0   & \ApplyGradient{41.0}{51.0}$^*$ & 63.0 & \ApplyGradient{63.0}{71.0} & 62.5   & \ApplyGradient{62.5}{82.0}$^*$ & 73.0 & \ApplyGradient{73.0}{87.0}$^*$ & 35.5   & \ApplyGradient{35.5}{41.5} & 51.5 & \ApplyGradient{51.5}{59.5} \\
Contreiver            & 68.0   & \ApplyGradient{68.0}{68.5} & 84.0 & \ApplyGradient{84.0}{85.0} & 66.0   & \ApplyGradient{66.0}{69.5} & 74.0 & \ApplyGradient{74.0}{80.0} & 68.0   & \ApplyGradient{68.0}{83.5}$^*$ & 80.5 & \ApplyGradient{80.5}{87.5} & 62.0   & \ApplyGradient{62.0}{65.0} & 77.5 & \ApplyGradient{77.5}{79.5} \\
LLM-Embedder          & 75.5  & \ApplyGradient{75.5}{75.5} & 86.5 & \ApplyGradient{86.5}{88.0} & 62.5  & \ApplyGradient{62.5}{72.5}$^*$ & 76.0 & \ApplyGradient{76.0}{79.5} & 67.5  & \ApplyGradient{67.5}{81.0}$^*$ & 77.5 & \ApplyGradient{77.5}{87.5}$^*$ & 70.0    & \ApplyGradient{70.0}{67.5} & 79.5 & \ApplyGradient{79.5}{82.0} \\
BGE$_{base}$              & 77.0 & \ApplyGradient{77.0}{73.0} & 86.0 & \ApplyGradient{86.0}{86.0} & 65.5 & \ApplyGradient{65.5}{71.5} & 77.0 & \ApplyGradient{77.0}{80.0} & 69.5 & \ApplyGradient{69.5}{81.5}$^*$ & 80.0 & \ApplyGradient{80.0}{87.5}$^*$ & 72.0 & \ApplyGradient{72.0}{70.0} & 83.0 & \ApplyGradient{83.0}{84.5} \\
BM25+UPR              & 63.0 & \ApplyGradient{63.0}{66.5} & 73.5 & \ApplyGradient{73.5}{78.0} & 57.0 & \ApplyGradient{57.0}{68.0}$^*$ & 68.5 & \ApplyGradient{68.5}{75.0} & 71.5 & \ApplyGradient{71.5}{83.0}$^*$ & 78.0 & \ApplyGradient{78.0}{89.0}$^*$ & 57.5 & \ApplyGradient{57.5}{61.5} & 60.0 & \ApplyGradient{60.0}{67.0} \\
BM25+MonoT5           & 66.5 & \ApplyGradient{66.5}{69.0} & 74.5 & \ApplyGradient{74.5}{80.5} & 62.0 & \ApplyGradient{62.0}{67.5} & 69.5 & \ApplyGradient{69.5}{76.0} & 72.0 & \ApplyGradient{72.0}{83.5}$^*$ & 78.0 & \ApplyGradient{78.0}{88.0}$^*$ & 53.5 & \ApplyGradient{53.5}{58.5} & 59.5 & \ApplyGradient{59.5}{66.5} \\
BM25+BGE$_{reranker}$     & 68.0 & \ApplyGradient{68.0}{69.5} & 76.5 & \ApplyGradient{76.5}{81.0} & 64.5 & \ApplyGradient{64.5}{68.5} & 71.0 & \ApplyGradient{71.0}{76.0} & 72.5 & \ApplyGradient{72.5}{84.0}$^*$ & 78.0 & \ApplyGradient{78.0}{88.5}$^*$ & 54.0 & \ApplyGradient{54.0}{61.0} & 60.0 & \ApplyGradient{60.0}{67.5} \\
BGE$_{base}$+UPR          & 75.5 & \ApplyGradient{75.5}{71.5} & 87.5 & \ApplyGradient{87.5}{88.0} & 64.0 & \ApplyGradient{64.0}{69.0} & 77.0 & \ApplyGradient{77.0}{79.5} & 76.0 & \ApplyGradient{76.0}{84.0}$^*$ & 84.5 & \ApplyGradient{84.5}{89.5} & 76.0 & \ApplyGradient{76.0}{71.0} & 84.5 & \ApplyGradient{84.5}{84.5} \\
BGE$_{base}$+MonoT5       & 75.0 & \ApplyGradient{75.0}{70.5} & 86.5 & \ApplyGradient{86.5}{86.5} & 68.5 & \ApplyGradient{68.5}{72.0} & 78.0 & \ApplyGradient{78.0}{81.5} & 77.0 & \ApplyGradient{77.0}{83.5} & 83.5 & \ApplyGradient{83.5}{89.5} & 72.0 & \ApplyGradient{72.0}{72.5} & 85.5 & \ApplyGradient{85.5}{86.0} \\
BGE$_{base}$+BGE$_{reranker}$ & 69.0 & \ApplyGradient{69.0}{68.0} & 84.0 & \ApplyGradient{84.0}{84.5} & 67.5 & \ApplyGradient{67.5}{70.5} & 78.0 & \ApplyGradient{78.0}{81.5} & 72.5 & \ApplyGradient{72.5}{83.5}$^*$ & 82.0 & \ApplyGradient{82.0}{88.0} & 73.0 & \ApplyGradient{73.0}{70.0} & 84.0 & \ApplyGradient{84.0}{85.0} \\

\bottomrule
\end{tabular}}
\caption{Short-term retrieval performance. A \colorbox{myblue}{blue} background indicates a decrease in retrieval results after the incorporation of LLM-generated text, while a \colorbox{myred}{purple} background signifies an increase. The deeper the color, the larger the discrepancy from the original results. Statistical significance at 0.05 relative to origin is marked with $^*$.}
\label{tab:combined_ret}
\end{table*}

%% file: parts/tables/short_term_figure_sim.tex
\begin{figure*} [h]
  \centering
  \setlength{\belowcaptionskip}{-5pt}
  \begin{subfigure}[t]{0.24\textwidth}
    \includegraphics[width=\textwidth]{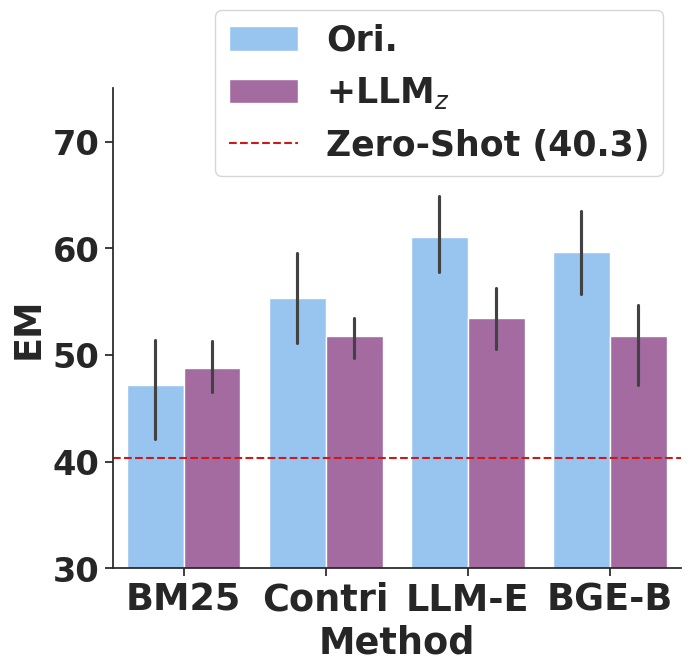}
    \caption{NQ}
    \label{fig:nq_sim}
  \end{subfigure}
  \hfill
  \begin{subfigure}[t]{0.24\textwidth}
    \includegraphics[width=\textwidth]{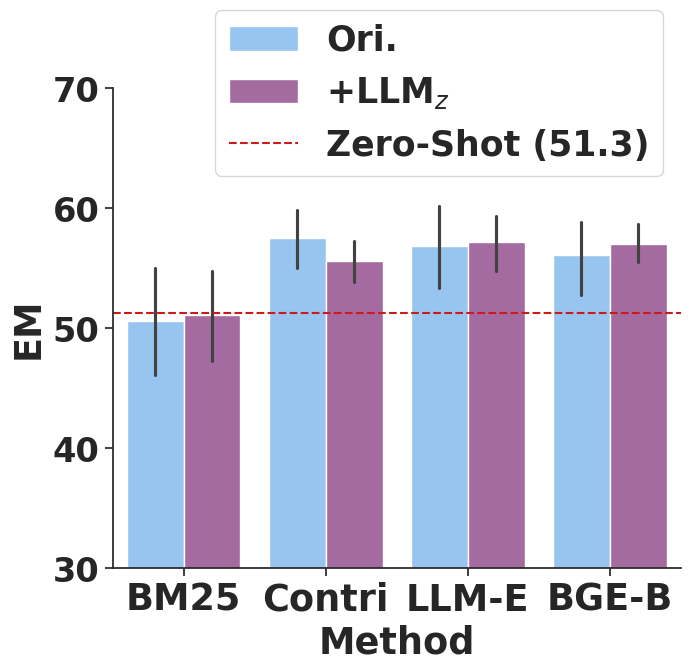}
    \caption{WebQ}
    \label{fig:webq_sim}
  \end{subfigure}
  \hfill
  \begin{subfigure}[t]{0.24\textwidth}
    \includegraphics[width=\textwidth]{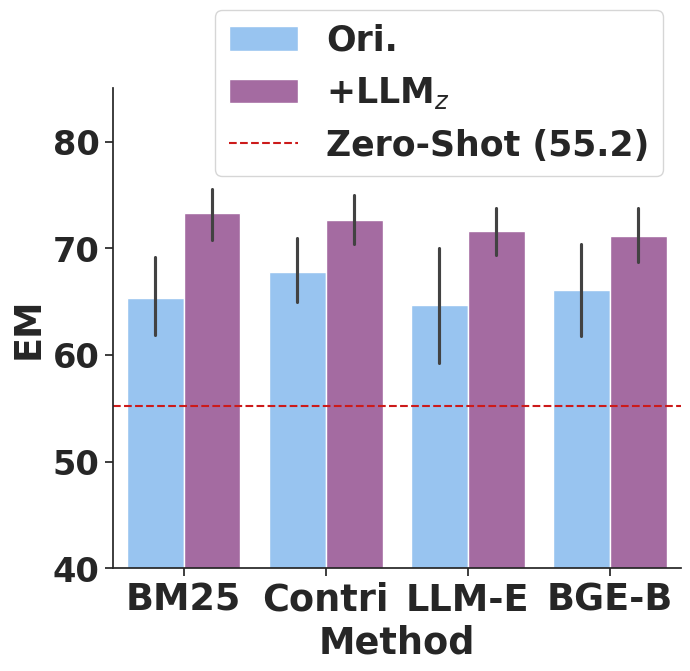}
    \caption{TriviaQA}
    \label{fig:tqa_sim}
  \end{subfigure}
  \hfill
  \begin{subfigure}[t]{0.24\textwidth}
    \includegraphics[width=\textwidth]{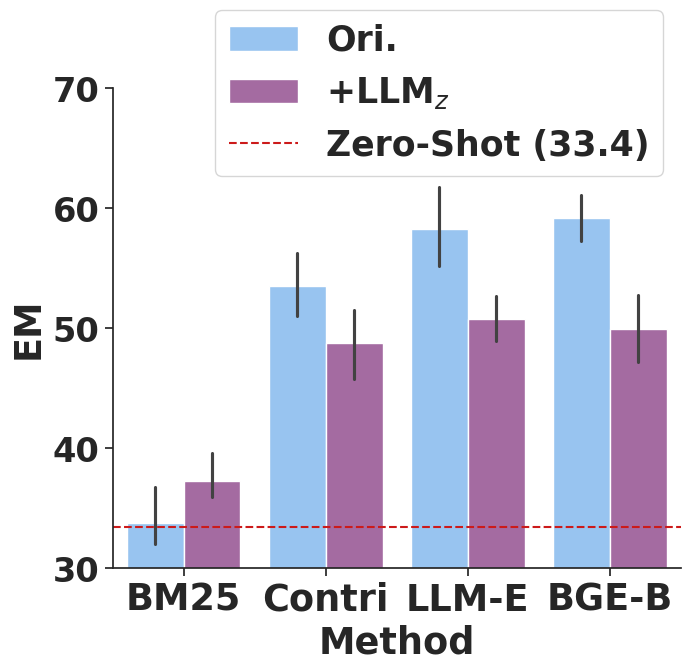}
    \caption{PopQA}
    \label{fig:pop_sim}
  \end{subfigure}
  \caption{Short-Term QA performance. For each retrieval method, we present both the average performance and the range of variation exhibited by five LLMs. A red dashed line symbolizes the average EM score for zero-shot question generation by LLMs. ``Ori.'' and ``+LLM$_Z$'' represent the average EM values when models use the original dataset or a dataset enhanced with LLM-generated texts as context, respectively. Retrieval methods are abbreviated: ``Contri'' for Contriever, ``LLM-E'' for LLM-Embedder, and ``BGE-B'' for BGE$_{base}$.}
  \label{fig:Short-Term}
\end{figure*}

%% file: parts/tables/long_term_figure.tex
{
\begin{figure}[h]
  \centering
  \begin{subfigure}[t]{0.19\textwidth}
    \includegraphics[width=\textwidth]{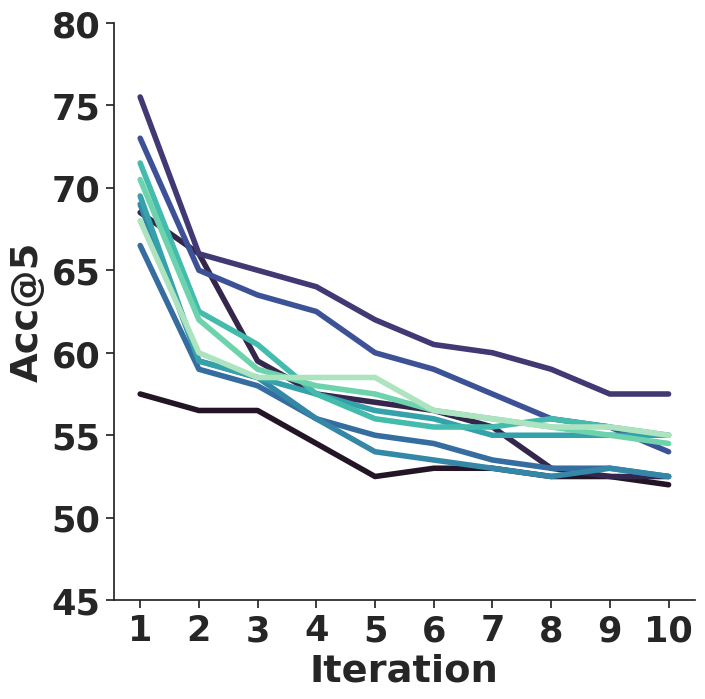}
    \caption{NQ}
    \label{fig:nq_loop_retrieval}
  \end{subfigure}
  \hfill
  \begin{subfigure}[t]{0.28\textwidth}
    \includegraphics[width=\textwidth]{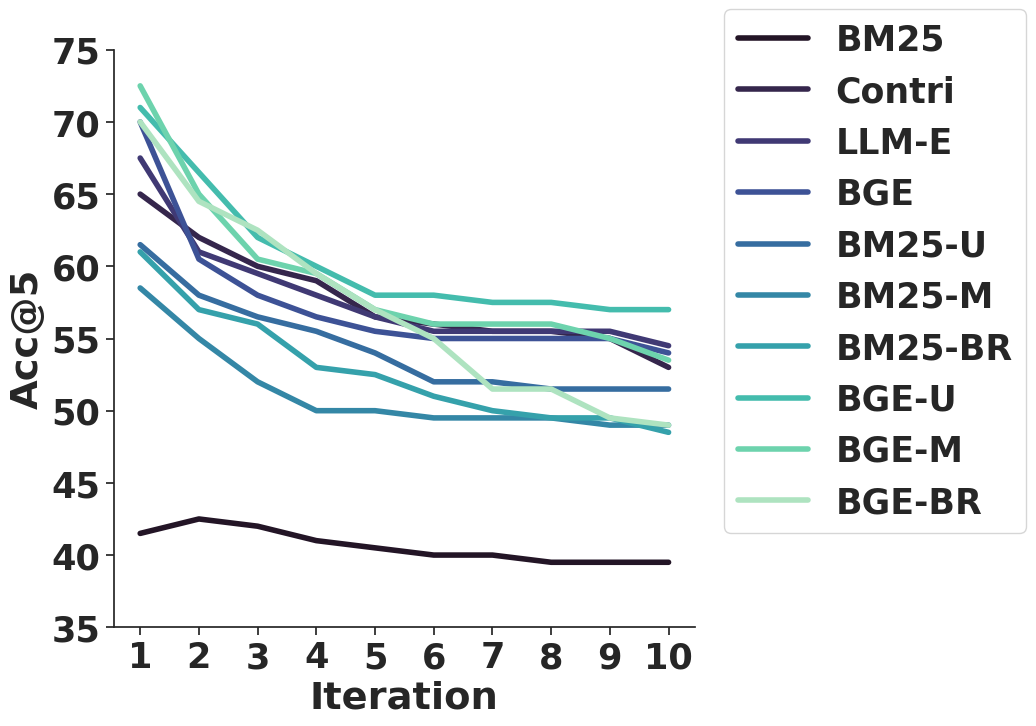}
    \caption{PopQA}
    \label{fig:pop_loop_retrieval}
  \end{subfigure}

  \vspace{0.17cm}
  \begin{subfigure}[t]{0.19\textwidth}
    \includegraphics[width=\textwidth]{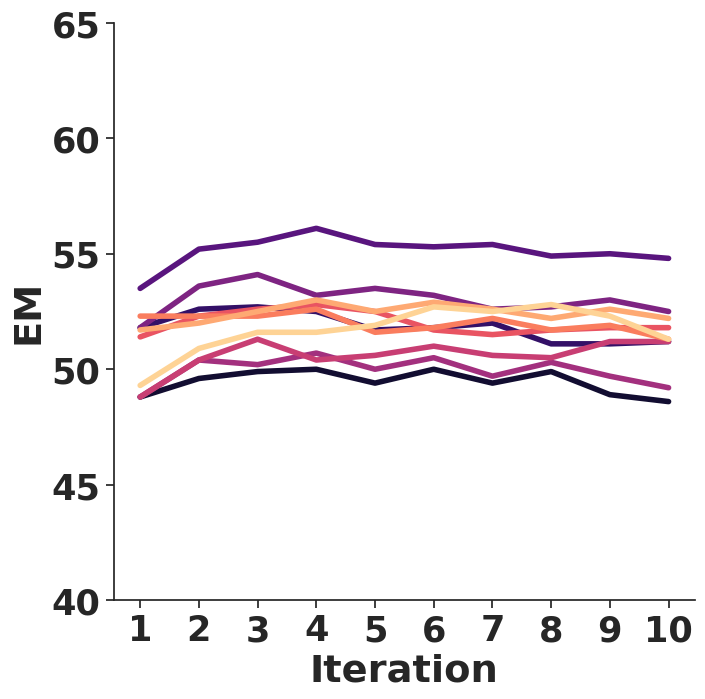}
    \caption{NQ}
    \label{fig:nq_loop_qa}
  \end{subfigure}
  \hfill
  \begin{subfigure}[t]{0.28\textwidth}
    \includegraphics[width=\textwidth]{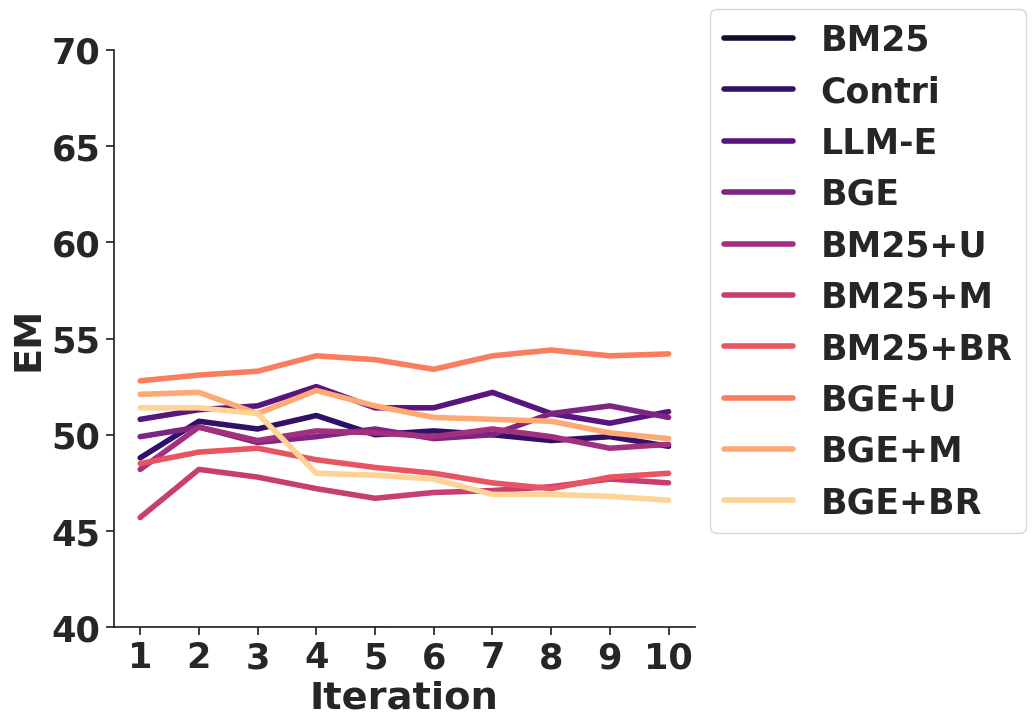}
    \caption{PopQA}
    \label{fig:pop_loop_qa}
  \end{subfigure}
  \setlength{\belowcaptionskip}{-4pt}
  \caption{Long-Term RAG performance. The upper section illustrates the retrieval outcomes for various methods, while the lower section depicts the average EM across LLMs. Iteration 1 represents the results following the incorporation of zero-shot LLM-generated text.  Abbreviated re-ranking methods in the legend are: +U for UPR, +M for MonoT5, and +BR for BGE-Reranker.}
  \label{fig:Long-Term}
\end{figure}}

%% file: parts/tables/dominance.tex
{\setlength{\belowcaptionskip}{-5pt}
\begin{table}[t]
\centering
\resizebox{0.47\textwidth}{!}{
\begin{tabular}{@{}lcccc@{}}
\toprule
\textbf{Method} & \textbf{NQ} & \textbf{WebQ} & \textbf{TriviaQA} & \textbf{PopQA} \\
\midrule
BM25 & \ApplyGradientd{50.0}{34.1} & \ApplyGradientd{50.0}{19.6} & \ApplyGradientd{50.0}{57.6} & \ApplyGradientd{50.0}{23.9} \\
Contriever & \ApplyGradientd{50.0}{72.8} & \ApplyGradientd{50.0}{75.2} & \ApplyGradientd{50.0}{80.1} & \ApplyGradientd{50.0}{67.0} \\
LLM-Embedder & \ApplyGradientd{50.0}{68.2} & \ApplyGradientd{50.0}{64.6} & \ApplyGradientd{50.0}{75.3} & \ApplyGradientd{50.0}{70.0} \\
BGE$_{base}$  & \ApplyGradientd{50.0}{80.7} & \ApplyGradientd{50.0}{84.1} & \ApplyGradientd{50.0}{85.6} & \ApplyGradientd{50.0}{81.5} \\
BM25+UPR & \ApplyGradientd{50.0}{62.3} & \ApplyGradientd{50.0}{49.8} & \ApplyGradientd{50.0}{75.7} & \ApplyGradientd{50.0}{47.1} \\
BM25+MonoT5 & \ApplyGradientd{50.0}{66.2} & \ApplyGradientd{50.0}{55.8} & \ApplyGradientd{50.0}{83.0} & \ApplyGradientd{50.0}{47.1} \\
BM25+BGE$_{reranker}$ & \ApplyGradientd{50.0}{64.4} & \ApplyGradientd{50.0}{55.2} & \ApplyGradientd{50.0}{81.6} & \ApplyGradientd{50.0}{46.6} \\
BGE$_{base}$+UPR & \ApplyGradientd{50.0}{74.4} & \ApplyGradientd{50.0}{69.3} & \ApplyGradientd{50.0}{79.1} & \ApplyGradientd{50.0}{71.2} \\
BGE$_{base}$+MonoT5 & \ApplyGradientd{50.0}{81.4} & \ApplyGradientd{50.0}{84.0} & \ApplyGradientd{50.0}{88.4} & \ApplyGradientd{50.0}{74.3} \\
BGE$_{base}$+BGE$_{reranker}$ & \ApplyGradientd{50.0}{67.2} & \ApplyGradientd{50.0}{74.2} & \ApplyGradientd{50.0}{83.2} & \ApplyGradientd{50.0}{72.8} \\
\bottomrule
\end{tabular}
}
\caption{Percentage of LLM-generated documents occupying the top 5 retrieval results, after augmenting each query with five LLM-generated documents. 
The \colorbox{myblue}{blue} background indicates a majority presence of human-generated documents, while the \colorbox{myred}{purple} background denotes a predominance of LLM-generated documents.
}
\label{tab:llm_dominance}
\end{table}}

%% file: parts/tables/percentage_figure.tex
\begin{figure}[t]
  \centering
  \begin{subfigure}[t]{0.19\textwidth}
    \includegraphics[width=\textwidth]{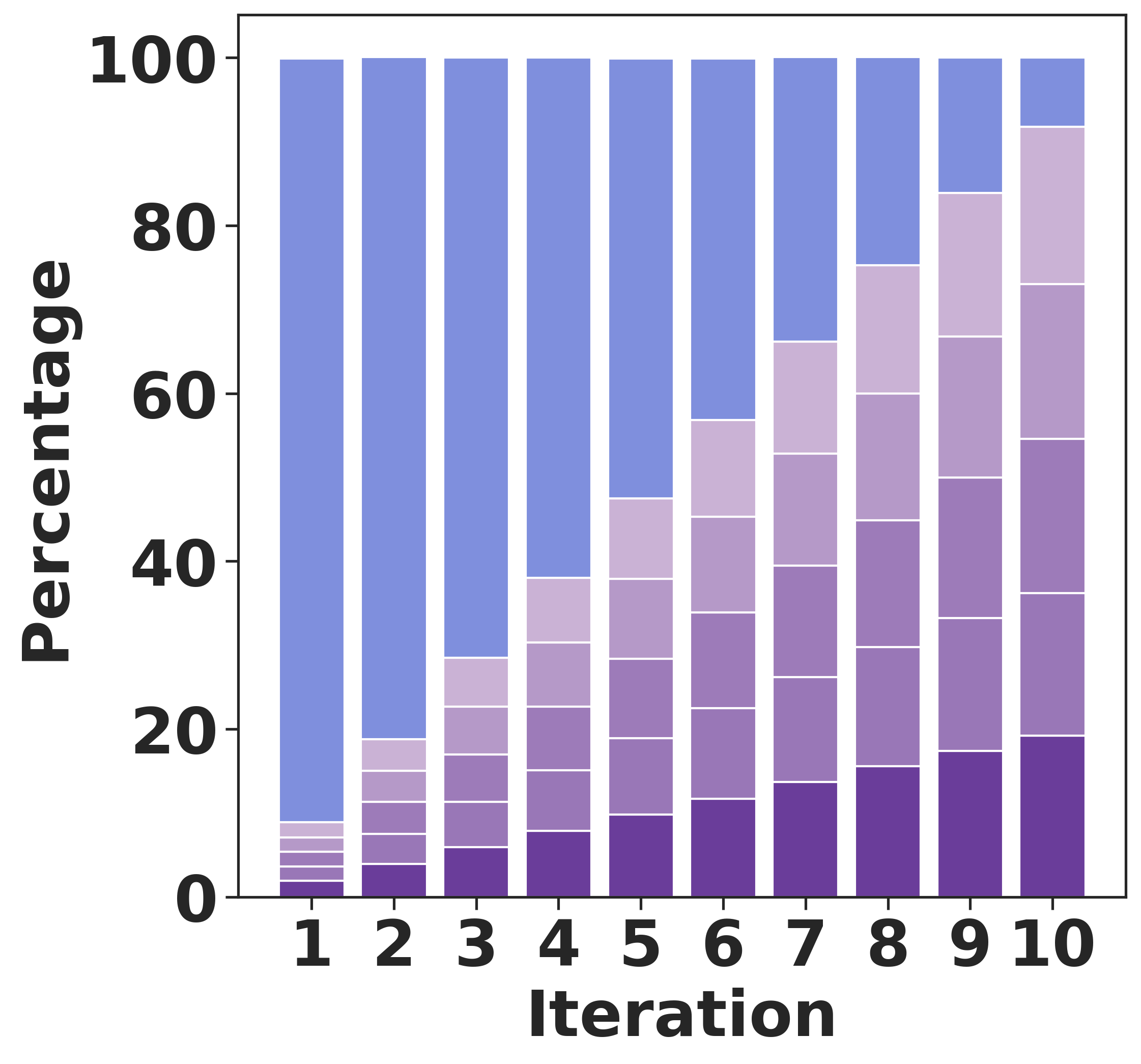}
    \caption{NQ}
    \label{fig:nq_per}
  \end{subfigure}
  \hfill
  \begin{subfigure}[t]{0.28\textwidth}
    \includegraphics[width=\textwidth]{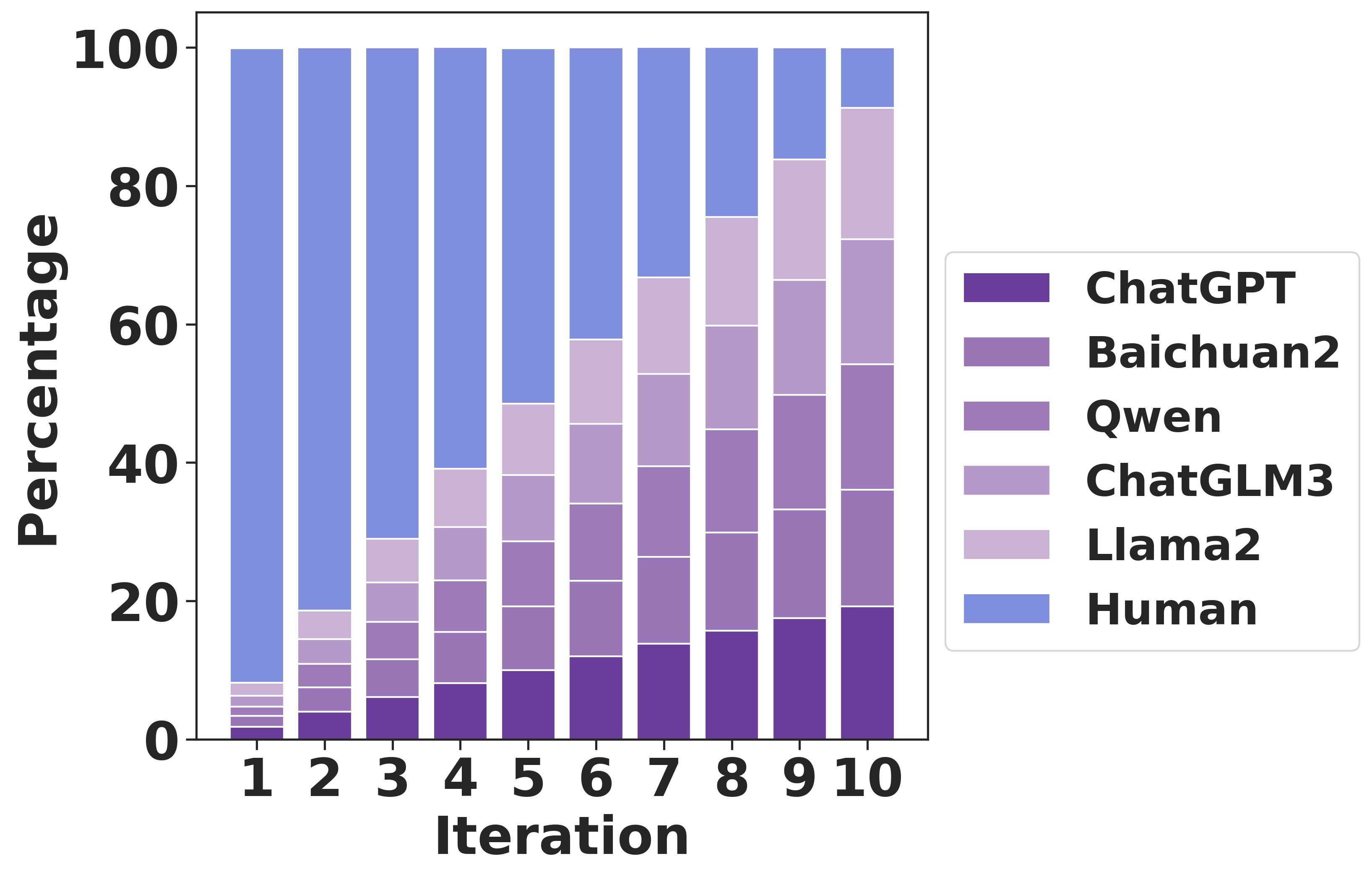}
    \caption{PopQA}
    \label{fig:pop_per}
  \end{subfigure}
  \setlength{\belowcaptionskip}{-7pt}
  \caption{Average percentage of texts from various sources within the top 50 search results over multiple iterations across different search methods.
  For results on WebQ and TriviaQA, please refer to Figure~\ref{fig:Percentage_app} in Appendix~\ref{sec:app_fig_wt}.}
  \label{fig:Percentage}
\end{figure}

%% file: parts/tables/bleu.tex
{\setlength{\belowcaptionskip}{-4pt}
\begin{figure}[t]
  \centering
  \begin{subfigure}[t]{0.18\textwidth}
    \includegraphics[width=\textwidth]{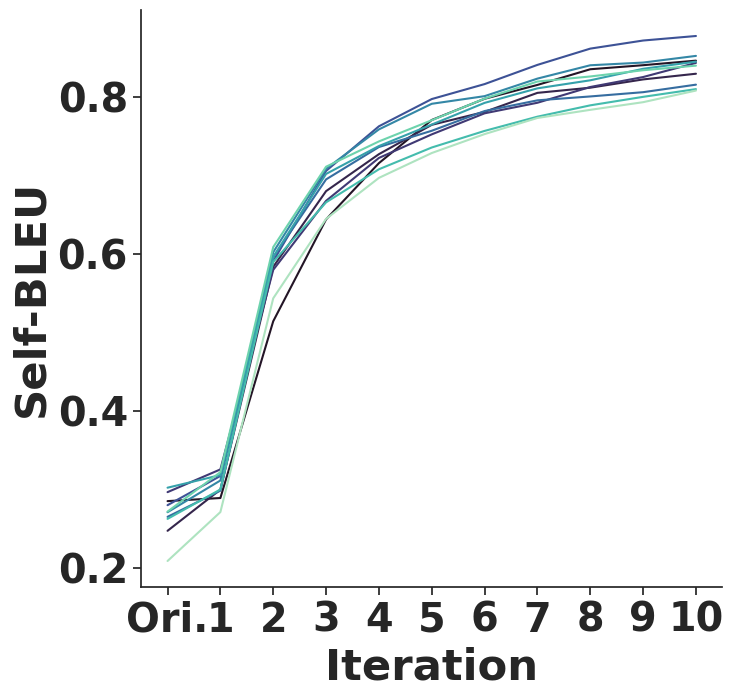}
    \caption{NQ}
    \label{fig:nq_bleu}
  \end{subfigure}
  \hfill
  \begin{subfigure}[t]{0.29\textwidth}
    \includegraphics[width=\textwidth]{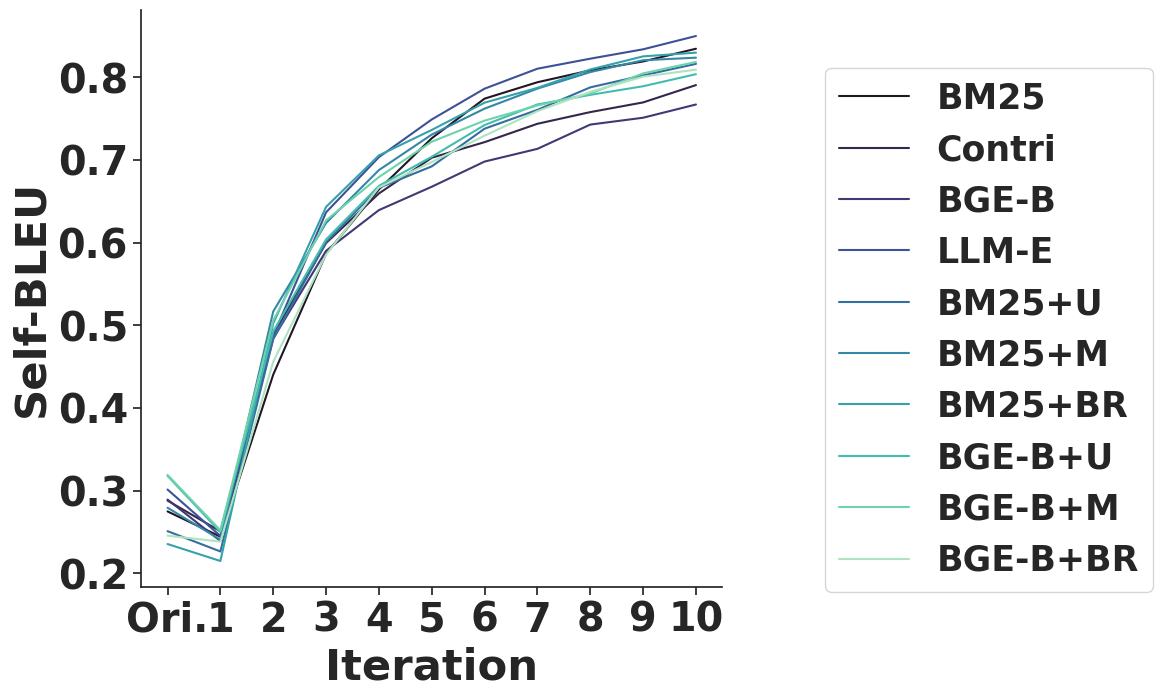}
    \caption{PopQA}
    \label{fig:pop_bleu}
  \end{subfigure}
  \caption{3-gram Self-BLEU score for the top 5 search results over iterations, from the original dataset (Ori.) to subsequent iterations including LLM-generated texts. 
  For results on WebQ and TriviaQA, please refer to Figure~\ref{fig:BLEU_app} in Appendix~\ref{sec:app_fig_wt}.}
  \label{fig:BLEU}
\end{figure}}

%% file: parts/tables/context.tex
\begin{figure*}[htbp]
  \centering
    \includegraphics[width=\textwidth]{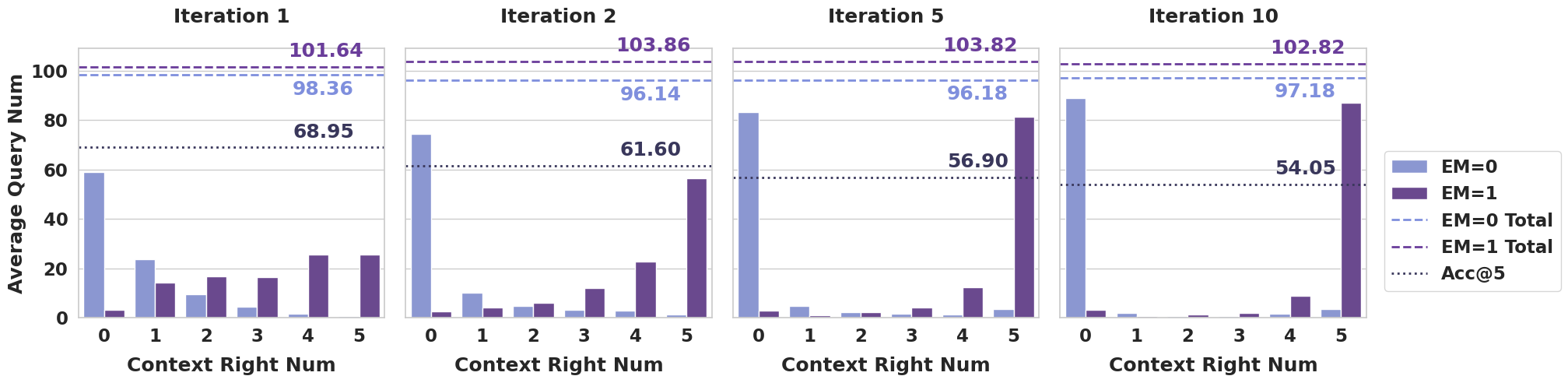}
    \caption{Correlation between the number of top 5 search results containing the correct answer (``Context Right Num'') and the accuracy of responses given by LLMs on the NQ dataset. The responses are categorized based on Exact Match (EM) score: EM=1 for correct and EM=0 for incorrect. The overall number of queries that the LLMs answered correctly (EM=1 Total) and incorrectly (EM=0 Total), along with the average retrieval accuracy (Acc@5)
    are shown by dashed lines. The results are averaged across different LLMs, retrieval, and ranking methods.}
    \label{fig:nq_context}
\end{figure*}

%% file: parts/tables/change_index.tex

\begin{figure*}[h]
  \centering
  \begin{subfigure}[t]{0.37\textwidth}
    \includegraphics[width=\textwidth]{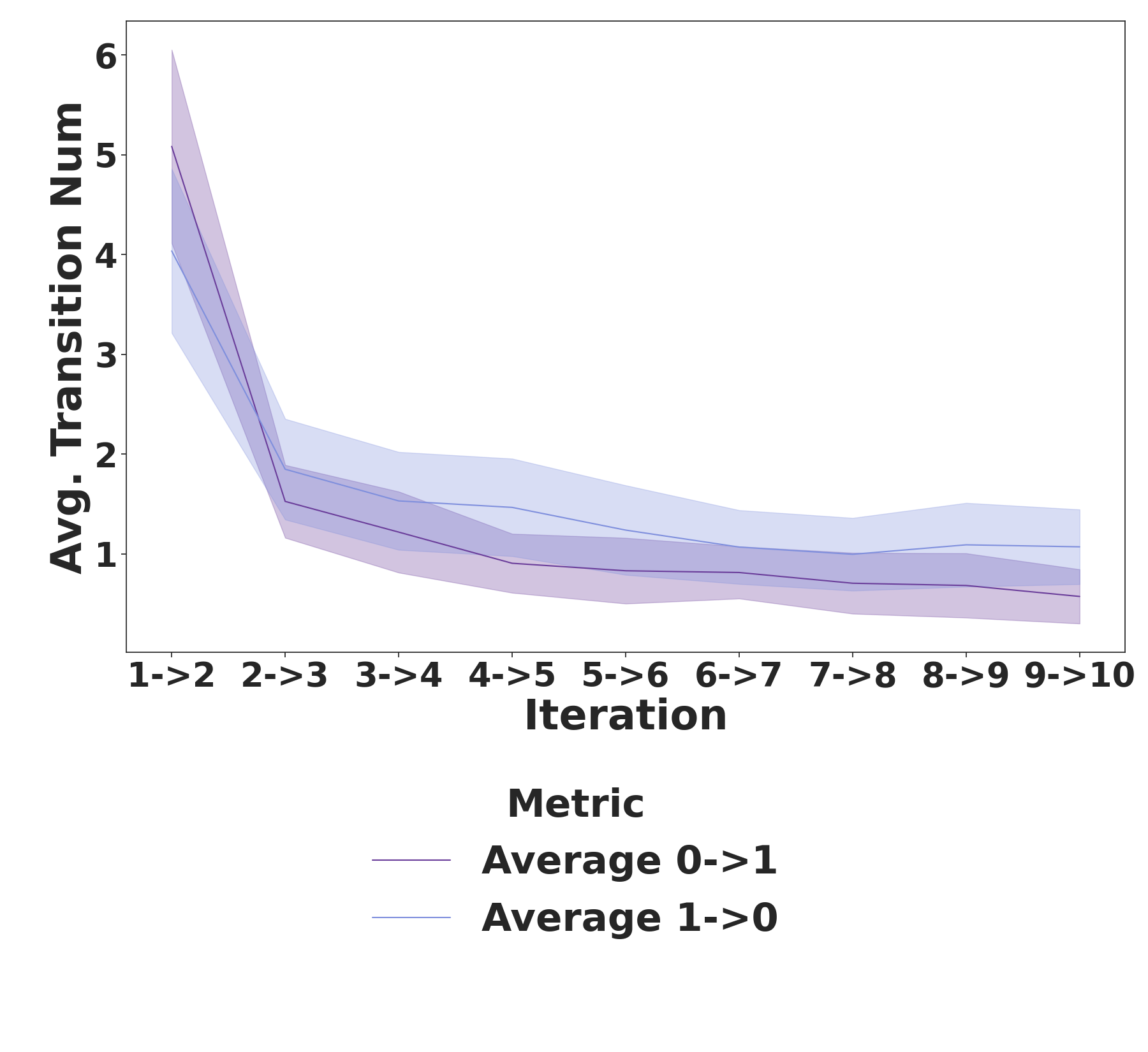}
    \caption{Average Query State Transition Number from incorrect to correct (``Average 0->1'') and correct to incorrect (``Average 1->0'') between consecutive iterations, aggregated across all IR methods and datasets.}
    \label{fig:change_index}
  \end{subfigure}
  \hfill
  \begin{subfigure}[t]{0.59\textwidth}
    \includegraphics[width=\textwidth]{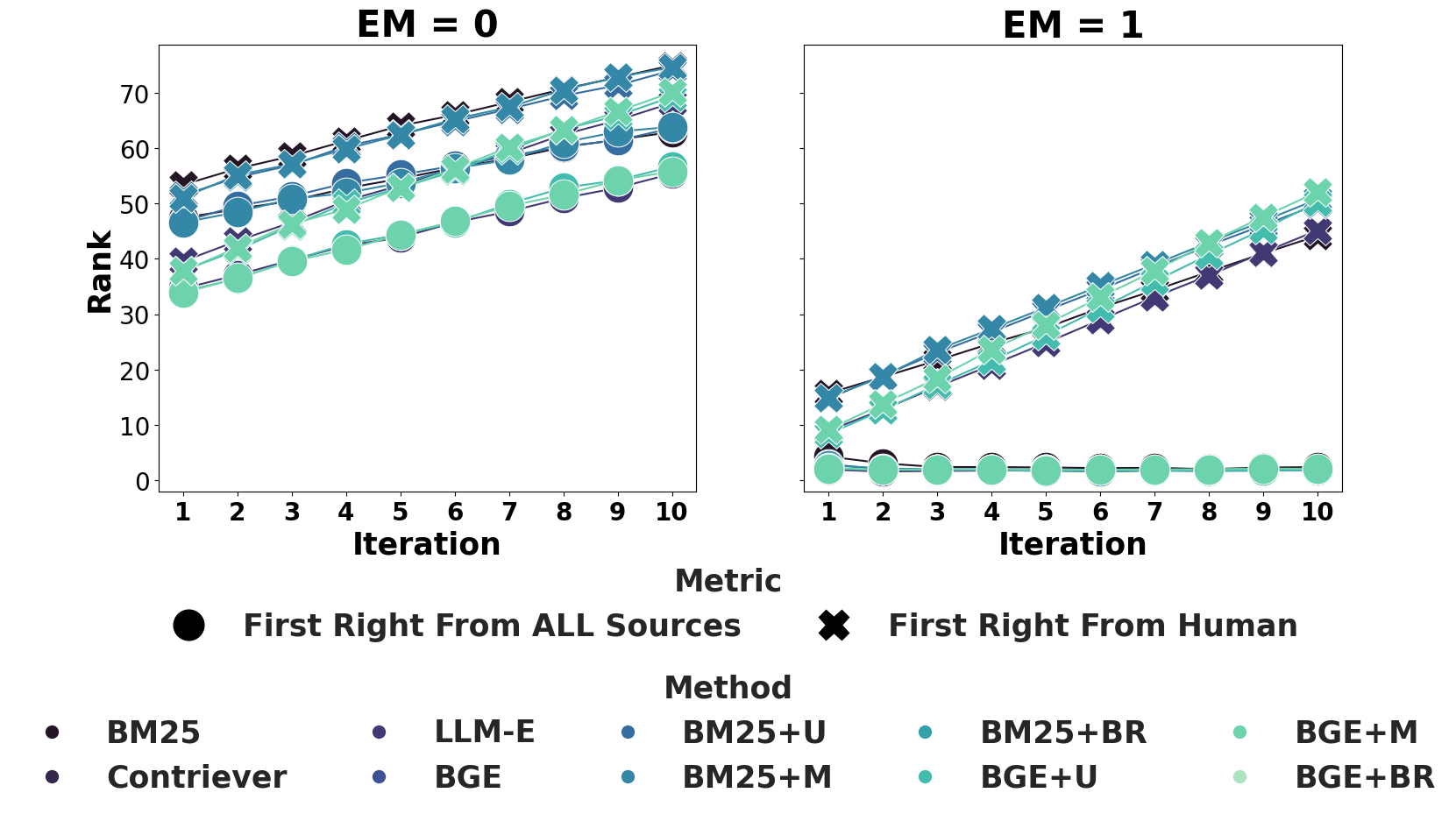}
    \caption{Ranking of the first retrieved document containing the correct answer in each iteration, with LLM responses being incorrect (EM=0) or correct (EM=1). ``First Right From All Sources'' refers to the average rank for the first text containing the correct answer, where the source could be either LLM or human; ``First Right From Human'' is for human-only sources, both considered across datasets and LLMs.}
    \label{fig:rank}
  \end{subfigure}
  
  \caption{Effects of ``Spiral of Silence'' on query and document level of RAG systems.}
  \label{fig:filter_Percentage}
\end{figure*}

%% file: parts/analysis.tex
\section{Analysis}
To further comprehend the ``Spiral of Silence'' effect, we illustrate its interaction with misinformation introduced by adversaries using LLMs. Moreover, we test two information filtering mechanisms to alleviate the progression of the effect. For more information on the experimental setup and results, refer to Appendix~\ref{sec:misinfo} and Appendix~\ref{sec:alleviate}.

%% file: parts/limit.tex
\section*{Limitations}
This study aims to present a new perspective on the impact of LLM-generated texts entering the internet on RAG systems. However, the complexity of reality means that it is impossible to account for all variables. 
The methods of LLM text generation and the mechanisms by which this content enters the retrieval set are constantly changing, which could affect the performance of RAG systems.
While ODQA serves as an insightful approach to evaluate the progression of RAG systems, it is necessary to recognize that ODQA assessments are not exhaustive in capturing the full spectrum of information retrieval scenarios. Nonetheless, the simulation framework proposed in this research is readily adaptable to other tasks that employ RAG systems. Our discussion introduces the ``Spiral of Silence'' as a potential outcome of the proliferation of LLM-generated texts. Although such a development is not predetermined, given the myriad of factors at play in the real world, this work aims to foster a deeper investigation into the phenomenon and its prospective implications for information diversity in AI-mediated environments.

\section*{Ethical Considerations}

This paper only explores the potential impact of the LLM-generated text, without involving the release of the generated text and the intervention of social progress, so the possibility of ethical risks is small.
We used publicly available LLMs and datasets to conduct experiments that did not involve any ethical issues.
In the appendix, we analyze the potential interplay between harmful information and the phenomena outlined in our paper, with a principal objective to draw attention to this issue to advocate for its resolution.

%% file: parts/appendix.tex
\section{Appendix}
\label{sec:appendix}

\subsection{Discussion of Application on ``Spiral of Silence''}\label{sec:discussion}
\input{parts/more_info_about_sos}

\subsection{Pseudo-Code of Simulation Process}\label{code:pseudo}
\input{parts/alg}
The pseudo-code of the simulation process in section~\ref{sec:simulation} is shown in Algotirhm~\ref{alg:simulation}.

\subsection{Post-Process Details}\label{sec:postprocess}
During the experimental process, we observe that the response texts from LLMs occasionally contain specific phrases at the beginning that indicate their identity. These phrases are difficult to remove through prompts and are irrelevant to the topic at hand. Examples include sentences such as:

\begin{itemize} \item \texttt{``I'd be happy to assist you with your question.''}  \item \texttt{``According to my knowledge...''}  \item \texttt{``As an AI language model...''}  \end{itemize}

We collect over 40 such sentences using a manual annotation approach and filter each LLM-generated text through string matching. If a matching string is found, the corresponding sentence or fragment is removed.

\subsection{Implementation Details.}\label{sec:implement}
\input{parts/implementation}

\subsection{Results on WebQ and TriviaQA}~\label{sec:app_fig_wt}
Figure~\ref{fig:Long-Term-app} shows long-term RAG performance on WebQ and TriviaQA. 

The percentage from various sources and the Self-BLEU of the retrieval results on WebQ and TriviaQA are shown in Figure~\ref{fig:Percentage_app} and Figure~\ref{fig:BLEU_app}.
\input{parts/tables/long_term_appendix}
\input{parts/tables/percentage_appendix}
\input{parts/tables/bleu_appendix}
\input{parts/evolution_appendix}
\input{parts/analysis_appendix}
\input{parts/prompts}

%% file: parts/more_info_about_sos.tex
 In aligning the ``Spiral of Silence'' theory with the focus of this study, emphasis on the aspect of the ``individual's will to express'' inherent in the original theory is purposefully diminished. The factors influencing the ``Spiral of Silence'' phenomenon, as mentioned in ~\citet{scheufle2000twenty}, with media and temporality being the principal elements within RAG systems, directly affect the relative standing of LLM and human texts as the system evolves. While the individual's desire to express may be indirectly affected by media and temporality, these are not the primary drivers of the ``Spiral of Silence'' within RAG systems. In RAG systems, we hypothesize that texts generated by LLMs will increasingly be favored in the hierarchy of information retrieval, whereas texts authored by humans might be systematically marginalized, resulting in a structural form of ``passive silencing".

%% file: parts/alg.tex
\begin{algorithm}[t]
\caption{Simulation Process}
\label{alg:simulation}
\begin{algorithmic}

\Function{RunRAG}{$D$, $Q$}
    \State $Res \gets$ empty list
    \For{$q \in Q$}
        \State $D' \gets$ \Call{Retrieve}{$q, D$}
        \State $p \gets$ \Call{GenPrompt}{$q$}
        \State $S \gets$ \Call{GenAnswer}{$p, D', q$}
        \State $S' \gets$ \Call{PostProc}{$S$}
        \State Add $(q, D', S')$ to $Res$
    \EndFor
    \State \Return $Res$
\EndFunction
\Statex 
\State $D_0 \gets$ \Call{LoadData}{}

\State $initRes \gets$ \Call{RunRAG}{$D_0, Q$}
\State $basePerf \gets$ \Call{EvalRAG}{$initRes$}

\State $T \gets$ \Call{GenZeroShot}{}
\State $D_1 \gets D_0$ \textbf{combined with} $T$

\State $t \gets$ number of iterations
\For{$i \gets 1$ \textbf{to} $t$}
    \State $iterRes \gets$ \Call{RunRAG}{$D_i, Q$}
    \State $perf_i \gets$ \Call{EvalRAG}{$iterRes$}
    \State $D_{i+1} \gets$ \Call{UpdateData}{$D_i, iterRes$}
\EndFor
\end{algorithmic}
\end{algorithm}

%% file: parts/implementation.tex
To construct and execute the simulated iterative framework, we adopt a diverse array of tools and technologies to facilitate real-time interaction between various retrieval methods, indexing architectures, and LLMs. We implement the APIs of various LLMs relying on api-for-open-llm\footnote{\url{https://github.com/xusenlinzy/api-for-open-llm}}.
With integration of LangChain\footnote{\url{https://github.com/langchain-ai/langchain}} with Faiss~\citep{faiss} and Elasticsearch\footnote{\url{https://github.com/elastic/elasticsearch}}, we execute batched incremental updates of LLM-generated documents in each iteration, thus simulating the process of document index updating by search engines in real-world scenarios. To maintain the diversity of the generated texts, we set the temperature at 0.7 for all LLMs.
In each iteration of the experiment, except for the zero-shot setting, 
we keep the size of the context document set \( D_i' \) fixed at 5. We rerank the first 100 documents recalled by the retrieval method when the step is applied. 
We apply the LLMs to generate response text, post-process via rules, and then merge their outputs into the index for each query in every iteration.
Therefore, for each iteration, we will add 4k new samples to the index. The total number of iterations \(t\) is set at 10, which results in a total of 40k invocations of the LLMs for each experimental run.

%% file: parts/tables/long_term_appendix.tex
\begin{figure}[t]
  \centering

  \begin{subfigure}[t]{0.19\textwidth}
    \includegraphics[width=\textwidth]{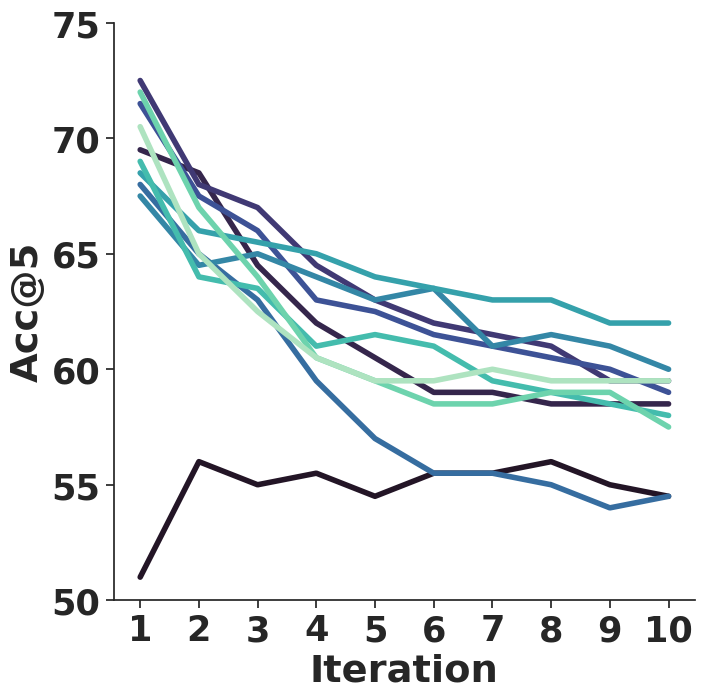}
    \caption{WebQ}
    \label{fig:webq_loop_retrieval}
  \end{subfigure}
  \hfill
  \begin{subfigure}[t]{0.28\textwidth}
    \includegraphics[width=\textwidth]{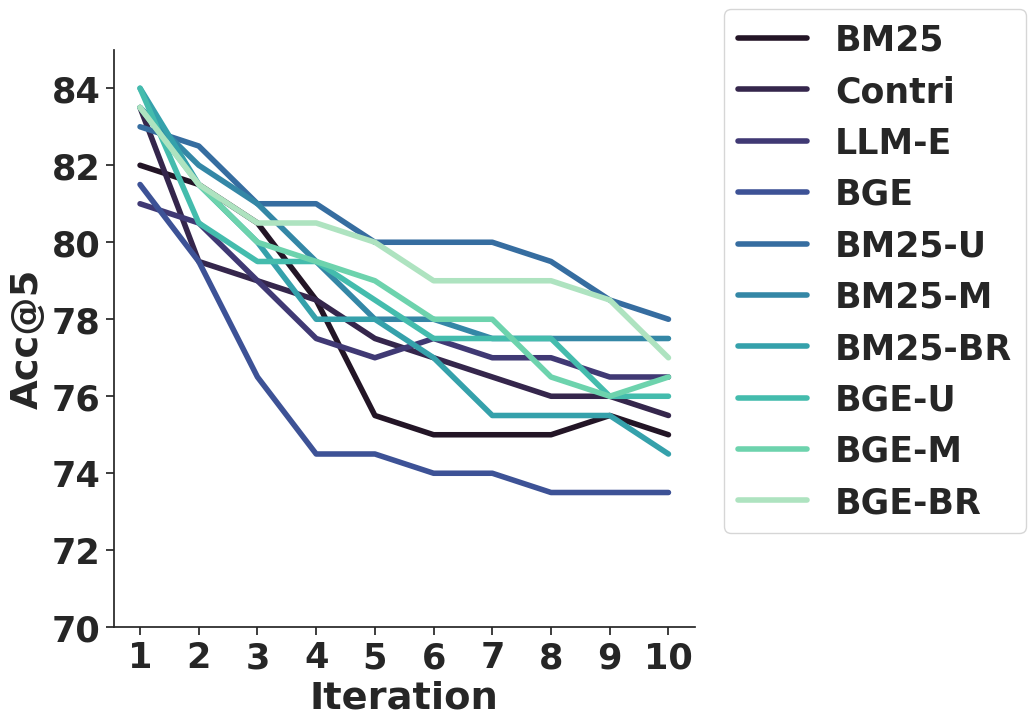}
    \caption{TriviaQA}
    \label{fig:tqa_loop_retrieval}
  \end{subfigure}

  \vspace{0.2cm}
  
  \begin{subfigure}[t]{0.19\textwidth}
    \includegraphics[width=\textwidth]{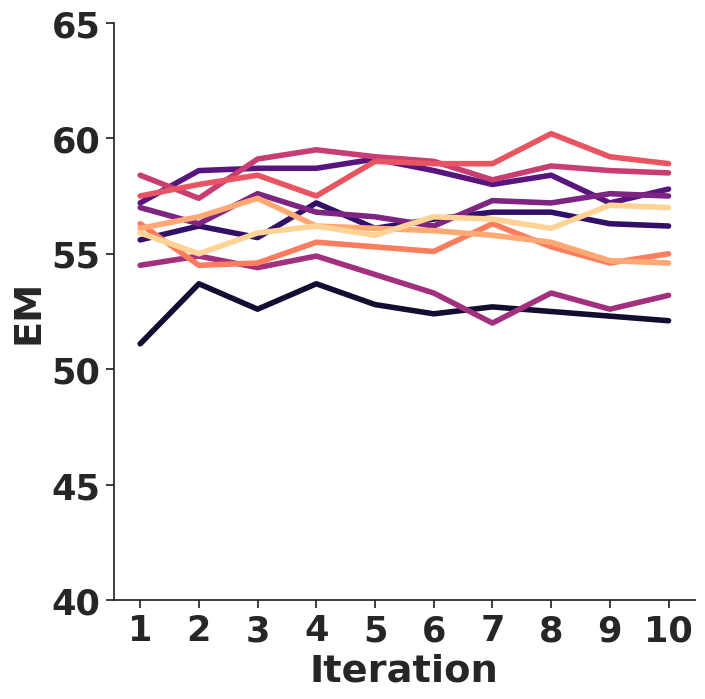}
    \caption{WebQ}
    \label{fig:webq_loop_qa}
  \end{subfigure}
  \hfill
  \begin{subfigure}[t]{0.28\textwidth}
    \includegraphics[width=\textwidth]{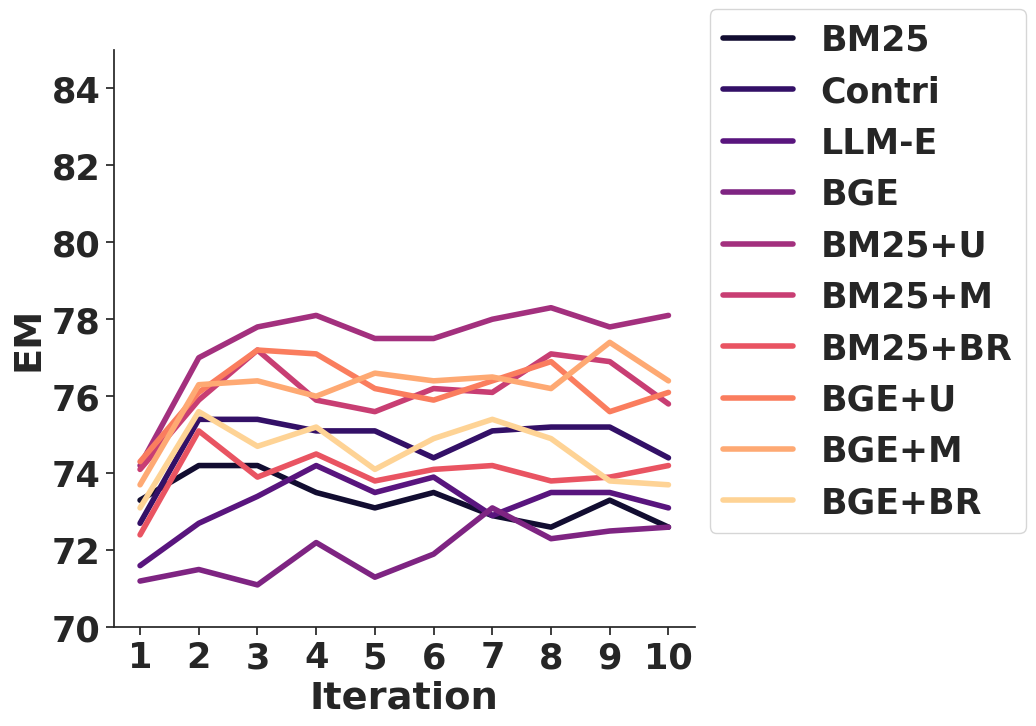}
    \caption{TriviaQA}
    \label{fig:tqa_loop_qa}
  \end{subfigure}
  
  \caption{Long-Term RAG performance for WebQ and TriviaQA. The upper section illustrates the retrieval outcomes for various methods, while the lower section depicts the average EM across LLMs. Iteration 1 represents the results following the incorporation of zero-shot LLM-generated text.  Abbreviated re-ranking methods in the legend are: +U for UPR, +M for MonoT5, and +BR for BGE-Reranker.}
  \label{fig:Long-Term-app}
\end{figure}

%% file: parts/tables/percentage_appendix.tex
\begin{figure}[t]
  \centering
    \begin{subfigure}[t]{0.19\textwidth}
    \includegraphics[width=\textwidth]{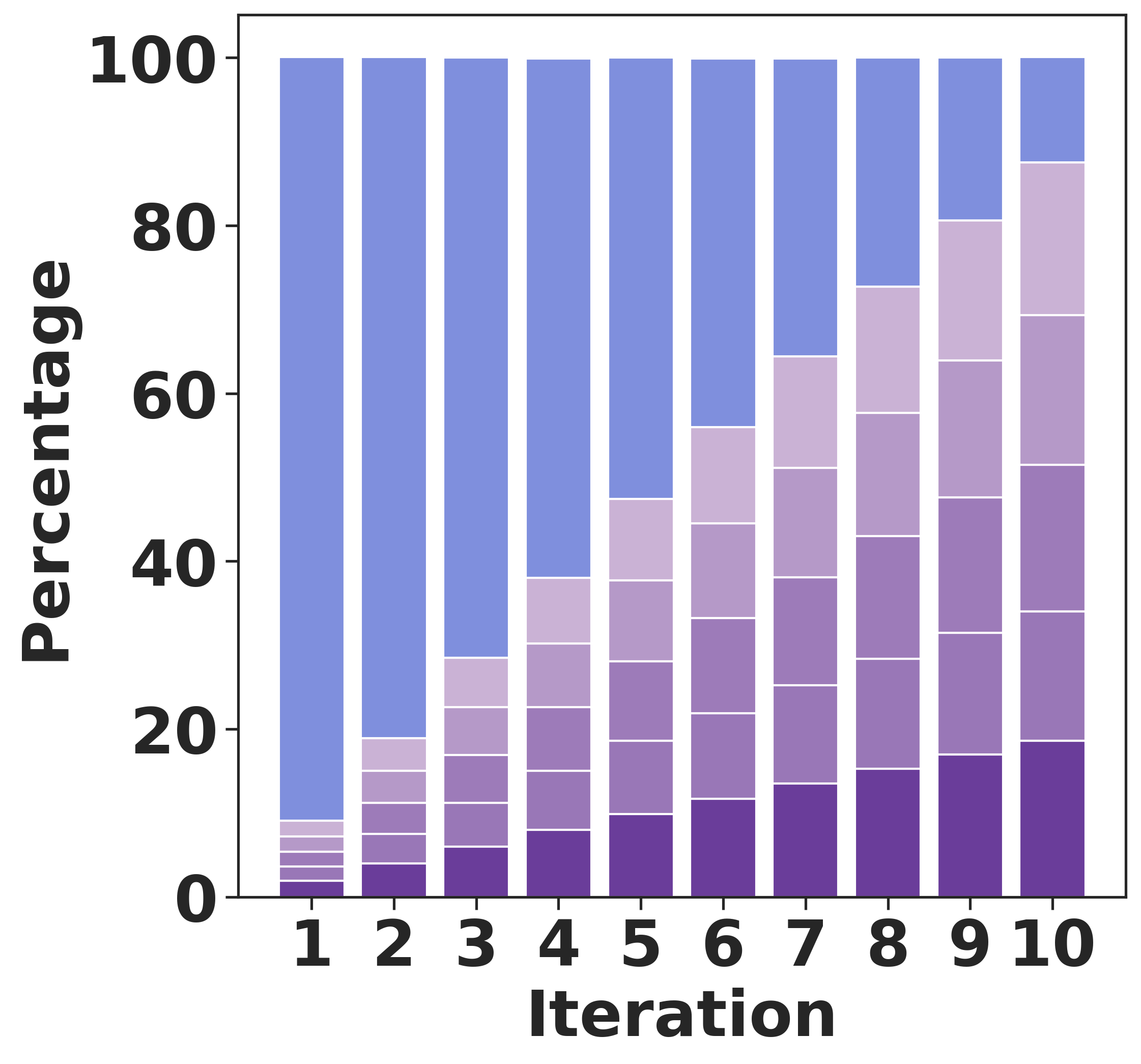}
    \caption{WebQ}
    \label{fig:webq_per}
  \end{subfigure}
  \hfill
    \begin{subfigure}[t]{0.28\textwidth}
    \includegraphics[width=\textwidth]{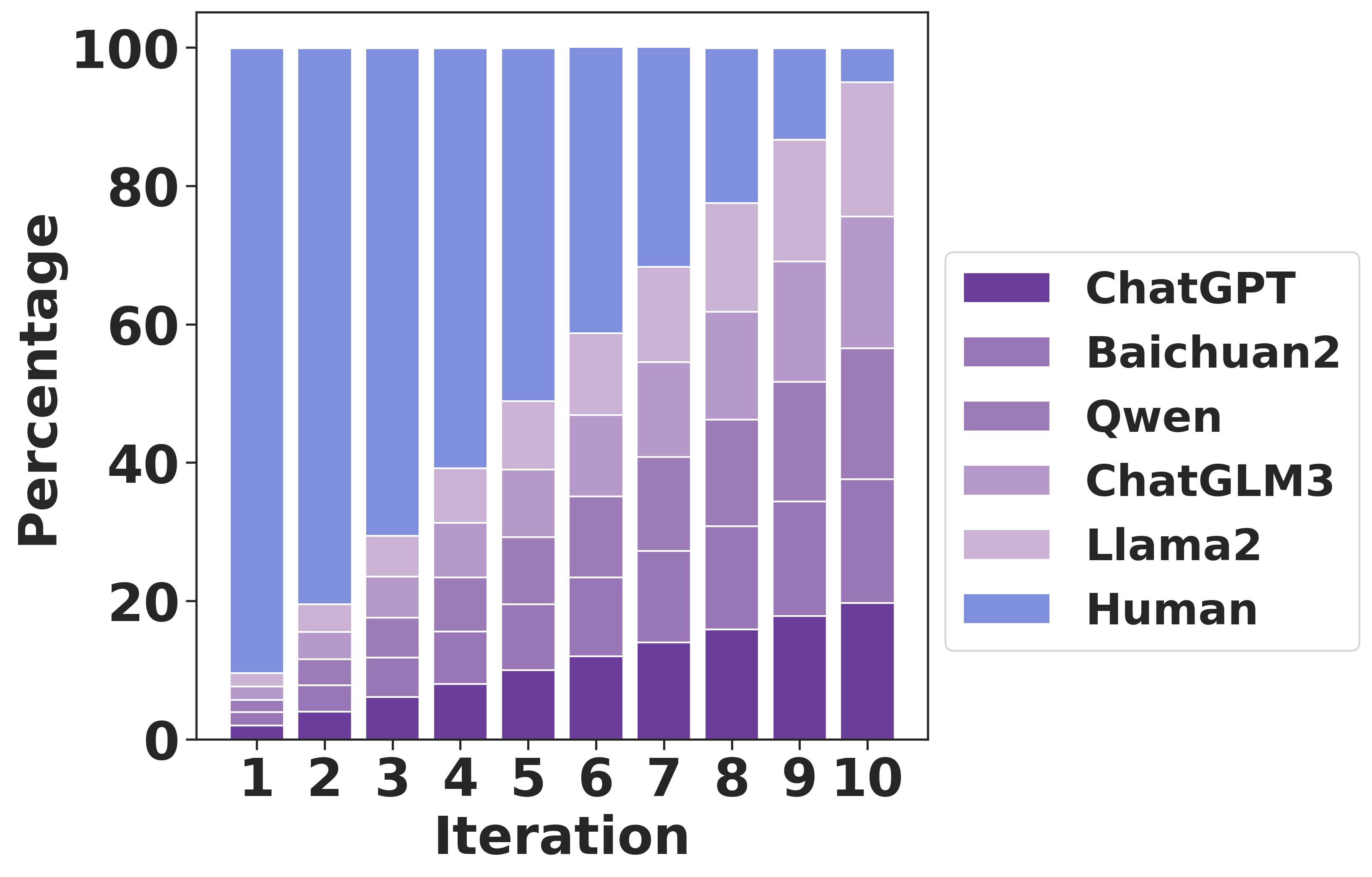}
    \caption{TriviaQA}
    \label{fig:tqa_per}
  \end{subfigure}
  \caption{Average Percentage of texts from various sources within the top 50 search results over multiple iterations across different search and methods for WebQ and TriviaQA.}
  \label{fig:Percentage_app}
\end{figure}

%% file: parts/tables/bleu_appendix.tex
{\setlength{\belowcaptionskip}{-5pt}
\begin{figure}[tbp]
  \centering
  \begin{subfigure}[t]{0.18\textwidth}
    \includegraphics[width=\textwidth]{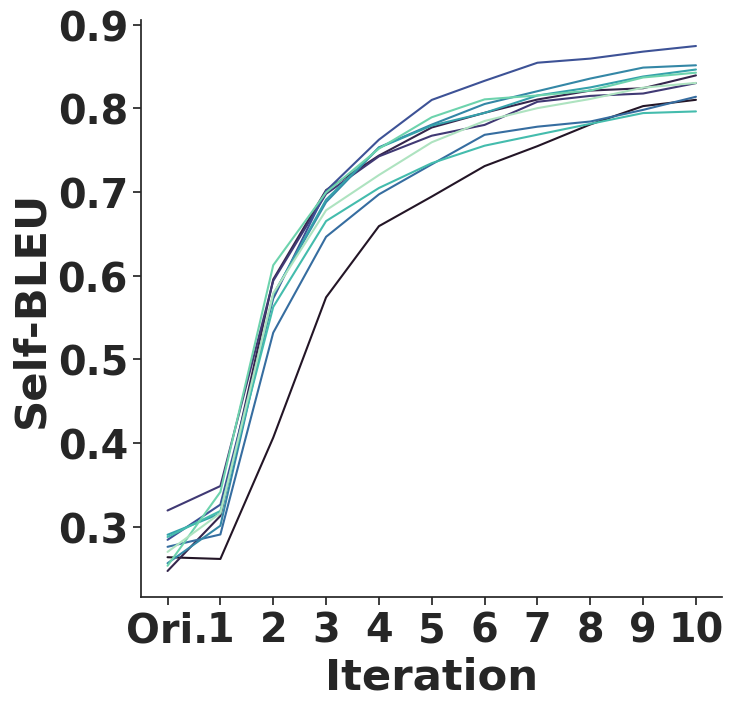}
    \caption{WebQ}
    \label{fig:webq_bleu}
  \end{subfigure}
  \hfill
  \begin{subfigure}[t]{0.29\textwidth}
    \includegraphics[width=\textwidth]{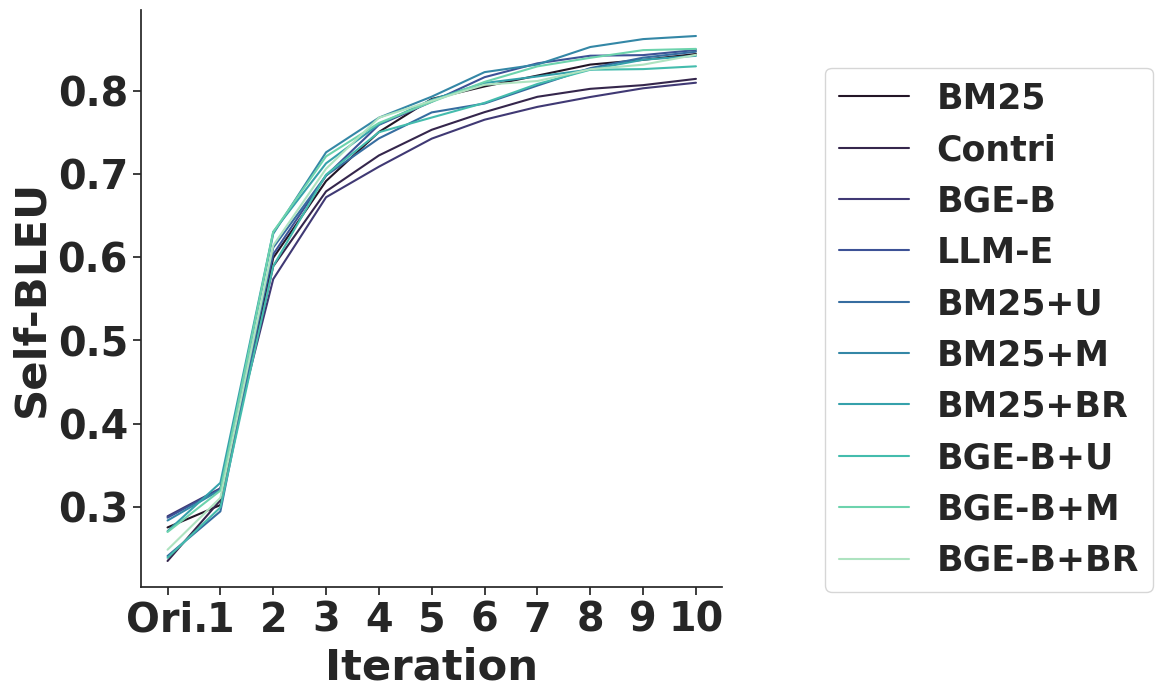}
    \caption{TriviaQA}
    \label{fig:tqa_bleu}
  \end{subfigure}
  \caption{3-gram Self-BLEU score for the top 5 search results over iterations for WebQ and TriviaQA, from the original dataset (Ori.) to subsequent iterations including LLM-generated texts. }
  \label{fig:BLEU_app}
\end{figure}}

%% file: parts/evolution_appendix.tex
\subsection{Simulation with Model Evolution}\label{sec:evolution}
In this section, we explore the impact of the continuous influx of text generated by evolving LLMs on RAG systems over time. To this end, we experiment by using a series of LLMs, ranging from weak to strong, to observe the phenomenon.

\textbf{Experimental Setup:} We conduct experiments based on BM25 on the NQ and WebQ datasets. For the text generation models, we use progressively larger versions of the Qwen1.5 model~\citep{qwen} across different iterations: 0.5b, 1.8b, and 4b versions for the first three iterations. For iterations four to seven, we employ Qwen1.5-7b-Chat~\citep{qwen}, LLaMA2-7b-Chat~\citep{DBLP:journals/corr/abs-2302-13971}, and Baichuan2-7b-Chat~\citep{baichuan}. Finally, for the eighth to tenth iterations, we utilize GPT-3.5-Turbo~\citep{openai_chatgpt}, LLaMA2-13b-Chat~\citep{DBLP:journals/corr/abs-2302-13971}, and Qwen1.5-14b-Chat~\citep{qwen}. This setup simulates the impact of increasing model capabilities on the RAG system over time.

\begin{table}[t]
\centering
\resizebox{0.44\textwidth}{!}{
\begin{tabular}{lcccc}
\toprule
Dataset & \multicolumn{2}{c}{NQ} & \multicolumn{2}{c}{WebQ} \\
\cmidrule(lr){2-3} \cmidrule(lr){4-5}
Iteration & Acc@20 & EM & Acc@20 & EM \\
\midrule
1  & 68.5 & 29.5 & 66.0 & 35.5 \\
2  & 67.5 & 28.5 & 65.0 & 38.7 \\
3  & 64.0 & 29.3 & 63.5 & 38.0 \\
4  & 63.0 & 32.5 & 62.5 & 43.5 \\
5  & 60.5 & 32.3 & 62.5 & 41.8 \\
6  & 58.0 & 32.0 & 61.5 & 43.3 \\
7  & 55.5 & 30.7 & 61.5 & 43.2 \\
8  & 52.0 & 33.7 & 60.5 & 44.0 \\
9  & 51.5 & 32.8 & 61.0 & 44.3 \\
10 & 47.5 & 33.7 & 58.5 & 44.2 \\
\bottomrule
\end{tabular}}
\caption{BM25 retrieval performance (Acc@20) and QA performance (EM) across iterations for NQ and WebQ under the evolving LLM setting.}
\label{tab:evo_ret_em}
\end{table}

\begin{table}[t]
\centering
\resizebox{0.27\textwidth}{!}{
\begin{tabular}{lcc}
\toprule
Iteration & NQ & WebQ \\
\midrule
1  & 77.7 & 87.0 \\
2  & 45.9 & 64.3 \\
3  & 27.2 & 48.6 \\
4  & 18.0 & 40.2 \\
5  & 14.6 & 36.2 \\
6  & 12.3 & 32.6 \\
7  & 10.1 & 28.9 \\
8  & 8.6 & 25.6 \\
9  & 6.7 & 21.9 \\
10 & 5.8 & 19.2 \\
\bottomrule
\end{tabular}}
\caption{Percentage of human text in the top 5 BM25 retrieval results across iterations for NQ and WebQ under the evolving LLM setting.}
\label{tab:evo_human}
\end{table}

\begin{table}[t]
\centering
\begin{tabular}{lcc}
\toprule
Model Name          & NQ    & WebQ  \\
\midrule
Qwen1.5-0.5b-Chat      & 12.5 & 6.9 \\
Qwen1.5-1.8b-Chat      & 7.4 & 4.4 \\
Qwen1.5-4b-Chat        & 8.5 & 6.3 \\
Qwen-7b-Chat        & 1.1 & 1.7 \\
LLaMA2-7b-Chat      & 20.9 & 18.5 \\
Baichuan2-7b-Chat   & 10.3 & 7.6 \\
LLaMA2-13B-Chat     & 19.1 & 19.9 \\
Qwen1.5-14b-chat       & 0.9 & 1.8 \\
GPT-3.5-Turbo       & 13.5 & 13.7 \\
Human               & 5.8 & 19.2 \\
\bottomrule
\end{tabular}
\caption{Percentage of generated text by different LLMs in the top 5 BM25 retrieval results at the end of the simulation (Iteration 10) for NQ and WebQ under the evolving LLM setting.}
\label{tab:evo_llm}
\end{table}

\begin{figure*}[htbp]
  \centering
    \includegraphics[width=\textwidth]{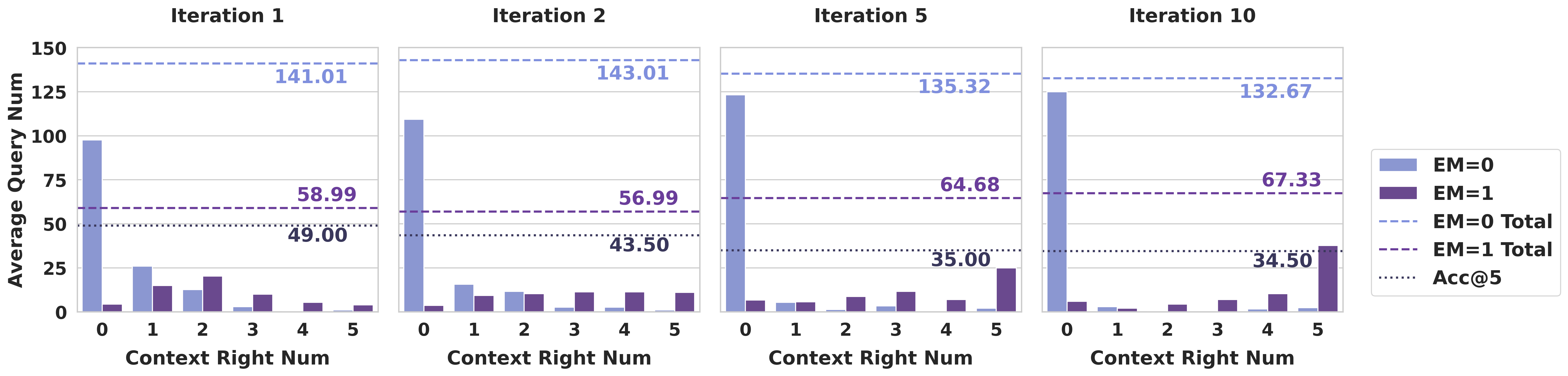}
    \caption{Correlation between the number of top 5 BM25 search results containing the correct answer ("Context Right Num") and the accuracy of responses given by evolving LLMs on the NQ dataset. The responses are categorized based on n Exact Match (EM) score: EM=1 for correct and EM=0 for incorrect. The overall number of queries that the LLMs answered correctly (EM=1 Total) and incorrectly (EM=0 Total), along with the average retrieval accuracy (Acc@5)
    are shown by dashed lines. The results are averaged across different LLMs.}
    \label{fig:evo_nq_context}
\end{figure*}

\begin{table}[t]
\centering
\begin{tabular}{lcc}
\toprule
Iteration & NQ & WebQ \\
\midrule
1  & 27.5 & 25.4 \\
2  & 49.8 & 39.5 \\
3  & 65.6 & 53.3 \\
4  & 74.4 & 60.9 \\
5  & 70.8 & 59.2 \\
6  & 72.3 & 60.9 \\
7  & 74.8 & 63.6 \\
8  & 75.5 & 64.5 \\
9  & 75.3 & 64.7 \\
10 & 78.2 & 68.4 \\
\bottomrule
\end{tabular}
\caption{3-gram Self-BLEU score for the top 5 BM25 search
results over iterations for NQ and WebQ under the evolving LLM setting.}
\label{tab:bleu-evo}
\end{table}

\textbf{Analysis of Experimental Results:}
From the experimental results, we can observe that: 
\textbf{1) Retrieval Performance Continues to Decline:} According to the Acc@20 results reported in Table~\ref{tab:evo_ret_em}, even with the gradual improvement of LLM performance in the simulation, the retrieval performance exhibits a general downward trend over time. Compared to the results in Table~\ref{tab:combined_ret}, the improvements in retrieval performance are also limited for weaker LLMs in the initial stage (73.5 $\rightarrow$ 68.5, 71.0 $\rightarrow$ 66.0).
\textbf{2) QA Results Show Improvement:} Based on the EM results reported in Table~\ref{tab:evo_ret_em}, we can see that with the expansion of the LLM scale, the QA results have improved, especially from the third to the fourth iteration, where the model expanded from less than 4B to 7B, resulting in more than a 10\% improvement in QA results on both datasets. Considering that QA results remain stable in Figure~\ref{fig:Long-Term} when LLMs are unchanged, the current improvement likely stems from the enhanced capabilities of the LLM itself.
\textbf{3) The Existence of the ``Spiral of Silence'' Phenomenon:} In Table~\ref{tab:evo_human}, we present the proportion of human text in the top 5 retrieval results during the iterations. It can be observed that LLM-generated text still dominates the retrieval results, and the influence of human text continues to decline. In Table~\ref{tab:evo_llm}, we report the proportion of different LLM-generated texts in the top 5 retrieval results after the tenth iteration. Overall, texts generated by larger-scale LLMs occupy a slightly higher proportion, but this is not absolute.
Table~\ref{tab:bleu-evo} presents the diversity of retrieval results, while Figure~\ref{fig:evo_nq_context} illustrates the relationship between retrieval accuracy and the generated answers. As the model evolves, the experimental results remain consistent with the conclusion in Section~\ref{sec:sos}: the diversity of retrieval results continues to decline, and polarization occurs. In summary, the experimental results indicate that the evolution of LLMs will not prevent the occurrence of the ``Spiral of Silence'' phenomenon.

%% file: parts/analysis_appendix.tex
\subsection{Effects of Misinformation}\label{sec:misinfo}
\input{parts/tables/mis_zero-shot}
In previous sections, we explored the impact of non-maliciously LLM-generated texts on the evolution of the RAG system over time. In this section, we will discuss the persistence of the ``Spiral of Silence'' when attackers deliberately inject specific misinformation into the RAG system, how misinformation could affect the system over time, and the feasibility of targeted misinformation injection.

\textbf{Experimental Setup}: Our experiment follows the CTRLGEN method detailed in ~\citet{DBLP:conf/emnlp/PanPCNKW23}, which aligns well with the zero-shot setting used in our trials and simulates the intent of malicious actors to create and propagate false information. Specifically, for each query, we generate five incorrect answers using GPT-3.5-Turbo and then randomly select one to guide five different LLMs to each create a document supporting that incorrect response. These documents replace the zero-shot data in the index from the experiments in Section~\ref{sec:pipeline} and are used for simulated iterative experiments. Details of the prompts used are provided in Appendix~\ref{sec:prompt}. For the sake of conciseness, we report only the experimental results for four retrieval methods.

\textbf{Experimental Evaluation}: When generating texts containing misinformation using LLMs, we face two primary challenges. First, the model may ignore the instructions, thus inadvertently generating texts that only contain the correct answer. Second, even if the LLM-generated text includes the provided incorrect answer, the content may not genuinely support that answer. To address these issues, we utilize GPT-3.5-Turbo to evaluate the alignment of text $t$ with the given answer $a$, which could be either correct or incorrect. This evaluation complements the calculation of the EM metric for texts generated by LLMs. We define the $\text{EM}_{llm}$ metric as follows:
\begin{equation} 
\text{EM}_{llm}(t, a) = \begin{cases} 
1, & \text{if } t \text{ contains and supports } a \\
0, & \text{otherwise} 
\end{cases} 
\end{equation}
Here, $a$ represents an answer that the text $t$ is being evaluated against. The text $t$ is deemed to contribute positively to this metric if it encompasses $a$ and is validated by GPT-3.5-Turbo as being supportive of $a$. The $\text{EM}_{llm}$ for the correct answers and specific incorrect answers, as generated by the CTRLGEN method containing misinformation, are presented in Table~\ref{tab:mis_zero-shot}.

Moreover,  we substantiate the rationality of employing EM as a QA evaluation metric in the experiments of Section~\ref{sec:exp} by calculating the Pearson correlation coefficient between $\text{EM}_{llm}$ and EM based on the experimental results in this section. We observe that the $\text{EM}_{llm}$ values for the texts generated by the other four LLMs, as verified by GPT-3.5-Turbo, have an average correlation exceeding 0.5 across four datasets when compared to the EM values obtained through direct string matching, as shown in Table~\ref{tab:pearson}. This demonstrates a significant correlation between the two metrics. For incorrect answers, the correlation is relatively lower, indicating the necessity of using GPT-3.5-Turbo to further filter texts in the exploration of misinformation. Considering the higher efficiency of direct EM calculation over $\text{EM}_{llm}$, we use EM to evaluate the QA quality of the RAG system for experiments in other sections.
\input{parts/tables/pearson}

\input{parts/tables/mis_dominance}
\textbf{Analysis of Experimental Results}: From the experimental results, we can observe that: \textbf{1) The ``Spiral of Silence'' Still Exists:} We first investigate the presence of the ``Spiral of Silence'' phenomenon when misinformation targeting the objective is injected into the corpus. As shown in Table~\ref{tab:mis_llm_dominance}, although the majority of the injected information is misleading, the content generated by the LLMs is still quickly ranked at the top by the retrieval systems, taking a dominant position. When comparing four different retrieval methods, the BM25 algorithm shows greater robustness than the others, being least affected by the LLM-generated content. However, it is noteworthy that approximately 20\% of the content generated by the LLM could still be quickly placed in the forefront of the search results by the BM25 algorithm. 
\input{parts/tables/mis_percentage_figure}
\input{parts/tables/mis_context}
Figure~\ref{fig:mis_Percentage} illustrates that over time, human-written texts are gradually excluded from the searchable range, and as depicted in Figure~\ref{fig:mis_nq_context}, the phenomenon of homogenization of opinions in search results persists. This further indicates that regardless of the accuracy of the LLM-generated information, the ``Spiral of Silence'' phenomenon remains present.
\textbf{2) The RAG System has a Limited Degree of Self-Correction Capability:} LLM-generated texts containing misinformation lead to a significant decline in retrieval and QA performance based on the RAG system compared to results in Section~\ref{sec:long-term}. With the continuation of the iteration process, the number of correct answers increased and the number of incorrect answers decreased, albeit by a small margin. This suggests that the RAG system has a certain degree of self-correction capability, which may stem from the model's knowledge or the human-written texts containing correct information retrieved in the initial stages. 
\input{parts/tables/mis_QA}
\textbf{3) The Introduction of a Small Amount of the LLM-generated Texts with Specific Misleading Information during the Iterative Process could Inject such Information into the RAG Output:} In Figure~\ref{fig:mis_QA}, we quantify the $\text{EM}_{llm}$ metric of the original and specific misleading answers generated by the RAG system based on four retrieval methods at various iteration stages. The results show that after the purposeful addition of misleading information (before the first iteration), the proportion of RAG system-generated answers containing specific misleading information significantly increases, especially on NQ and PopQA, where the proportion of incorrect answers exceeds that of correct ones, and the influence of misleading answers persisted over time. However, the BM25 algorithm exhibits relatively higher robustness, and the $\text{EM}_{llm}$ of incorrect answers output by the RAG system based on it remains lower than the other three retrieval methods. The experimental results of this section reveal that despite the presence of self-correcting mechanisms, the injection of specific misleading information can still severely compromise the system's accuracy and enable the manipulation of the RAG system to consistently output specific misinformation in response to certain questions. Therefore, without timely intervention, the ``Spiral of Silence'' phenomenon could marginalize accurate information, leading to severe misinformation consequences.

\subsection{Attempts to Alleviate the ``Spiral of Silence"}\label{sec:alleviate}
The ``Spiral of Silence'' effect could lead to the marginalization of human-generated text expression and further enhance the homogeneity of retrieval outcomes. If left unaddressed, this phenomenon could precipitate a series of adverse repercussions. To mitigate or eliminate the influence of the ``Spiral of Silence'' effect, this section initiates a discussion on two fronts. First, from the perspective of the authenticity of sources, we employ the widely used AIGC detection technologies to filter out and exclude all non-human-produced texts at the top of the search results. Second, addressing the validity of content, we strive to maintain diversity among the top search results to overcome potential issues caused by excessive homogenization.
\input{parts/tables/filter_retrieval}
\input{parts/tables/Filter_QA}

\textbf{Experimental Setup}: To balance the efficiency and effectiveness of the retrieval system, for each set of search results returned by the system, we post-process to acquire the top 5 qualifying documents that are visible to the LLMs. In the \textbf{source filtering} experiment, we employ the Hello-SimpleAI/chatgpt-qa-detector-roberta\footnote{\url{https://github.com/Hello-SimpleAI/chatgpt-comparison-detection}} model to authenticate the origins of the texts within the search results, aiming to retain the first 5 documents identified as human-generated and supply them as input to the LLM's context. For the \textbf{content filtering} part of the experiment, we apply a selection process based on computing the 3-gram Self-BLEU scores. The specific procedure is as follows: For the top 5 documents returned for each search query, we initially calculate their Self-BLEU scores; if the score exceeds a predetermined threshold (set at 0.4 for this experiment), we then compute the Self-BLEU scores for all possible combinations of 4 documents and select the minimum value among them. This minimum value indicates the maximum individual document contribution to the Self-BLEU score not included in the calculation. Subsequently, we exclude the document contributing the most to the Self-BLEU score and incorporate the next ranked document into the combination, repeating this filtering process until the combination's Self-BLEU score meets the preset threshold criteria.

\textbf{Analysis of Experimental Results}:
From the experimental results, we can observe that:
\textbf{1) Both Approaches Yield more Stable Retrieval Outcomes; however, the Source Filtering Method Incurs a Performance Cost:}
Figure~\ref{fig:filter_retrieval} and Figure~\ref{fig:filter_qa} illustrate the variations in the average retrieval outcomes and QA performance across datasets, before and after the application of two distinct filtering strategies, compared to an unfiltered condition. Observations indicate that by implementing source-based and diversity-based filtering methods, the fluctuation range of the top 5 retrieval results is reduced compared to the non-intervention scenario, suggesting that the filtering mechanisms can bring a more stable retrieval performance for RAG systems. Across the four datasets, the retrieval performance following SELF-BLEU value filtering generally surpasses the unfiltered condition; conversely, the source-based filtering strategy results in an overall performance degradation. This could be attributed to the discriminating model erroneously excluding valid human-generated texts while aiming to eliminate those generated by LLMs. Moreover, in QA tasks, diversity filtering either enhances or maintains QA performance, whereas source-based document filtering leads to a decline in QA performance across all datasets. For instance, on the TriviaQA dataset, the average EM score drops by over 14\%. 
\textbf{2) Both Methods can only Alleviate the ``Spiral of Silence'' Phenomenon to Varying Degrees but Cannot Eliminate it:}
Figure~\ref{fig:filter_Percentage} displays the proportion of documents from different sources within the top 5 retrieval results in each iteration under three filtering setups on the NQ dataset. It is observable that without any filtering strategy, human-generated texts rapidly vanish from the top 5 documents in the initial iterations. The SELF-BLEU value filtering method retains human-generated texts to a small extent; source filtering, on the other hand, maximally filters out LLM-generated texts, especially those produced by GPT-3.5-Turbo, Qwen, and ChatGLM3, with over 30\% of human-generated texts remaining in the top 5 by the end of the tenth iteration. However, despite both filtering strategies slowing the disappearance of human texts, the proportion of human-generated content continues to exhibit a declining trend.
Figure~\ref{fig:filter_bleu_nq_context} and Figure~\ref{fig:filter_source_nq_context} demonstrate that compared to the absence of filtering strategies, both filtering methods slow down the polarization speed of top document accuracy in retrieval performance. Overall, we discovered that filtering based on the source of documents and their diversity can, to some extent, slow down the emergence of the “Spiral of Silence” phenomenon. Source-based filtering has a more pronounced effect in terms of preserving the proportion of human-generated texts and mitigating viewpoint polarization; however, this benefit comes at the expense of the performance of the RAG system. Text filtering based on diversity shows superior performance in maintaining RAG system functionality, but it has a weaker impact on preserving the ratio of human texts and alleviating viewpoint polarization. Despite these findings, neither method can completely eradicate the ``Spiral of Silence'' effect, indicating the imperative to explore additional solutions. For example, there is a need to investigate retrieval models that can effectively balance between LLM-generated documents and human-generated documents to address this issue.
\input{parts/tables/top5_filter_percentage}
\input{parts/tables/filter_bleu_context}
\input{parts/tables/filter_source_context}

%% file: parts/tables/mis_zero-shot.tex
\begin{table*}[htbp]
\centering
\resizebox{0.97\textwidth}{!}{
\begin{tabular}{@{}lcccccccc@{}}
\toprule
Model        & \multicolumn{2}{c}{NQ} & \multicolumn{2}{c}{WebQ} & \multicolumn{2}{c}{TriviaQA} & \multicolumn{2}{c}{PopQA} \\ 
             & Correct    & Incorrect         & Correct     & Incorrect         & Correct        & Incorrect           & Correct      & Incorrect         \\ 
\midrule
GPT-3.5-Turbo & 0.015    & 0.7           & 0.075     & 0.57          & 0.105        & 0.57            & 0.045      & 0.71          \\
Baichuan2-13B-Chat      & 0.11     & 0.595         & 0.21      & 0.44          & 0.16         & 0.455           & 0.1        & 0.65          \\
Qwen-14B-Cha          & 0.065    & 0.61          & 0.11      & 0.565         & 0.165        & 0.535           & 0.05       & 0.7           \\
ChatGLM3-6B       & 0.085    & 0.605         & 0.195     & 0.435         & 0.245        & 0.415           & 0.105      & 0.61          \\
LLaMA2-13B-Chat         & 0.04     & 0.43          & 0.085     & 0.385         & 0.125        & 0.405           & 0.03       & 0.55          \\
\midrule
Avg           & 0.063    & 0.588         & 0.135     & 0.479         & 0.16         & 0.476           & 0.066      & 0.644         \\
\bottomrule
\end{tabular}}
\caption{EM$_{llm}$ of different models for Correct answers and specific Incorrect answers.}
\label{tab:mis_zero-shot}
\end{table*}

%% file: parts/tables/pearson.tex
\begin{table}[t]
\centering
\resizebox{0.45\textwidth}{!}{
\begin{tabular}{lcccc}
\hline
\textbf{Answer Type}        & NQ  & WebQ & TriviaQA  & PopQA  \\ \hline
Correct & 0.740  & 0.568 & 0.836 & 0.529 \\
Incorrect & -0.574 & -0.389 & -0.385 & -0.121 \\ \hline
\end{tabular}}
\caption{Evaluation of the Average Pearson Correlation Coefficient between EM and $\text{EM}_{llm}$ for Correct and Incorrect Answers.}
\label{tab:pearson}
\end{table}

%% file: parts/tables/mis_dominance.tex
\begin{table}[t]
\centering
\resizebox{0.45\textwidth}{!}{
\begin{tabular}{@{}lcccc@{}}
\toprule
\textbf{Method} & \textbf{NQ} & \textbf{WebQ} & \textbf{TriviaQA} & \textbf{PopQA} \\
\midrule
BM25 & \ApplyGradientd{50.0}{26.7} & \ApplyGradientd{50.0}{17.7} & \ApplyGradientd{50.0}{24.7} & \ApplyGradientd{50.0}{47.4} \\
Contriever & \ApplyGradientd{50.0}{60.7} & \ApplyGradientd{50.0}{62.5} & \ApplyGradientd{50.0}{64.7} & \ApplyGradientd{50.0}{67.2} \\
LLM-Embedder & \ApplyGradientd{50.0}{65.8} & \ApplyGradientd{50.0}{70.4} & \ApplyGradientd{50.0}{73.3} & \ApplyGradientd{50.0}{74.2} \\
BGE$_{base}$  & \ApplyGradientd{50.0}{50.7} & \ApplyGradientd{50.0}{48.6} & \ApplyGradientd{50.0}{63.2} & \ApplyGradientd{50.0}{60.6} \\

\bottomrule
\end{tabular}
}
\caption{Percentage of LLM-generated documents with \textbf{Misinformation} occupying the top 5 retrieval results, after augmenting each query with five documents generated by LLMs. Data entries framed by a \colorbox{myblue}{blue} background indicate a majority presence of human-generated documents, while entries with a \colorbox{myred}{purple} background denote a predominance of LLM-generated documents.}
\label{tab:mis_llm_dominance}
\end{table}

%% file: parts/tables/mis_percentage_figure.tex
\begin{figure*}[htbp]
  \centering
  \begin{subfigure}[t]{0.22\textwidth}
    \includegraphics[width=\textwidth]{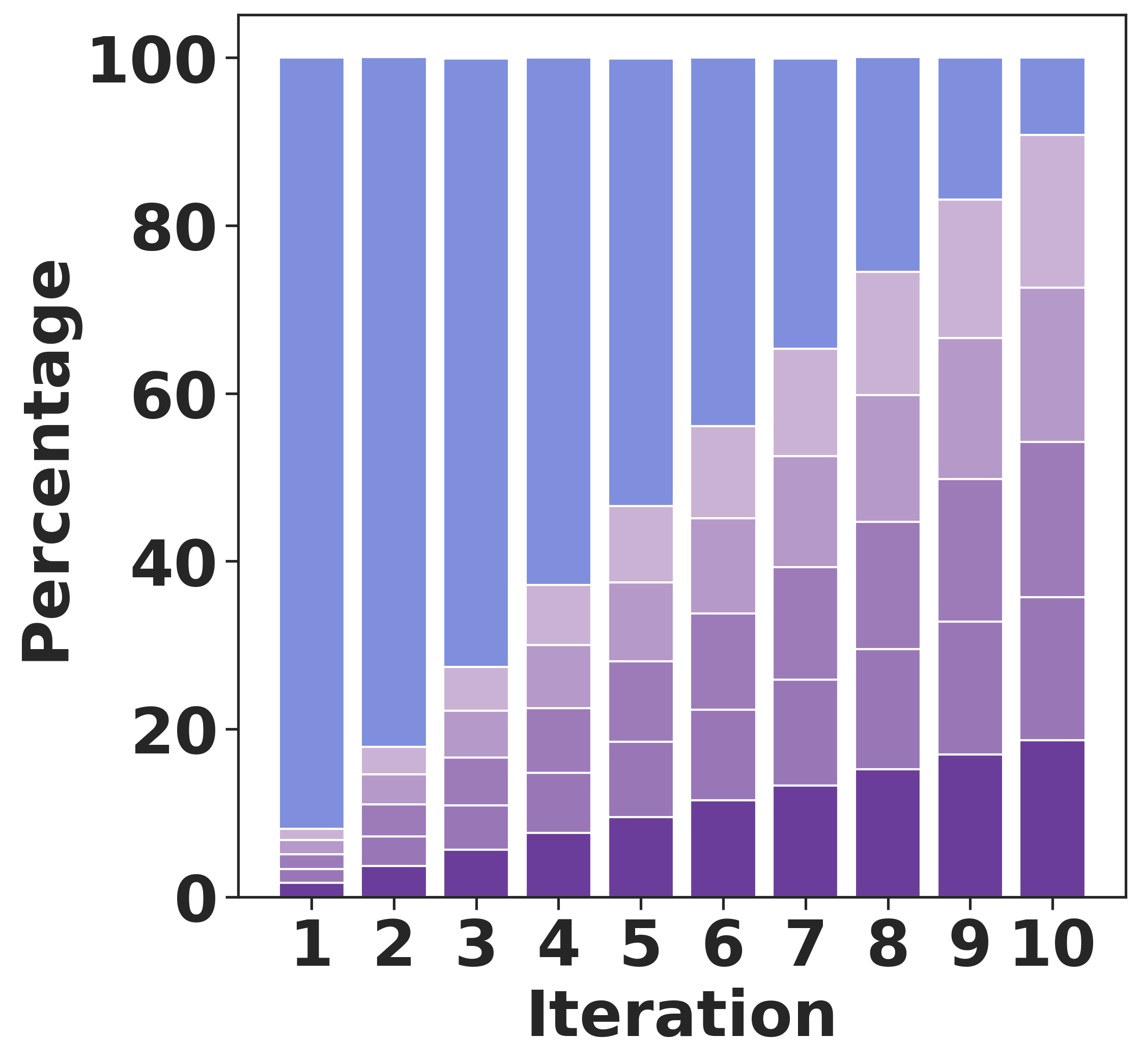}
    \caption{NQ}
    \label{fig:mis_nq_per}
  \end{subfigure}
  \hfill
  \begin{subfigure}[t]{0.22\textwidth}
    \includegraphics[width=\textwidth]{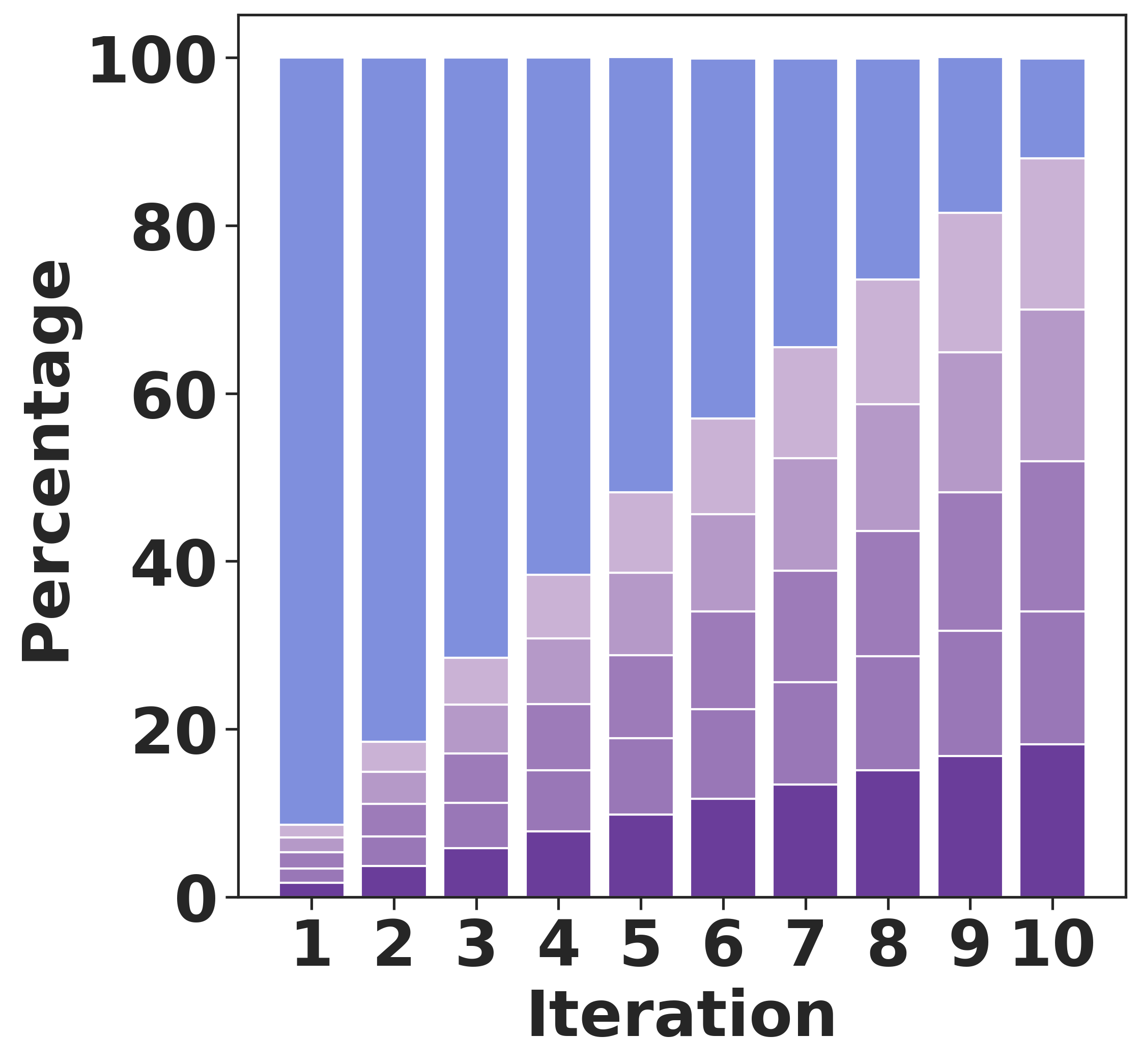}
    \caption{WebQ}
    \label{fig:mis_webq_per}
  \end{subfigure}
  \hfill
  \begin{subfigure}[t]{0.22\textwidth}
    \includegraphics[width=\textwidth]{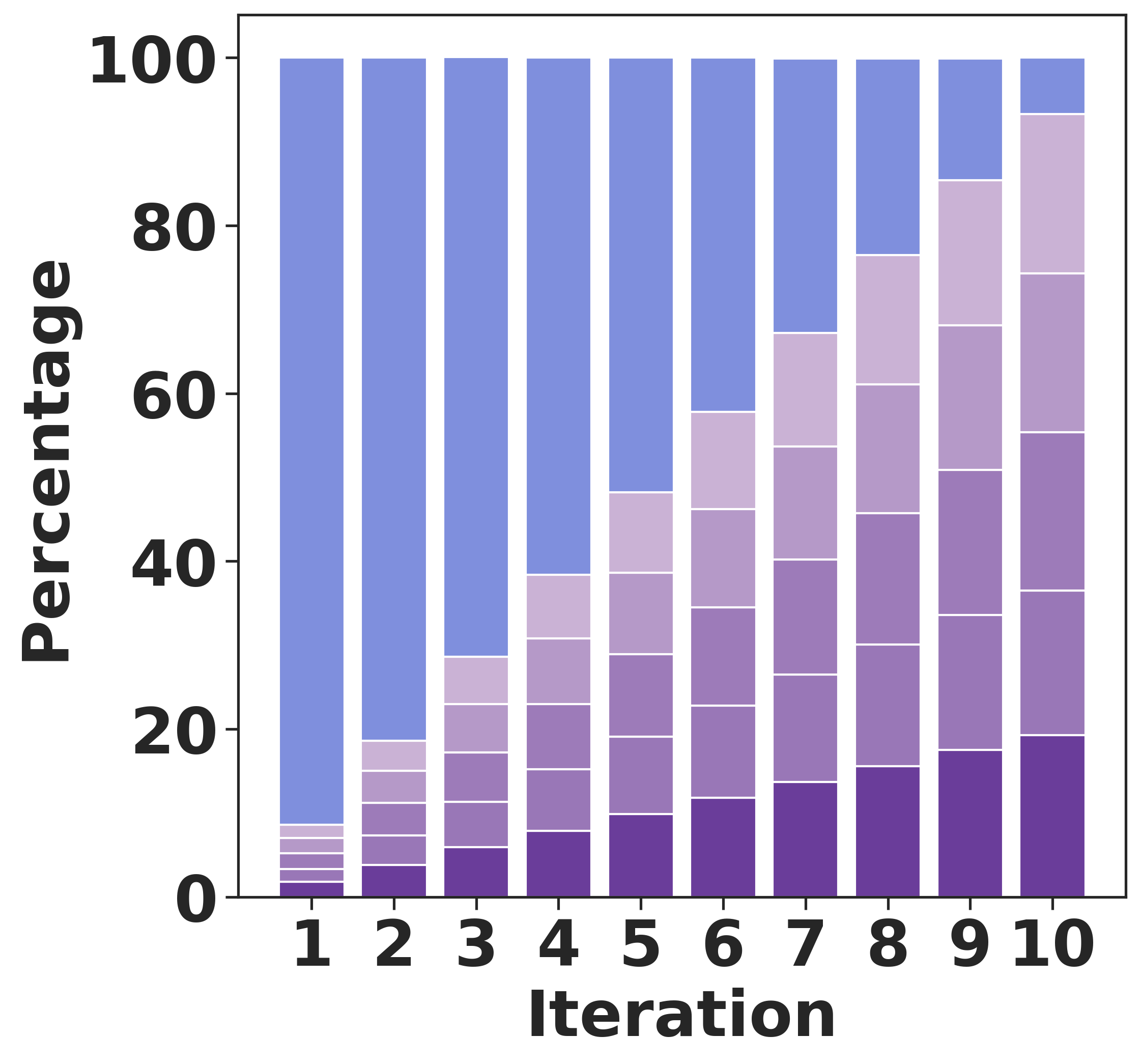}
    \caption{TriviaQA}
    \label{fig:mis_tqa_per}
  \end{subfigure}
  \hfill
  \begin{subfigure}[t]{0.32\textwidth}
    \includegraphics[width=\textwidth]{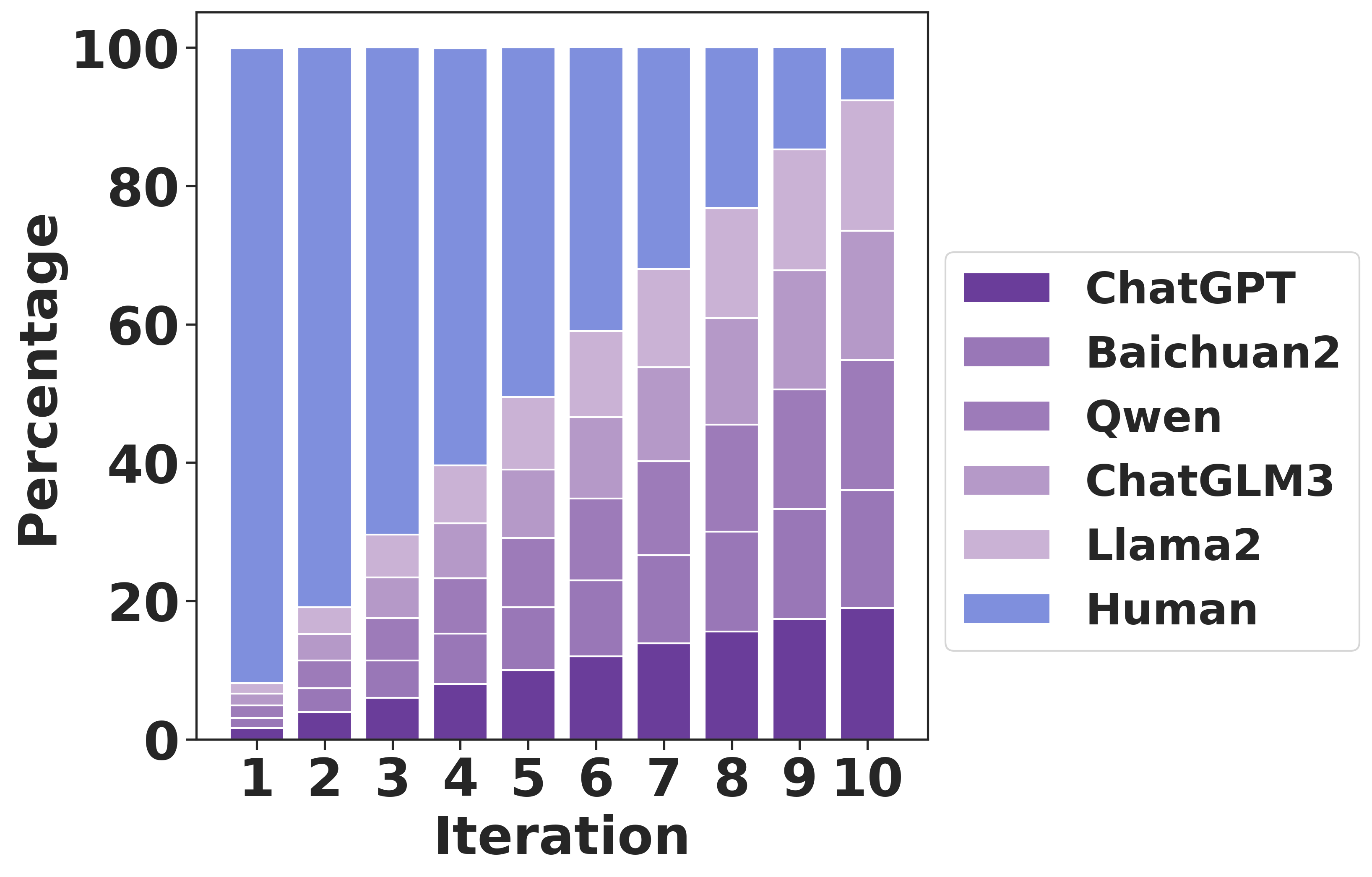}
    \caption{PopQA}
    \label{fig:mis_pop_per}
  \end{subfigure}
  \caption{Average percentage of texts from various sources within the top 50 search results over multiple iterations when adding \textbf{Misinformation} across different search methods.}
  \label{fig:mis_Percentage}
\end{figure*}

%% file: parts/tables/mis_context.tex
\begin{figure*}[htbp]
  \centering
    \includegraphics[width=\textwidth]{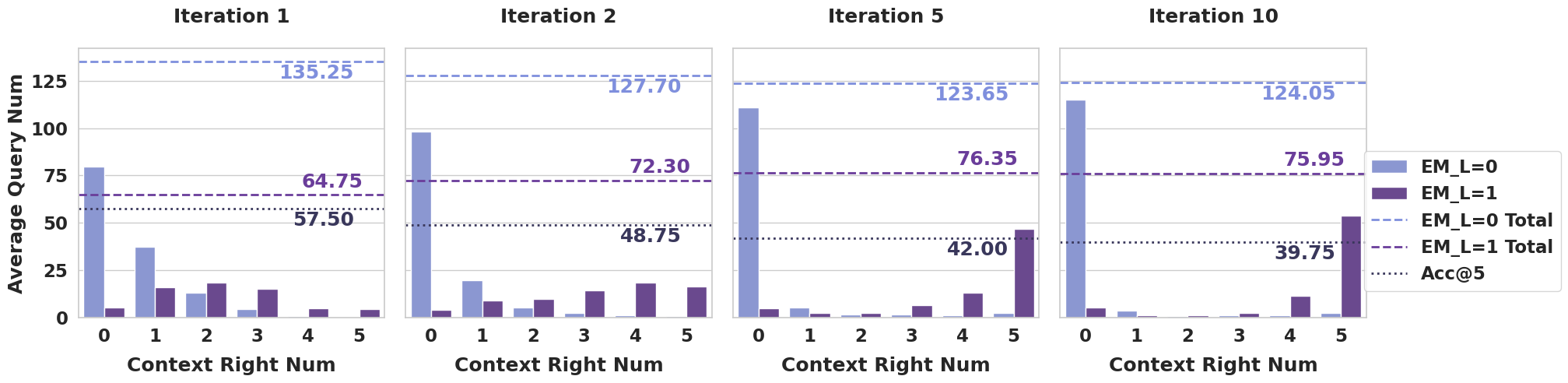}
    \caption{Correlation between the number of top 5 search results containing the correct answer (``Context Right Num'') and the accuracy of responses given by LLMs on the NQ dataset when adding \textbf{Misinformation}. The responses are categorized based on EM$_{llm}$ score: EM$\text{\_L}$=1 for correct and EM$\text{\_L}$=0 for incorrect. The overall number of queries that the LLMs answered correctly (EM$\text{\_L}$=1 Total) and incorrectly (EM$\text{\_L}$=0 Total), along with the average retrieval accuracy (Acc@5)
    are shown by dashed lines. The results are averaged across different LLMs, retrieval and ranking methods.}
    \label{fig:mis_nq_context}
\end{figure*}

%% file: parts/tables/mis_QA.tex
\begin{figure*}[htbp]
  \centering
    \includegraphics[width=\textwidth]{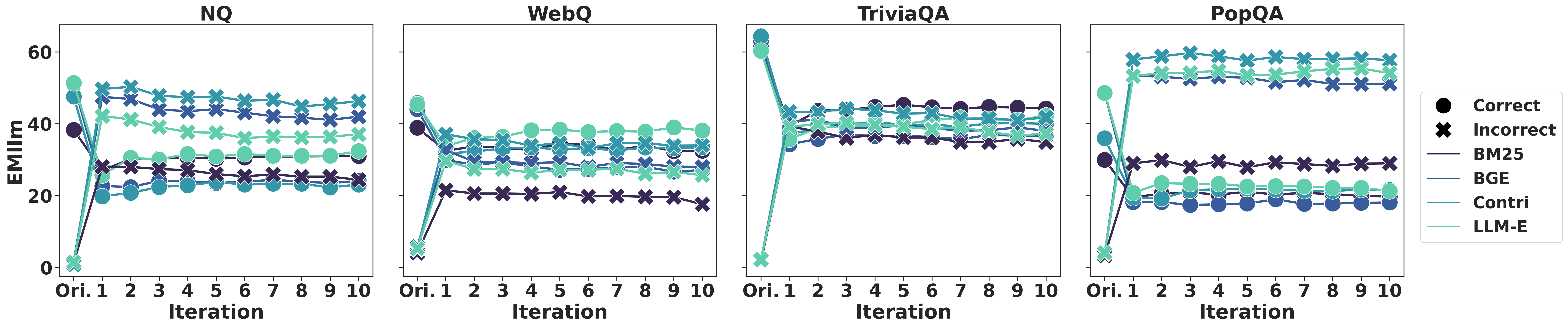}
    \caption{The average EM$_\text{llm}$ scores of the correct and incorrect answers of the RAG system in the simulation across datasets and LLMs.}
    \label{fig:mis_QA}
\end{figure*}

%% file: parts/tables/filter_retrieval.tex
\begin{figure*}[tbp]
  \centering
  \begin{subfigure}[t]{0.21\textwidth}
    \includegraphics[width=\textwidth]{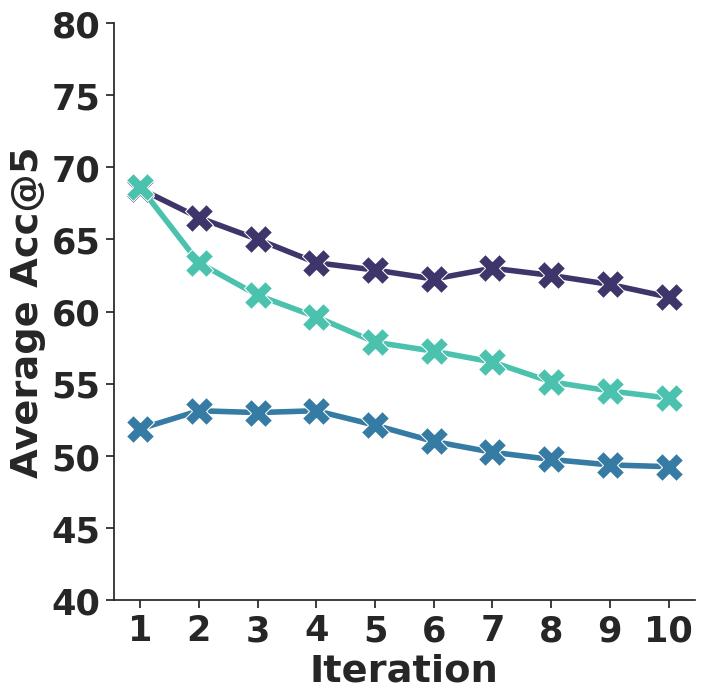}
    \caption{NQ}
    \label{fig:nq_filter_retrieval}
  \end{subfigure}
  \hfill
  \begin{subfigure}[t]{0.21\textwidth}
    \includegraphics[width=\textwidth]{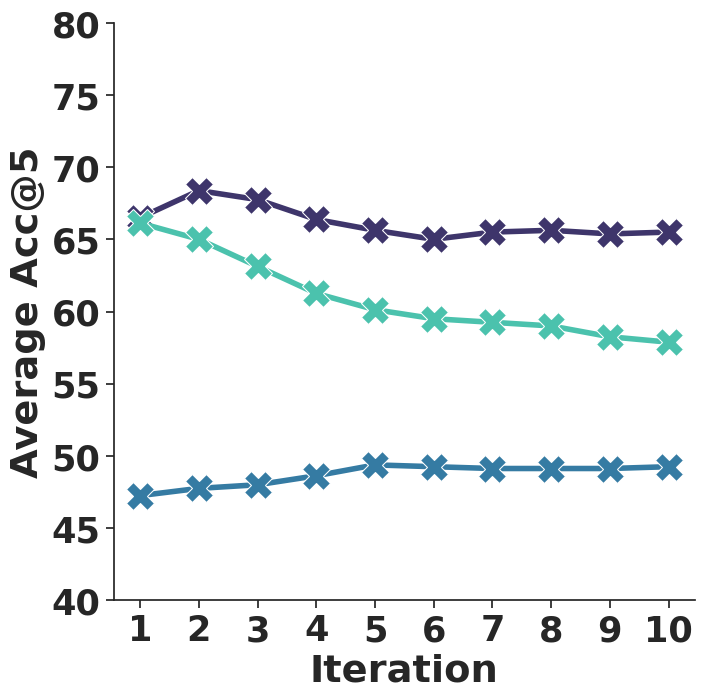}
    \caption{WebQ}
    \label{fig:webq_filter_retrieval}
  \end{subfigure}
  \hfill
  \begin{subfigure}[t]{0.22\textwidth}
    \includegraphics[width=\textwidth]{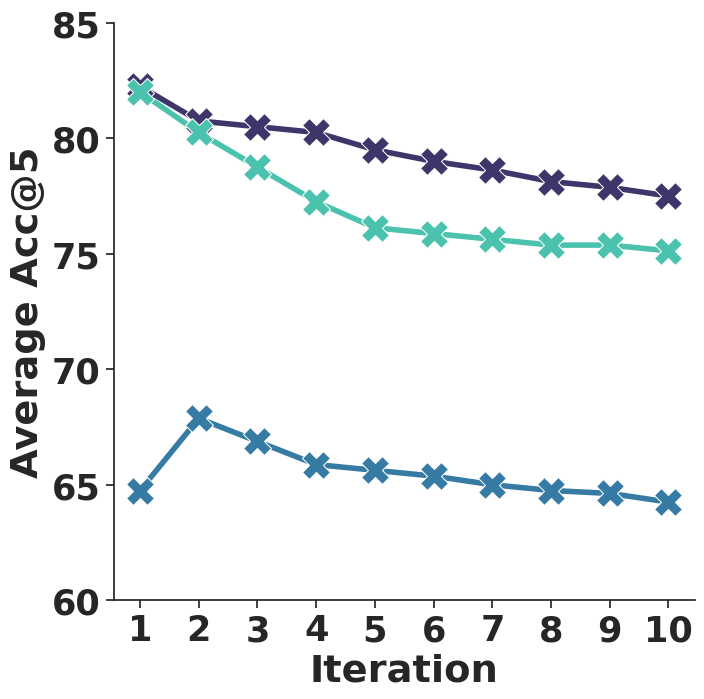}
    \caption{TriviaQA}
    \label{fig:tqa_filter_retrieval}
  \end{subfigure}
  \hfill
  \begin{subfigure}[t]{0.31\textwidth}
    \includegraphics[width=\textwidth]{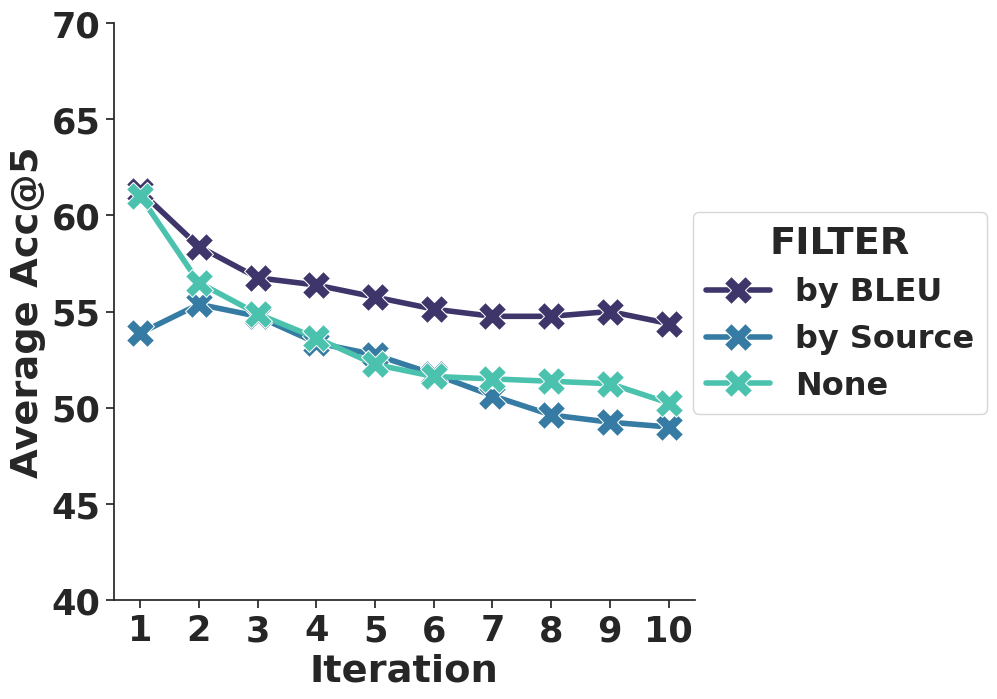}
    \caption{PopQA}
    \label{fig:pop_filter_retrieval}
  \end{subfigure}
  \caption{Average long-term retrieval performance of different filtering strategies.}
  \label{fig:filter_retrieval}
\end{figure*}

%% file: parts/tables/Filter_QA.tex
\begin{figure*}[tbp]
  \centering
  \begin{subfigure}[t]{0.21\textwidth}
    \includegraphics[width=\textwidth]{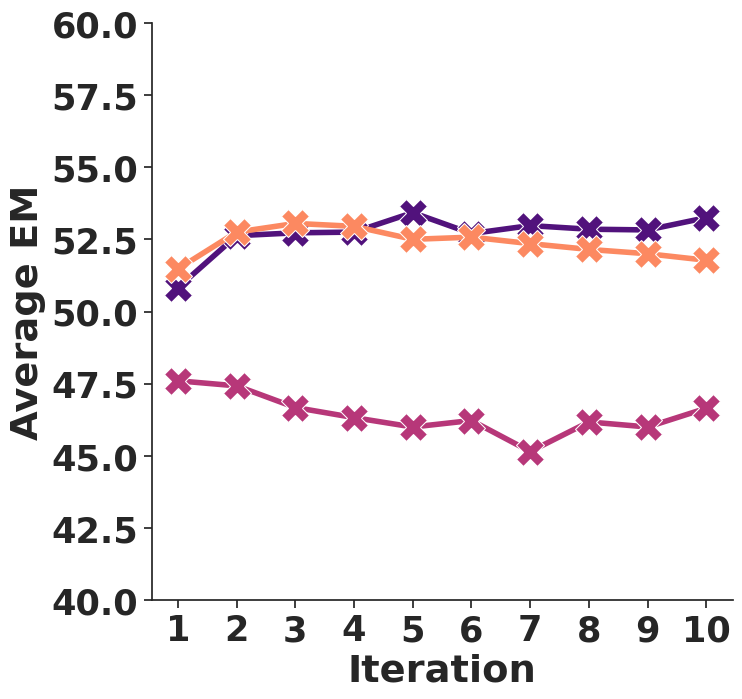}
    \caption{NQ}
    \label{fig:nq_filter_qa}
  \end{subfigure}
  \hfill
  \begin{subfigure}[t]{0.21\textwidth}
    \includegraphics[width=\textwidth]{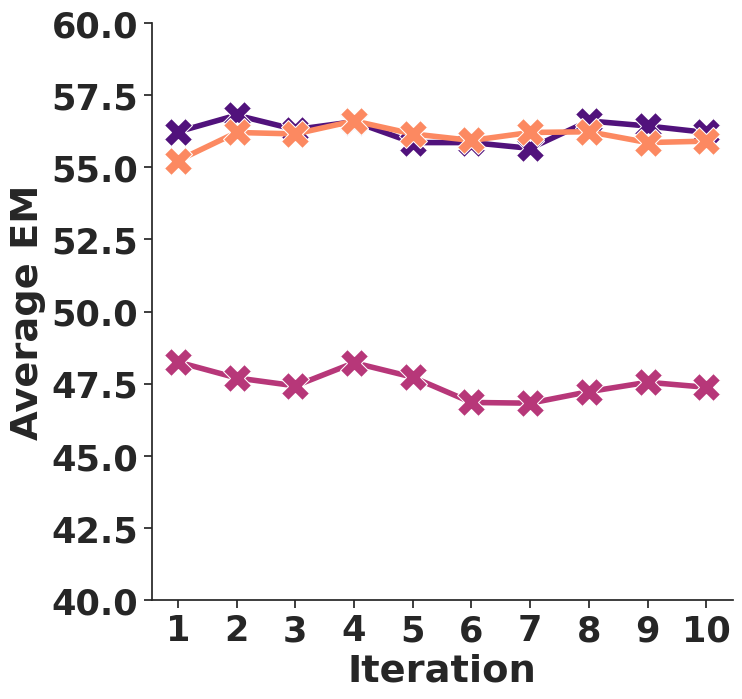}
    \caption{WebQ}
    \label{fig:webq_filter_qa}
  \end{subfigure}
  \hfill
  \begin{subfigure}[t]{0.22\textwidth}
    \includegraphics[width=\textwidth]{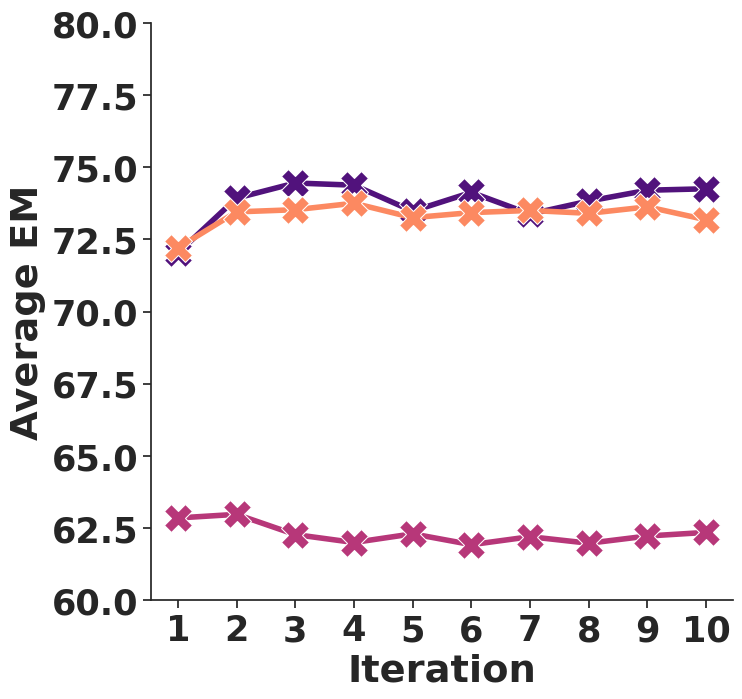}
    \caption{TriviaQA}
    \label{fig:tqa_filter_qa}
  \end{subfigure}
  \hfill
  \begin{subfigure}[t]{0.3\textwidth}
    \includegraphics[width=\textwidth]{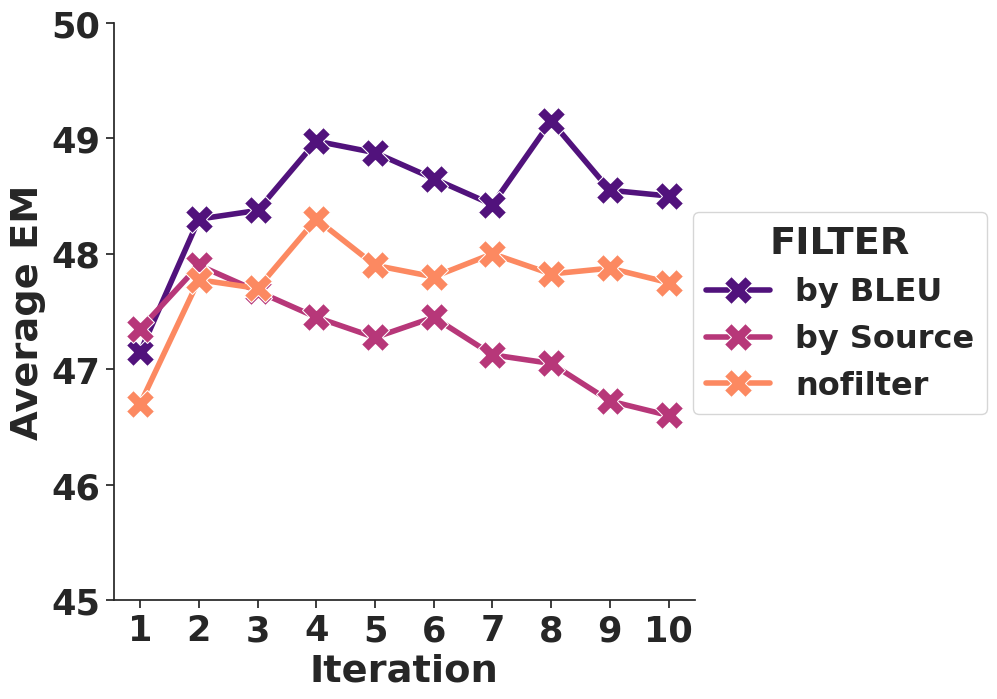}
    \caption{PopQA}
    \label{fig:pop_filter_qa}
  \end{subfigure}
  \caption{Average long-term QA performance of different filtering strategies.}
  \label{fig:filter_qa}
\end{figure*}

%% file: parts/tables/top5_filter_percentage.tex
\begin{figure*}[htbp]
  \centering
  \begin{subfigure}[t]{0.27\textwidth}
    \includegraphics[width=\textwidth]{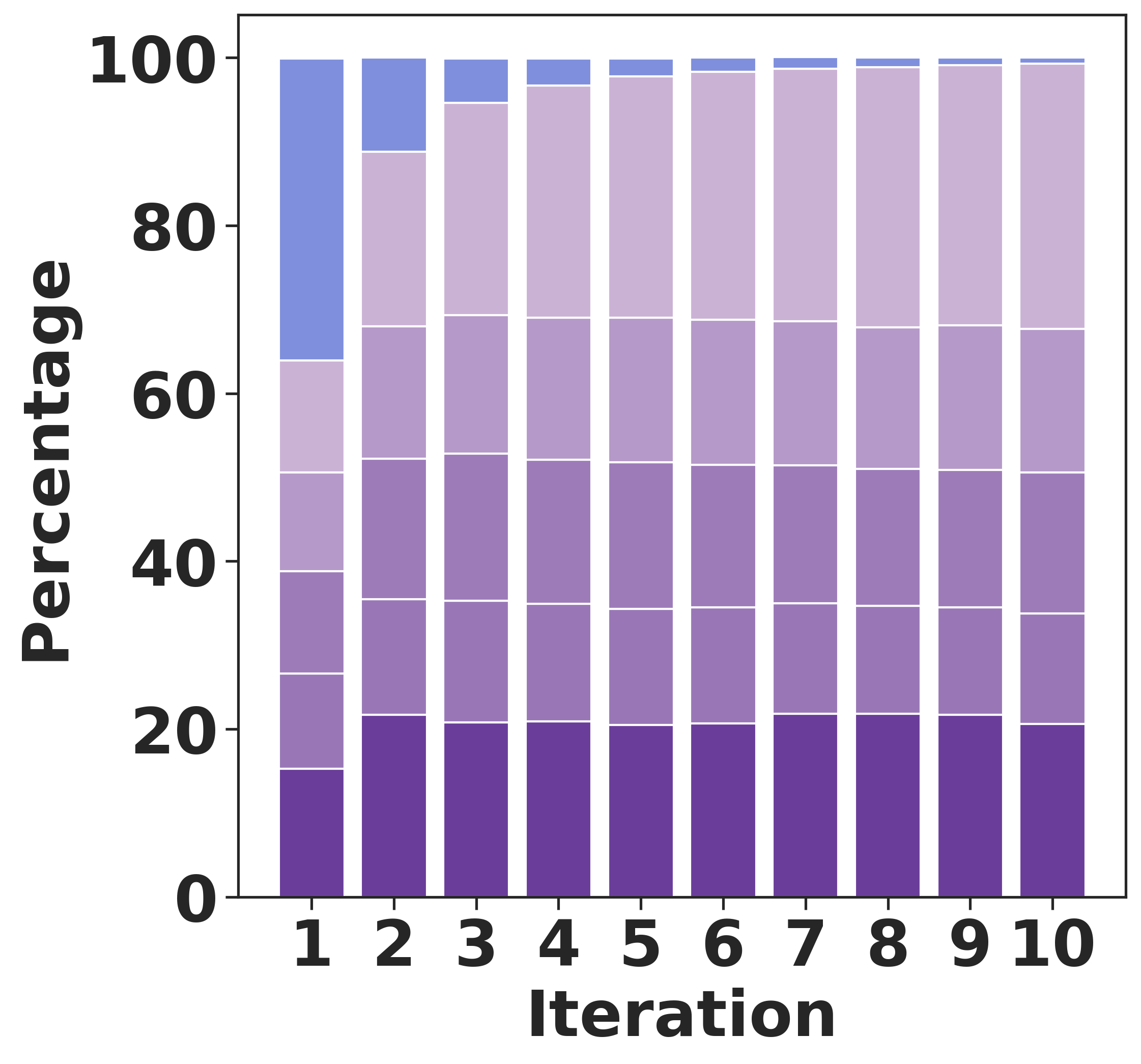}
    \caption{None}
    \label{fig:top5_nq_no_filter}
  \end{subfigure}
  \hfill
  \begin{subfigure}[t]{0.27\textwidth}
    \includegraphics[width=\textwidth]{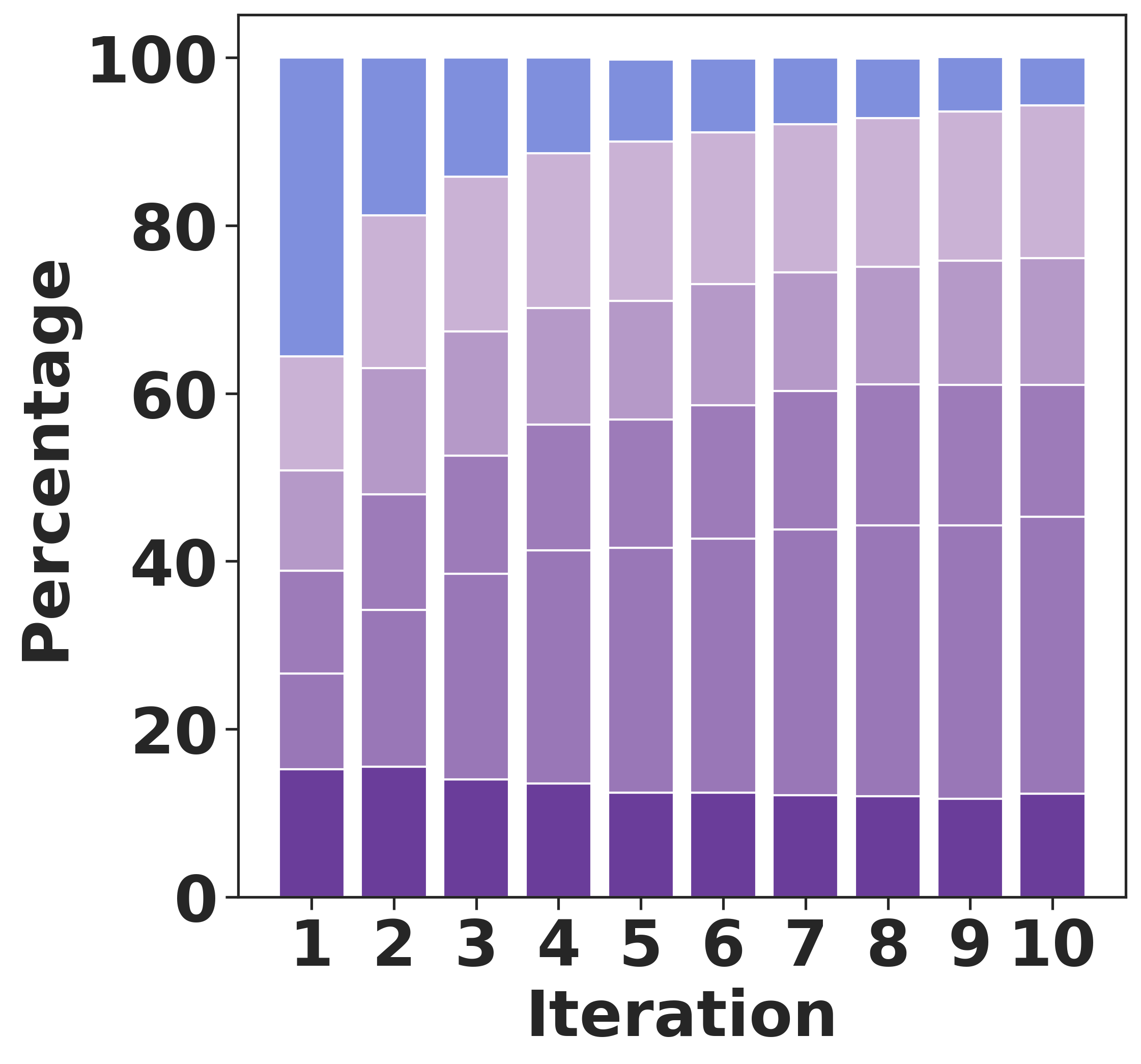}
    \caption{Content Filtering}
    \label{fig:top5_nq_filter_bleu}
  \end{subfigure}
  \hfill
  \begin{subfigure}[t]{0.4\textwidth}
    \includegraphics[width=\textwidth]{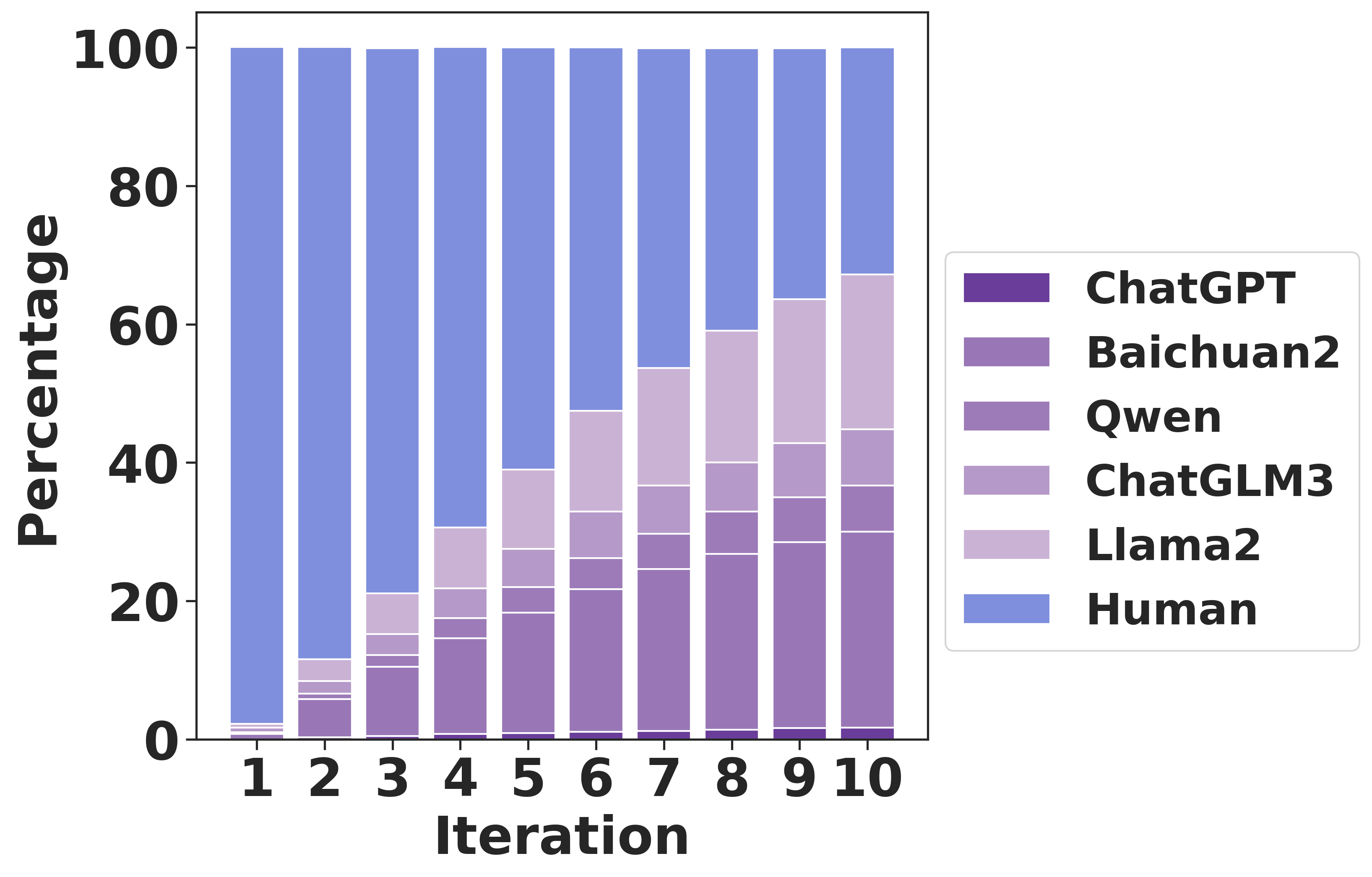}
    \caption{Source Filtering}
    \label{fig:top5_nq_filter_source}
  \end{subfigure}
  \caption{Average percentage of texts from various sources within the top 5 search results over multiple iterations on NQ when using different filtering strategies across different search methods.}
  \label{fig:filter_Percentage}
\end{figure*}

%% file: parts/tables/filter_bleu_context.tex
\begin{figure*}[htbp]
  \centering
    \includegraphics[width=\textwidth]{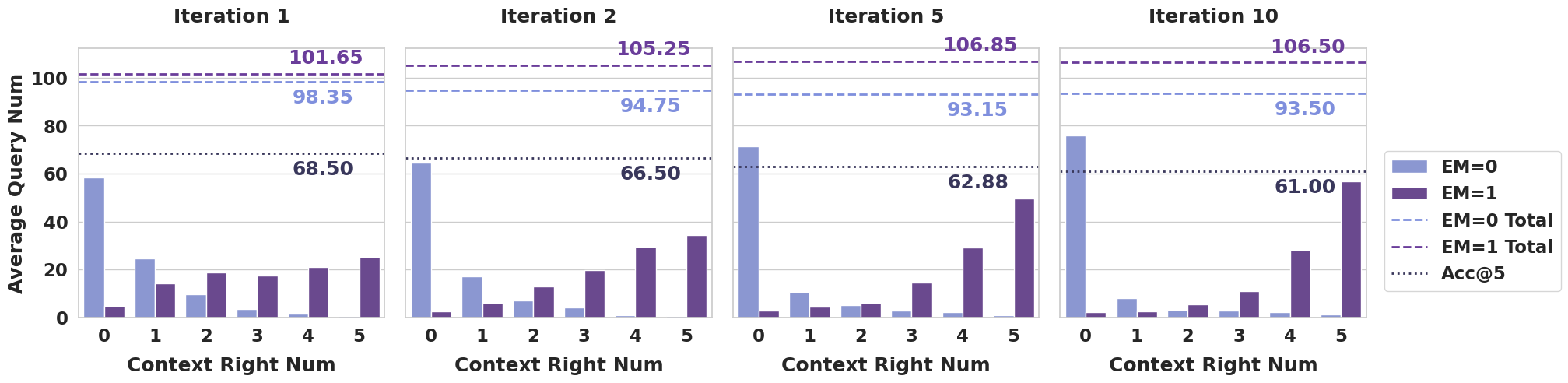}
    \caption{Correlation between the number of top 5 search results containing the correct answer ("Context Right
Num") and the accuracy of responses given by LLMs on the NQ dataset when using \textbf{Content Filtering}. The responses are categorized based on
Exact Match (EM) score: EM=1 for correct and EM=0 for incorrect. The overall number of queries that the LLMs
answered correctly (EM=1 Total) and incorrectly (EM=0 Total), along with the average retrieval accuracy (Acc@5)
are shown by dashed lines. The results are averaged across different LLMs, retrieval, and ranking methods.}
    \label{fig:filter_bleu_nq_context}
\end{figure*}

%% file: parts/tables/filter_source_context.tex
\begin{figure*}[htbp]
  \centering
    \includegraphics[width=\textwidth]{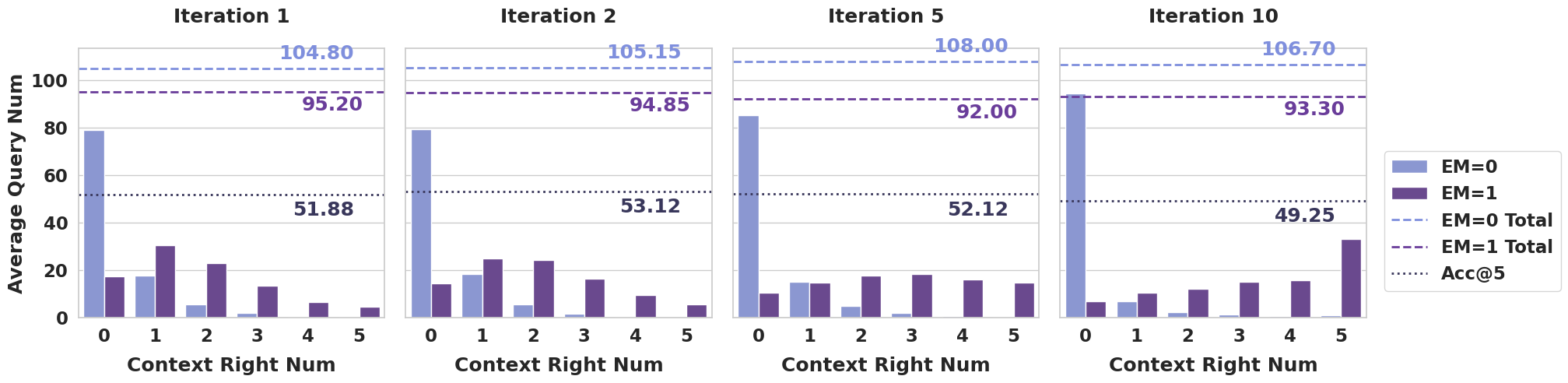}
    \caption{Correlation between the number of top 5 search results containing the correct answer (``Context Right
Num'') and the accuracy of responses given by LLMs on the NQ dataset when using \textbf{Source Filtering}. The responses are categorized based on
Exact Match (EM) score: EM=1 for correct and EM=0 for incorrect. The overall number of queries that the LLMs
answered correctly (EM=1 Total) and incorrectly (EM=0 Total), along with the average retrieval accuracy (Acc@5)
are shown by dashed lines. The results are averaged across different LLMs, retrieval and ranking methods.}
    \label{fig:filter_source_nq_context}
\end{figure*}

%% file: parts/prompts.tex
\subsection{Prompts}\label{sec:prompt}
The prompts used in the experiment are shown in Table~\ref{tab:prompt}.
\begin{table*}[htbp]
\centering
\label{tab:prompts}
\resizebox{0.95\textwidth}{!}{
\begin{tabular}{p{0.25\textwidth}|p{0.7\textwidth}}
\toprule
\textbf{Task} & \textbf{Prompts} \\
\hline
Zero-Shot Generation & \texttt{Provide a background document in 100 words according to your knowledge from Wikipedia to answer the given question. \newline \newline
Question:\{\textbf{question\_str}\} \newline \newline
Background Document:} \\
\hline
Generation with Contexts & \texttt{Context information is below. \newline
--------------------- \newline
[Context 1]:\{\textbf{context\_str1}\}\newline
...\newline
[Context 5]:\{\textbf{context\_str5}\}\newline
--------------------- \newline
Using both the context information and also using your own knowledge, answer the following question with a background document in 100 words. \newline \newline
Question:\{\textbf{question\_str}\} \newline \newline
Background Document:} \\
\hline
Mis\_Answer Generation & \texttt{Generate a false answer to the given question. It should be short (less than ten words in general) and look plausible, compared to the reference answer. \newline \newline
Question:\{\textbf{question\_str}\} \newline \newline
Reference Answers:\{\textbf{ref\_answer\_str}\}, \newline \newline
False answer:} \\
\hline
Mis\_Passage Generation & \texttt{Suppose you are generating realistic-looking claims for a quiz contest. You need to generate a background document less than 100 words in support of the answer to the given question. Don't contain any word in the original answers in \{ref\_answer\_str\}. The background document must contain the following given answers with their original form. \newline \newline
Question:\{\textbf{question\_str}\} \newline \newline
Answers:\{\textbf{false\_answer\_str}\}, \newline \newline
Background document: }\\
\hline
Answer Check & \texttt{Does the following response support the answer to the question? \newline 
Question: \{\textbf{question\_str}\} \newline 
Response: \{\textbf{response\_str}\} \newline 
Answer: \{\textbf{ref\_answer\_str}\} / \{\textbf{false\_answer\_str}\} \newline
Just answer 'yes' or 'no'. }\\
\bottomrule
\end{tabular}}
\caption{Prompts for different tasks.}
\label{tab:prompt}
\end{table*}

%% file: SOS.bbl
\begin{thebibliography}{58}
\expandafter\ifx\csname natexlab\endcsname\relax\def\natexlab#1{#1}\fi

\bibitem[{Adlakha et~al.(2023)Adlakha, BehnamGhader, Lu, Meade, and Reddy}]{DBLP:journals/corr/abs-2307-16877}
Vaibhav Adlakha, Parishad BehnamGhader, Xing~Han Lu, Nicholas Meade, and Siva Reddy. 2023.
\newblock \href {https://doi.org/10.48550/ARXIV.2307.16877} {Evaluating correctness and faithfulness of instruction-following models for question answering}.
\newblock \emph{CoRR}, abs/2307.16877.

\bibitem[{Alatawi et~al.(2021)Alatawi, Cheng, Tahir, Karami, Jiang, Black, and Liu}]{DBLP:journals/corr/abs-2112-05084}
Faisal Alatawi, Lu~Cheng, Anique Tahir, Mansooreh Karami, Bohan Jiang, Tyler Black, and Huan Liu. 2021.
\newblock \href {http://arxiv.org/abs/2112.05084} {A survey on echo chambers on social media: Description, detection and mitigation}.
\newblock \emph{CoRR}, abs/2112.05084.

\bibitem[{Alemohammad et~al.(2023)Alemohammad, Casco{-}Rodriguez, Luzi, Humayun, Babaei, LeJeune, Siahkoohi, and Baraniuk}]{DBLP:journals/corr/abs-2307-01850}
Sina Alemohammad, Josue Casco{-}Rodriguez, Lorenzo Luzi, Ahmed~Imtiaz Humayun, Hossein Babaei, Daniel LeJeune, Ali Siahkoohi, and Richard~G. Baraniuk. 2023.
\newblock \href {https://doi.org/10.48550/ARXIV.2307.01850} {Self-consuming generative models go {MAD}}.
\newblock \emph{CoRR}, abs/2307.01850.

\bibitem[{Bai et~al.(2023)Bai, Bai, Chu, Cui, Dang, Deng, Fan, Ge, Han, Huang, Hui, Ji, Li, Lin, Lin, Liu, Liu, Lu, Lu, Ma, Men, Ren, Ren, Tan, Tan, Tu, Wang, Wang, Wang, Wu, Xu, Xu, Yang, Yang, Yang, Yang, Yao, Yu, Yuan, Yuan, Zhang, Zhang, Zhang, Zhang, Zhou, Zhou, Zhou, and Zhu}]{qwen}
Jinze Bai, Shuai Bai, Yunfei Chu, Zeyu Cui, Kai Dang, Xiaodong Deng, Yang Fan, Wenbin Ge, Yu~Han, Fei Huang, Binyuan Hui, Luo Ji, Mei Li, Junyang Lin, Runji Lin, Dayiheng Liu, Gao Liu, Chengqiang Lu, Keming Lu, Jianxin Ma, Rui Men, Xingzhang Ren, Xuancheng Ren, Chuanqi Tan, Sinan Tan, Jianhong Tu, Peng Wang, Shijie Wang, Wei Wang, Shengguang Wu, Benfeng Xu, Jin Xu, An~Yang, Hao Yang, Jian Yang, Shusheng Yang, Yang Yao, Bowen Yu, Hongyi Yuan, Zheng Yuan, Jianwei Zhang, Xingxuan Zhang, Yichang Zhang, Zhenru Zhang, Chang Zhou, Jingren Zhou, Xiaohuan Zhou, and Tianhang Zhu. 2023.
\newblock \href {https://doi.org/10.48550/ARXIV.2309.16609} {Qwen technical report}.
\newblock \emph{CoRR}, abs/2309.16609.

\bibitem[{Berant et~al.(2013)Berant, Chou, Frostig, and Liang}]{webq}
Jonathan Berant, Andrew Chou, Roy Frostig, and Percy Liang. 2013.
\newblock \href {https://aclanthology.org/D13-1160/} {Semantic parsing on freebase from question-answer pairs}.
\newblock In \emph{Proceedings of the 2013 Conference on Empirical Methods in Natural Language Processing, {EMNLP} 2013, 18-21 October 2013, Grand Hyatt Seattle, Seattle, Washington, USA, {A} meeting of SIGDAT, a Special Interest Group of the {ACL}}, pages 1533--1544. {ACL}.

\bibitem[{Borgeaud et~al.(2022)Borgeaud, Mensch, Hoffmann, Cai, Rutherford, Millican, van~den Driessche, Lespiau, Damoc, Clark, de~Las~Casas, Guy, Menick, Ring, Hennigan, Huang, Maggiore, Jones, Cassirer, Brock, Paganini, Irving, Vinyals, Osindero, Simonyan, Rae, Elsen, and Sifre}]{Retro}
Sebastian Borgeaud, Arthur Mensch, Jordan Hoffmann, Trevor Cai, Eliza Rutherford, Katie Millican, George van~den Driessche, Jean{-}Baptiste Lespiau, Bogdan Damoc, Aidan Clark, Diego de~Las~Casas, Aurelia Guy, Jacob Menick, Roman Ring, Tom Hennigan, Saffron Huang, Loren Maggiore, Chris Jones, Albin Cassirer, Andy Brock, Michela Paganini, Geoffrey Irving, Oriol Vinyals, Simon Osindero, Karen Simonyan, Jack~W. Rae, Erich Elsen, and Laurent Sifre. 2022.
\newblock \href {https://proceedings.mlr.press/v162/borgeaud22a.html} {Improving language models by retrieving from trillions of tokens}.
\newblock In \emph{International Conference on Machine Learning, {ICML} 2022, 17-23 July 2022, Baltimore, Maryland, {USA}}, volume 162 of \emph{Proceedings of Machine Learning Research}, pages 2206--2240. {PMLR}.

\bibitem[{Cai et~al.(2019)Cai, Wang, Bi, Tu, Liu, Lam, and Shi}]{DBLP:conf/naacl/CaiWBTLLS19}
Deng Cai, Yan Wang, Wei Bi, Zhaopeng Tu, Xiaojiang Liu, Wai Lam, and Shuming Shi. 2019.
\newblock \href {https://doi.org/10.18653/V1/N19-1124} {Skeleton-to-response: Dialogue generation guided by retrieval memory}.
\newblock In \emph{Proceedings of the 2019 Conference of the North American Chapter of the Association for Computational Linguistics: Human Language Technologies, {NAACL-HLT} 2019, Minneapolis, MN, USA, June 2-7, 2019, Volume 1 (Long and Short Papers)}, pages 1219--1228. Association for Computational Linguistics.

\bibitem[{Chen and Shu(2023)}]{DBLP:journals/corr/abs-2311-05656}
Canyu Chen and Kai Shu. 2023.
\newblock \href {https://doi.org/10.48550/ARXIV.2311.05656} {Combating misinformation in the age of llms: Opportunities and challenges}.
\newblock \emph{CoRR}, abs/2311.05656.

\bibitem[{Chen et~al.(2023)Chen, Lin, Han, and Sun}]{DBLP:journals/corr/abs-2309-01431}
Jiawei Chen, Hongyu Lin, Xianpei Han, and Le~Sun. 2023.
\newblock \href {https://doi.org/10.48550/ARXIV.2309.01431} {Benchmarking large language models in retrieval-augmented generation}.
\newblock \emph{CoRR}, abs/2309.01431.

\bibitem[{Chitra and Musco(2020)}]{DBLP:conf/wsdm/ChitraM20}
Uthsav Chitra and Christopher Musco. 2020.
\newblock \href {https://doi.org/10.1145/3336191.3371825} {Analyzing the impact of filter bubbles on social network polarization}.
\newblock In \emph{{WSDM} '20: The Thirteenth {ACM} International Conference on Web Search and Data Mining, Houston, TX, USA, February 3-7, 2020}, pages 115--123. {ACM}.

\bibitem[{Conneau et~al.(2020)Conneau, Khandelwal, Goyal, Chaudhary, Wenzek, Guzm{\'{a}}n, Grave, Ott, Zettlemoyer, and Stoyanov}]{xlm-roberta}
Alexis Conneau, Kartikay Khandelwal, Naman Goyal, Vishrav Chaudhary, Guillaume Wenzek, Francisco Guzm{\'{a}}n, Edouard Grave, Myle Ott, Luke Zettlemoyer, and Veselin Stoyanov. 2020.
\newblock \href {https://doi.org/10.18653/V1/2020.ACL-MAIN.747} {Unsupervised cross-lingual representation learning at scale}.
\newblock In \emph{Proceedings of the 58th Annual Meeting of the Association for Computational Linguistics, {ACL} 2020, Online, July 5-10, 2020}, pages 8440--8451. Association for Computational Linguistics.

\bibitem[{Dai et~al.(2023{\natexlab{a}})Dai, Zhou, Pang, Liu, Hu, Liu, Zhang, and Xu}]{DBLP:journals/corr/abs-2310-20501}
Sunhao Dai, Yuqi Zhou, Liang Pang, Weihao Liu, Xiaolin Hu, Yong Liu, Xiao Zhang, and Jun Xu. 2023{\natexlab{a}}.
\newblock \href {https://doi.org/10.48550/ARXIV.2310.20501} {Llms may dominate information access: Neural retrievers are biased towards llm-generated texts}.
\newblock \emph{CoRR}, abs/2310.20501.

\bibitem[{Dai et~al.(2023{\natexlab{b}})Dai, Lang, Zheng, Yu, Huang, and Li}]{DBLP:conf/acl/DaiLZ0HL23}
Yi~Dai, Hao Lang, Yinhe Zheng, Bowen Yu, Fei Huang, and Yongbin Li. 2023{\natexlab{b}}.
\newblock \href {https://doi.org/10.18653/V1/2023.FINDINGS-ACL.361} {Domain incremental lifelong learning in an open world}.
\newblock In \emph{Findings of the Association for Computational Linguistics: {ACL} 2023, Toronto, Canada, July 9-14, 2023}, pages 5844--5865. Association for Computational Linguistics.

\bibitem[{Du et~al.(2022)Du, Qian, Liu, Ding, Qiu, Yang, and Tang}]{chatglm}
Zhengxiao Du, Yujie Qian, Xiao Liu, Ming Ding, Jiezhong Qiu, Zhilin Yang, and Jie Tang. 2022.
\newblock Glm: General language model pretraining with autoregressive blank infilling.
\newblock In \emph{Proceedings of the 60th Annual Meeting of the Association for Computational Linguistics (Volume 1: Long Papers)}, pages 320--335.

\bibitem[{Goldstein et~al.(2023)Goldstein, Sastry, Musser, DiResta, Gentzel, and Sedova}]{DBLP:journals/corr/abs-2301-04246}
Josh~A. Goldstein, Girish Sastry, Micah Musser, Renee DiResta, Matthew Gentzel, and Katerina Sedova. 2023.
\newblock \href {https://doi.org/10.48550/ARXIV.2301.04246} {Generative language models and automated influence operations: Emerging threats and potential mitigations}.
\newblock \emph{CoRR}, abs/2301.04246.

\bibitem[{{Google}(2023)}]{gemini}
{Google}. 2023.
\newblock Introducing gemini: our largest and most capable ai model.
\newblock \url{https://blog.google/technology/ai/google-gemini-ai/?utm_source=gdm&utm_medium=referral#sundar-note}.
\newblock Accessed: 2023-12-06.

\bibitem[{Guu et~al.(2020)Guu, Lee, Tung, Pasupat, and Chang}]{REALM}
Kelvin Guu, Kenton Lee, Zora Tung, Panupong Pasupat, and Ming{-}Wei Chang. 2020.
\newblock \href {http://arxiv.org/abs/2002.08909} {{REALM:} retrieval-augmented language model pre-training}.
\newblock \emph{CoRR}, abs/2002.08909.

\bibitem[{He et~al.(2023)He, Zhang, and Roth}]{DBLP:journals/corr/abs-2301-00303}
Hangfeng He, Hongming Zhang, and Dan Roth. 2023.
\newblock \href {https://doi.org/10.48550/ARXIV.2301.00303} {Rethinking with retrieval: Faithful large language model inference}.
\newblock \emph{CoRR}, abs/2301.00303.

\bibitem[{Huang et~al.(2023{\natexlab{a}})Huang, Yu, Ma, Zhong, Feng, Wang, Chen, Peng, Feng, Qin, and Liu}]{DBLP:journals/corr/abs-2311-05232}
Lei Huang, Weijiang Yu, Weitao Ma, Weihong Zhong, Zhangyin Feng, Haotian Wang, Qianglong Chen, Weihua Peng, Xiaocheng Feng, Bing Qin, and Ting Liu. 2023{\natexlab{a}}.
\newblock \href {https://doi.org/10.48550/ARXIV.2311.05232} {A survey on hallucination in large language models: Principles, taxonomy, challenges, and open questions}.
\newblock \emph{CoRR}, abs/2311.05232.

\bibitem[{Huang et~al.(2023{\natexlab{b}})Huang, Gupta, Xia, Li, and Chen}]{DBLP:journals/corr/abs-2310-06987}
Yangsibo Huang, Samyak Gupta, Mengzhou Xia, Kai Li, and Danqi Chen. 2023{\natexlab{b}}.
\newblock \href {https://doi.org/10.48550/ARXIV.2310.06987} {Catastrophic jailbreak of open-source llms via exploiting generation}.
\newblock \emph{CoRR}, abs/2310.06987.

\bibitem[{Izacard et~al.(2022)Izacard, Caron, Hosseini, Riedel, Bojanowski, Joulin, and Grave}]{contriever}
Gautier Izacard, Mathilde Caron, Lucas Hosseini, Sebastian Riedel, Piotr Bojanowski, Armand Joulin, and Edouard Grave. 2022.
\newblock \href {https://openreview.net/forum?id=jKN1pXi7b0} {Unsupervised dense information retrieval with contrastive learning}.
\newblock \emph{Trans. Mach. Learn. Res.}, 2022.

\bibitem[{Izacard and Grave(2021)}]{DBLP:conf/eacl/IzacardG21}
Gautier Izacard and Edouard Grave. 2021.
\newblock \href {https://doi.org/10.18653/V1/2021.EACL-MAIN.74} {Leveraging passage retrieval with generative models for open domain question answering}.
\newblock In \emph{Proceedings of the 16th Conference of the European Chapter of the Association for Computational Linguistics: Main Volume, {EACL} 2021, Online, April 19 - 23, 2021}, pages 874--880. Association for Computational Linguistics.

\bibitem[{Izacard et~al.(2023)Izacard, Lewis, Lomeli, Hosseini, Petroni, Schick, Dwivedi{-}Yu, Joulin, Riedel, and Grave}]{Atlas}
Gautier Izacard, Patrick S.~H. Lewis, Maria Lomeli, Lucas Hosseini, Fabio Petroni, Timo Schick, Jane Dwivedi{-}Yu, Armand Joulin, Sebastian Riedel, and Edouard Grave. 2023.
\newblock \href {http://jmlr.org/papers/v24/23-0037.html} {Atlas: Few-shot learning with retrieval augmented language models}.
\newblock \emph{J. Mach. Learn. Res.}, 24:251:1--251:43.

\bibitem[{Ji et~al.(2023)Ji, Lee, Frieske, Yu, Su, Xu, Ishii, Bang, Madotto, and Fung}]{DBLP:journals/csur/JiLFYSXIBMF23}
Ziwei Ji, Nayeon Lee, Rita Frieske, Tiezheng Yu, Dan Su, Yan Xu, Etsuko Ishii, Yejin Bang, Andrea Madotto, and Pascale Fung. 2023.
\newblock \href {https://doi.org/10.1145/3571730} {Survey of hallucination in natural language generation}.
\newblock \emph{{ACM} Comput. Surv.}, 55(12):248:1--248:38.

\bibitem[{Jiang et~al.(2019)Jiang, Chiappa, Lattimore, Gy{\"{o}}rgy, and Kohli}]{DBLP:conf/aies/JiangCLGK19}
Ray Jiang, Silvia Chiappa, Tor Lattimore, Andr{\'{a}}s Gy{\"{o}}rgy, and Pushmeet Kohli. 2019.
\newblock \href {https://doi.org/10.1145/3306618.3314288} {Degenerate feedback loops in recommender systems}.
\newblock In \emph{Proceedings of the 2019 {AAAI/ACM} Conference on AI, Ethics, and Society, {AIES} 2019, Honolulu, HI, USA, January 27-28, 2019}, pages 383--390. {ACM}.

\bibitem[{Johnson et~al.(2019)Johnson, Douze, and J{\'e}gou}]{faiss}
Jeff Johnson, Matthijs Douze, and Herv{\'e} J{\'e}gou. 2019.
\newblock Billion-scale similarity search with {GPUs}.
\newblock \emph{IEEE Transactions on Big Data}, 7(3):535--547.

\bibitem[{Joshi et~al.(2017)Joshi, Choi, Weld, and Zettlemoyer}]{tqa}
Mandar Joshi, Eunsol Choi, Daniel~S. Weld, and Luke Zettlemoyer. 2017.
\newblock \href {https://doi.org/10.18653/V1/P17-1147} {Triviaqa: {A} large scale distantly supervised challenge dataset for reading comprehension}.
\newblock In \emph{Proceedings of the 55th Annual Meeting of the Association for Computational Linguistics, {ACL} 2017, Vancouver, Canada, July 30 - August 4, Volume 1: Long Papers}, pages 1601--1611. Association for Computational Linguistics.

\bibitem[{Karpukhin et~al.(2020)Karpukhin, Oguz, Min, Lewis, Wu, Edunov, Chen, and Yih}]{DBLP:conf/emnlp/KarpukhinOMLWEC20}
Vladimir Karpukhin, Barlas Oguz, Sewon Min, Patrick S.~H. Lewis, Ledell Wu, Sergey Edunov, Danqi Chen, and Wen{-}tau Yih. 2020.
\newblock \href {https://doi.org/10.18653/V1/2020.EMNLP-MAIN.550} {Dense passage retrieval for open-domain question answering}.
\newblock In \emph{Proceedings of the 2020 Conference on Empirical Methods in Natural Language Processing, {EMNLP} 2020, Online, November 16-20, 2020}, pages 6769--6781. Association for Computational Linguistics.

\bibitem[{Kwiatkowski et~al.(2019)Kwiatkowski, Palomaki, Redfield, Collins, Parikh, Alberti, Epstein, Polosukhin, Devlin, Lee, Toutanova, Jones, Kelcey, Chang, Dai, Uszkoreit, Le, and Petrov}]{nq}
Tom Kwiatkowski, Jennimaria Palomaki, Olivia Redfield, Michael Collins, Ankur~P. Parikh, Chris Alberti, Danielle Epstein, Illia Polosukhin, Jacob Devlin, Kenton Lee, Kristina Toutanova, Llion Jones, Matthew Kelcey, Ming{-}Wei Chang, Andrew~M. Dai, Jakob Uszkoreit, Quoc Le, and Slav Petrov. 2019.
\newblock \href {https://doi.org/10.1162/TACL\_A\_00276} {Natural questions: a benchmark for question answering research}.
\newblock \emph{Trans. Assoc. Comput. Linguistics}, 7:452--466.

\bibitem[{Lewis et~al.(2020)Lewis, Perez, Piktus, Petroni, Karpukhin, Goyal, K{\"{u}}ttler, Lewis, Yih, Rockt{\"{a}}schel, Riedel, and Kiela}]{RAG}
Patrick S.~H. Lewis, Ethan Perez, Aleksandra Piktus, Fabio Petroni, Vladimir Karpukhin, Naman Goyal, Heinrich K{\"{u}}ttler, Mike Lewis, Wen{-}tau Yih, Tim Rockt{\"{a}}schel, Sebastian Riedel, and Douwe Kiela. 2020.
\newblock \href {https://proceedings.neurips.cc/paper/2020/hash/6b493230205f780e1bc26945df7481e5-Abstract.html} {Retrieval-augmented generation for knowledge-intensive {NLP} tasks}.
\newblock In \emph{Advances in Neural Information Processing Systems 33: Annual Conference on Neural Information Processing Systems 2020, NeurIPS 2020, December 6-12, 2020, virtual}.

\bibitem[{Liang et~al.(2021)Liang, Wu, Morency, and Salakhutdinov}]{DBLP:conf/icml/LiangWMS21}
Paul~Pu Liang, Chiyu Wu, Louis{-}Philippe Morency, and Ruslan Salakhutdinov. 2021.
\newblock \href {http://proceedings.mlr.press/v139/liang21a.html} {Towards understanding and mitigating social biases in language models}.
\newblock In \emph{Proceedings of the 38th International Conference on Machine Learning, {ICML} 2021, 18-24 July 2021, Virtual Event}, volume 139 of \emph{Proceedings of Machine Learning Research}, pages 6565--6576. {PMLR}.

\bibitem[{Lin et~al.(2022)Lin, Liu, Tong, and Xiao}]{DBLP:journals/tkde/LinLTX22}
Chen Lin, Dugang Liu, Hanghang Tong, and Yanghua Xiao. 2022.
\newblock \href {https://doi.org/10.1109/TKDE.2020.3013973} {Spiral of silence and its application in recommender systems}.
\newblock \emph{{IEEE} Trans. Knowl. Data Eng.}, 34(6):2934--2947.

\bibitem[{Liu et~al.(2023{\natexlab{a}})Liu, Li, Shang, Jiang, Liu, Lam, and Wong}]{DBLP:conf/emnlp/LiuLSJLLW23}
Chang Liu, Xiaoguang Li, Lifeng Shang, Xin Jiang, Qun Liu, Edmund~Y. Lam, and Ngai Wong. 2023{\natexlab{a}}.
\newblock \href {https://aclanthology.org/2023.findings-emnlp.961} {Gradually excavating external knowledge for implicit complex question answering}.
\newblock In \emph{Findings of the Association for Computational Linguistics: {EMNLP} 2023, Singapore, December 6-10, 2023}, pages 14405--14417. Association for Computational Linguistics.

\bibitem[{Liu et~al.(2019)Liu, Lin, Zhang, Xiao, and Tong}]{DBLP:conf/wsdm/LiuLZXT19}
Dugang Liu, Chen Lin, Zhilin Zhang, Yanghua Xiao, and Hanghang Tong. 2019.
\newblock \href {https://doi.org/10.1145/3289600.3291003} {Spiral of silence in recommender systems}.
\newblock In \emph{Proceedings of the Twelfth {ACM} International Conference on Web Search and Data Mining, {WSDM} 2019, Melbourne, VIC, Australia, February 11-15, 2019}, pages 222--230. {ACM}.

\bibitem[{Liu et~al.(2023{\natexlab{b}})Liu, Yavuz, Meng, Moorthy, Joty, Xiong, and Zhou}]{DBLP:journals/corr/abs-2308-12574}
Ye~Liu, Semih Yavuz, Rui Meng, Meghana Moorthy, Shafiq Joty, Caiming Xiong, and Yingbo Zhou. 2023{\natexlab{b}}.
\newblock \href {https://doi.org/10.48550/ARXIV.2308.12574} {Exploring the integration strategies of retriever and large language models}.
\newblock \emph{CoRR}, abs/2308.12574.

\bibitem[{Mallen et~al.(2022)Mallen, Asai, Zhong, Das, Hajishirzi, and Khashabi}]{popqa}
Alex Mallen, Akari Asai, Victor Zhong, Rajarshi Das, Hannaneh Hajishirzi, and Daniel Khashabi. 2022.
\newblock \href {https://doi.org/10.48550/ARXIV.2212.10511} {When not to trust language models: Investigating effectiveness and limitations of parametric and non-parametric memories}.
\newblock \emph{CoRR}, abs/2212.10511.

\bibitem[{Noelle-Neumann(1974)}]{noelle1974spiral}
Elisabeth Noelle-Neumann. 1974.
\newblock The spiral of silence a theory of public opinion.
\newblock \emph{Journal of communication}, 24(2):43--51.

\bibitem[{Nogueira et~al.(2020)Nogueira, Jiang, Pradeep, and Lin}]{Monot5}
Rodrigo~Frassetto Nogueira, Zhiying Jiang, Ronak Pradeep, and Jimmy Lin. 2020.
\newblock \href {https://doi.org/10.18653/V1/2020.FINDINGS-EMNLP.63} {Document ranking with a pretrained sequence-to-sequence model}.
\newblock In \emph{Findings of the Association for Computational Linguistics: {EMNLP} 2020, Online Event, 16-20 November 2020}, volume {EMNLP} 2020 of \emph{Findings of {ACL}}, pages 708--718. Association for Computational Linguistics.

\bibitem[{{OpenAI}(2022)}]{openai_chatgpt}
{OpenAI}. 2022.
\newblock {ChatGPT}.
\newblock \url{https://openai.com/blog/chatgpt}.
\newblock Accessed: January 10, 2024.

\bibitem[{OpenAI(2023)}]{DBLP:journals/corr/abs-2303-08774}
OpenAI. 2023.
\newblock \href {https://doi.org/10.48550/arXiv.2303.08774} {{GPT-4} technical report}.
\newblock \emph{CoRR}, abs/2303.08774.

\bibitem[{Pan et~al.(2023)Pan, Pan, Chen, Nakov, Kan, and Wang}]{DBLP:conf/emnlp/PanPCNKW23}
Yikang Pan, Liangming Pan, Wenhu Chen, Preslav Nakov, Min{-}Yen Kan, and William Wang. 2023.
\newblock \href {https://aclanthology.org/2023.findings-emnlp.97} {On the risk of misinformation pollution with large language models}.
\newblock In \emph{Findings of the Association for Computational Linguistics: {EMNLP} 2023, Singapore, December 6-10, 2023}, pages 1389--1403. Association for Computational Linguistics.

\bibitem[{Raffel et~al.(2020)Raffel, Shazeer, Roberts, Lee, Narang, Matena, Zhou, Li, and Liu}]{T5}
Colin Raffel, Noam Shazeer, Adam Roberts, Katherine Lee, Sharan Narang, Michael Matena, Yanqi Zhou, Wei Li, and Peter~J. Liu. 2020.
\newblock \href {http://jmlr.org/papers/v21/20-074.html} {Exploring the limits of transfer learning with a unified text-to-text transformer}.
\newblock \emph{J. Mach. Learn. Res.}, 21:140:1--140:67.

\bibitem[{Ram et~al.(2023)Ram, Levine, Dalmedigos, Muhlgay, Shashua, Leyton{-}Brown, and Shoham}]{DBLP:journals/corr/abs-2302-00083}
Ori Ram, Yoav Levine, Itay Dalmedigos, Dor Muhlgay, Amnon Shashua, Kevin Leyton{-}Brown, and Yoav Shoham. 2023.
\newblock \href {https://doi.org/10.48550/ARXIV.2302.00083} {In-context retrieval-augmented language models}.
\newblock \emph{CoRR}, abs/2302.00083.

\bibitem[{Sachan et~al.(2022)Sachan, Lewis, Joshi, Aghajanyan, Yih, Pineau, and Zettlemoyer}]{UPR}
Devendra~Singh Sachan, Mike Lewis, Mandar Joshi, Armen Aghajanyan, Wen{-}tau Yih, Joelle Pineau, and Luke Zettlemoyer. 2022.
\newblock \href {https://doi.org/10.18653/V1/2022.EMNLP-MAIN.249} {Improving passage retrieval with zero-shot question generation}.
\newblock In \emph{Proceedings of the 2022 Conference on Empirical Methods in Natural Language Processing, {EMNLP} 2022, Abu Dhabi, United Arab Emirates, December 7-11, 2022}, pages 3781--3797. Association for Computational Linguistics.

\bibitem[{Sanh et~al.(2022)Sanh, Webson, Raffel, Bach, Sutawika, Alyafeai, Chaffin, Stiegler, Raja, Dey, Bari, Xu, Thakker, Sharma, Szczechla, Kim, Chhablani, Nayak, Datta, Chang, Jiang, Wang, Manica, Shen, Yong, Pandey, Bawden, Wang, Neeraj, Rozen, Sharma, Santilli, F{\'{e}}vry, Fries, Teehan, Scao, Biderman, Gao, Wolf, and Rush}]{T0}
Victor Sanh, Albert Webson, Colin Raffel, Stephen~H. Bach, Lintang Sutawika, Zaid Alyafeai, Antoine Chaffin, Arnaud Stiegler, Arun Raja, Manan Dey, M~Saiful Bari, Canwen Xu, Urmish Thakker, Shanya~Sharma Sharma, Eliza Szczechla, Taewoon Kim, Gunjan Chhablani, Nihal~V. Nayak, Debajyoti Datta, Jonathan Chang, Mike~Tian{-}Jian Jiang, Han Wang, Matteo Manica, Sheng Shen, Zheng~Xin Yong, Harshit Pandey, Rachel Bawden, Thomas Wang, Trishala Neeraj, Jos Rozen, Abheesht Sharma, Andrea Santilli, Thibault F{\'{e}}vry, Jason~Alan Fries, Ryan Teehan, Teven~Le Scao, Stella Biderman, Leo Gao, Thomas Wolf, and Alexander~M. Rush. 2022.
\newblock \href {https://openreview.net/forum?id=9Vrb9D0WI4} {Multitask prompted training enables zero-shot task generalization}.
\newblock In \emph{The Tenth International Conference on Learning Representations, {ICLR} 2022, Virtual Event, April 25-29, 2022}. OpenReview.net.

\bibitem[{Scheufle and Moy(2000)}]{scheufle2000twenty}
Dietram~A Scheufle and Patricia Moy. 2000.
\newblock Twenty-five years of the spiral of silence: A conceptual review and empirical outlook.
\newblock \emph{International journal of public opinion research}, 12(1):3--28.

\bibitem[{Schick(2020)}]{schick2020deep}
Nina Schick. 2020.
\newblock \emph{Deep Fakes and the Infocalypse: What You Urgently Need To Know}.
\newblock Monoray.

\bibitem[{Shumailov et~al.(2023)Shumailov, Shumaylov, Zhao, Gal, Papernot, and Anderson}]{DBLP:journals/corr/abs-2305-17493}
Ilia Shumailov, Zakhar Shumaylov, Yiren Zhao, Yarin Gal, Nicolas Papernot, and Ross~J. Anderson. 2023.
\newblock \href {https://doi.org/10.48550/ARXIV.2305.17493} {The curse of recursion: Training on generated data makes models forget}.
\newblock \emph{CoRR}, abs/2305.17493.

\bibitem[{Touvron et~al.(2023)Touvron, Lavril, Izacard, Martinet, Lachaux, Lacroix, Rozi{\`{e}}re, Goyal, Hambro, Azhar, Rodriguez, Joulin, Grave, and Lample}]{DBLP:journals/corr/abs-2302-13971}
Hugo Touvron, Thibaut Lavril, Gautier Izacard, Xavier Martinet, Marie{-}Anne Lachaux, Timoth{\'{e}}e Lacroix, Baptiste Rozi{\`{e}}re, Naman Goyal, Eric Hambro, Faisal Azhar, Aur{\'{e}}lien Rodriguez, Armand Joulin, Edouard Grave, and Guillaume Lample. 2023.
\newblock \href {https://doi.org/10.48550/arXiv.2302.13971} {Llama: Open and efficient foundation language models}.
\newblock \emph{CoRR}, abs/2302.13971.

\bibitem[{Trivedi et~al.(2023)Trivedi, Balasubramanian, Khot, and Sabharwal}]{DBLP:conf/acl/TrivediBKS23}
Harsh Trivedi, Niranjan Balasubramanian, Tushar Khot, and Ashish Sabharwal. 2023.
\newblock \href {https://doi.org/10.18653/V1/2023.ACL-LONG.557} {Interleaving retrieval with chain-of-thought reasoning for knowledge-intensive multi-step questions}.
\newblock In \emph{Proceedings of the 61st Annual Meeting of the Association for Computational Linguistics (Volume 1: Long Papers), {ACL} 2023, Toronto, Canada, July 9-14, 2023}, pages 10014--10037. Association for Computational Linguistics.

\bibitem[{Xiao et~al.(2023)Xiao, Liu, Zhang, and Muennighof}]{BGE}
Shitao Xiao, Zheng Liu, Peitian Zhang, and Niklas Muennighof. 2023.
\newblock \href {https://doi.org/10.48550/ARXIV.2309.07597} {C-pack: Packaged resources to advance general chinese embedding}.
\newblock \emph{CoRR}, abs/2309.07597.

\bibitem[{Xu et~al.(2023)Xu, Fan, and Kankanhalli}]{DBLP:conf/mm/XuFK23}
Danni Xu, Shaojing Fan, and Mohan~S. Kankanhalli. 2023.
\newblock \href {https://doi.org/10.1145/3581783.3612704} {Combating misinformation in the era of generative {AI} models}.
\newblock In \emph{Proceedings of the 31st {ACM} International Conference on Multimedia, {MM} 2023, Ottawa, ON, Canada, 29 October 2023- 3 November 2023}, pages 9291--9298. {ACM}.

\bibitem[{Yang et~al.(2023)Yang, Xiao, Wang, Zhang, Bian, Yin, Lv, Pan, Wang, Yan, Yang, Deng, Wang, Liu, Ai, Dong, Zhao, Xu, Sun, Zhang, Liu, Ji, Xie, Dai, Fang, Su, Song, Liu, Ru, Ma, Wang, Liu, Lin, Nie, Guo, Sun, Zhang, Li, Li, Cheng, Chen, Zeng, Wang, Chen, Men, Yu, Pan, Shen, Wang, Li, Jiang, Gao, Zhang, Zhou, and Wu}]{baichuan}
Aiyuan Yang, Bin Xiao, Bingning Wang, Borong Zhang, Ce~Bian, Chao Yin, Chenxu Lv, Da~Pan, Dian Wang, Dong Yan, Fan Yang, Fei Deng, Feng Wang, Feng Liu, Guangwei Ai, Guosheng Dong, Haizhou Zhao, Hang Xu, Haoze Sun, Hongda Zhang, Hui Liu, Jiaming Ji, Jian Xie, Juntao Dai, Kun Fang, Lei Su, Liang Song, Lifeng Liu, Liyun Ru, Luyao Ma, Mang Wang, Mickel Liu, MingAn Lin, Nuolan Nie, Peidong Guo, Ruiyang Sun, Tao Zhang, Tianpeng Li, Tianyu Li, Wei Cheng, Weipeng Chen, Xiangrong Zeng, Xiaochuan Wang, Xiaoxi Chen, Xin Men, Xin Yu, Xuehai Pan, Yanjun Shen, Yiding Wang, Yiyu Li, Youxin Jiang, Yuchen Gao, Yupeng Zhang, Zenan Zhou, and Zhiying Wu. 2023.
\newblock \href {https://doi.org/10.48550/ARXIV.2309.10305} {Baichuan 2: Open large-scale language models}.
\newblock \emph{CoRR}, abs/2309.10305.

\bibitem[{Yu et~al.(2023)Yu, Iter, Wang, Xu, Ju, Sanyal, Zhu, Zeng, and Jiang}]{DBLP:conf/iclr/0002IWXJ000023}
Wenhao Yu, Dan Iter, Shuohang Wang, Yichong Xu, Mingxuan Ju, Soumya Sanyal, Chenguang Zhu, Michael Zeng, and Meng Jiang. 2023.
\newblock \href {https://openreview.net/pdf?id=fB0hRu9GZUS} {Generate rather than retrieve: Large language models are strong context generators}.
\newblock In \emph{The Eleventh International Conference on Learning Representations, {ICLR} 2023, Kigali, Rwanda, May 1-5, 2023}. OpenReview.net.

\bibitem[{Zhang et~al.(2023)Zhang, Xiao, Liu, Dou, and Nie}]{llm-embedder}
Peitian Zhang, Shitao Xiao, Zheng Liu, Zhicheng Dou, and Jian{-}Yun Nie. 2023.
\newblock \href {https://doi.org/10.48550/ARXIV.2310.07554} {Retrieve anything to augment large language models}.
\newblock \emph{CoRR}, abs/2310.07554.

\bibitem[{Zhou et~al.(2023)Zhou, Alon, Xu, Jiang, and Neubig}]{DBLP:conf/iclr/Zhou0XJN23}
Shuyan Zhou, Uri Alon, Frank~F. Xu, Zhengbao Jiang, and Graham Neubig. 2023.
\newblock \href {https://openreview.net/pdf?id=ZTCxT2t2Ru} {Docprompting: Generating code by retrieving the docs}.
\newblock In \emph{The Eleventh International Conference on Learning Representations, {ICLR} 2023, Kigali, Rwanda, May 1-5, 2023}. OpenReview.net.

\bibitem[{Zhu et~al.(2018)Zhu, Lu, Zheng, Guo, Zhang, Wang, and Yu}]{DBLP:conf/sigir/ZhuLZGZWY18}
Yaoming Zhu, Sidi Lu, Lei Zheng, Jiaxian Guo, Weinan Zhang, Jun Wang, and Yong Yu. 2018.
\newblock \href {https://doi.org/10.1145/3209978.3210080} {Texygen: {A} benchmarking platform for text generation models}.
\newblock In \emph{The 41st International {ACM} {SIGIR} Conference on Research {\&} Development in Information Retrieval, {SIGIR} 2018, Ann Arbor, MI, USA, July 08-12, 2018}, pages 1097--1100. {ACM}.

\bibitem[{Zhuo et~al.(2023)Zhuo, Huang, Chen, and Xing}]{DBLP:journals/corr/abs-2301-12867}
Terry~Yue Zhuo, Yujin Huang, Chunyang Chen, and Zhenchang Xing. 2023.
\newblock \href {https://doi.org/10.48550/ARXIV.2301.12867} {Exploring {AI} ethics of chatgpt: {A} diagnostic analysis}.
\newblock \emph{CoRR}, abs/2301.12867.

\end{thebibliography}
